\documentclass[aps,prd,onecolumn, showpacs,nofootinbib,amsmath,groupedaddress]{revtex4}
\usepackage{epsfig}
\usepackage{graphicx}
\usepackage{amsfonts}
\usepackage{latexsym}
\usepackage{amsmath,amssymb}
\usepackage{verbatim}
\usepackage{mathtools}
\usepackage{setspace}
\usepackage{slashed}
\usepackage[all]{xypic}

\usepackage{tikz}
\usepackage[hypertex]{hyperref}

\begin{document}
\title{The holographic spectral function in 
non-equilibrium states}
\author{Souvik Banerjee}
\affiliation{Institute of Physics, Sachivalaya Marg, Bhubaneswar -751 005, India}
\email{souvik@iopb.res.in}

\author{Ramakrishnan Iyer}
\affiliation{Department of Physics and Astronomy, University
of Southern California,
Los Angeles, California 90089-0484, USA}
\email{ramaiyer@usc.edu}

\author{Ayan Mukhopadhyay}
\affiliation{LPTHE,
UPMC -- Paris 6; CNRS UMR
7589, 
Tour 13-14, 4$^{\grave{e}me}$ \'etage, Boite 126, 4 Place
Jussieu, 75252 Paris Cedex 05, France 
}
\email{ayan@lpthe.jussieu.fr}

\date{\today}

\begin{abstract}
{\noindent We develop holographic prescriptions for  obtaining spectral functions in non-equilibrium states and space-time dependent non-equilibrium shifts in the energy and spin of quasi-particle like excitations. We reproduce strongly coupled versions of aspects of non-equilibrium dynamics of Fermi surfaces in Landau's Fermi-liquid theory. We find that the incoming wave boundary condition at the horizon does not suffice to obtain a well-defined perturbative expansion for non-equilibrium observables. Our prescription, based on analysis of regularity at the horizon, allows such a perturbative expansion to be achieved nevertheless and can be precisely formulated in a universal manner independent of the non-equilibrium state, provided the state thermalizes. We also find that the non-equilibrium spectral function furnishes information about the relaxation modes of the system. Along the way, we argue that in a typical non-supersymmetric theory with a gravity dual, there may exist a window of temperature and chemical potential at large $N$, in which a generic non-equilibrium 
state can be characterized by just a finitely few operators with low scaling dimensions, even far away from the hydrodynamic limit.}

\end{abstract}

\maketitle

\tableofcontents

\section{Introduction and an outline of results}

Holography has given us a new paradigm to deal with strongly coupled systems \cite{adscft}. One of the many attractive features of this paradigm is that we can deal with phenomena at strong coupling in real time. 

Though there has been substantial progress in using holography to study hydrodynamics \cite{fluidgravity1, Janik, fluidgravity2, Sayantani} and relaxation of strongly coupled systems \cite{myself1, myself2, myself3}, we still lack a systematic method for studying non-equilibrium Green's functions in holography. The latter turn out to be extremely useful in many applications such as understanding thermalization \footnote{Holographic non-equilibrium Green's functions as an aid for understanding thermalization have been studied earlier in \cite{thermalization} using geodesic approximation, etc.} and obtaining strongly coupled generalizations of quantum kinetic theories, to name a few. The importance of pursuing this direction can be readily illustrated by two examples.

Modeling the space-time evolution of matter formed by ultra-relativistic collisions of heavy ions at RHIC and ALICE is a great theoretical challenge. It is equally challenging to develop reliable methods of inference for deducing this space-time evolution \cite{Florkowski}. Ultimately, it is important to not only understand how the matter thermalizes incredibly fast in time $\leq$ $1$ fm at temperature  about 175 MeV (at RHIC) and subsequently undergoes hydrodynamic expansion, but also how hadrons and resonances are produced and transported in this so-called fireball before finally getting frozen chemically and thermally. Ultimately, we do infer the expansion of the fireball from the emitted hadrons. If the expansion of the RHIC fireball is indeed governed by strongly coupled physics, then we can expect that holography will not only help us in modeling the space-time evolution of the fireball, but also help us improve upon existing techniques like Hanbury-Brown-Twiss pion-interferometry used to deduce the expansion of the fireball. 

Quantum kinetic theories are already being employed to understand the dynamics of the hadron gas after the chemical and thermal freeze-out in the hydrodynamically expanding fireball \cite{Bleicher}. However, in order to understand the details of how the hadron gas comes to existence in the first place and its subsequent freeze-out, as also correlations in the emissions of hadrons, one needs quantum kinetic theories constructed using non-equilibrium Green's functions.  
Therefore, to understand such questions at strong coupling using holography, we need to develop formalism to systematically obtain non-equilibrium Green's functions.

The second example pertains to holographic models of non-Fermi liquids \cite{Lee, Liu, Cubrovic, Faulkner, Hartnoll} \footnote{For interesting holographic models of Fermi liquids see \cite{Sachdev}. Our comments are applicable to such models as well.}. Holography has been successful in reproducing some of the features of ARPES experiments in cuprates and other strongly correlated electron systems - the spectral function has a pole on a momentum shell at zero frequency and also shows non-trivial scaling for low energy excitations.  These results may be interpreted as holographic reproduction of Fermi surfaces different from that in Landau's Fermi liquid theory. In absence of a better way of dealing with strongly interacting fermions at finite density, holographic methods could provide us with useful qualitative insights.

Nevertheless, to test such holographic models, we need to see if we can also reproduce qualitative aspects of non-equilibrium dynamics in strongly interacting fermionic systems. Ultimately, when the electrons are weakly interacting, Landau's Fermi liquid theory gives a unified way of dealing with both equilibrium and non-equilibrium phenomena. It is reasonable to expect that holography can do a similar job at strong coupling. Once again, we need to understand how to obtain quantum kinetic theory from holography, and therefore a systematic method of obtaining non-equilibrium Green's functions.

There are two important issues associated with obtaining non-equilibrium Green's functions in field theory \cite{reviews}.
\begin{enumerate}
\item There is no partition function which plays the role of generating functional of non-equilibrium Green's functions. As we will review briefly later, these are obtained from a generalized effective action. The effective action technique guarantees the full hierarchy is consistently solved and Ward identities are preserved.
\item We cannot use conventional perturbation theory to obtain the behavior in time, like for instance, dependence of observables on hydrodynamic and relaxation modes. This is because usual time-dependent perturbation theory gives us the behavior in time in the form of a Taylor series, which fails to capture late time behavior like exponential decay.
\end{enumerate}
Therefore, even at weak coupling non-equilibrium field theory is hard and typically we need to make educated guesses, depending on the understanding of a specific system. It will be remarkable if, on the strong coupling side, holography can provide us with a good perturbation theory for the non-equilibrium observables we will deal with here. The lack of a generating functional for non-equilibrium correlation functions on the field theory side, nevertheless, makes it hard to use the holographic dictionary to translate such observables to the field theory side.

The observables of primary importance are two-point correlation functions. In the vacuum, once the Euclidean Green's function is specified, we can analytically continue to obtain the Feynman propagator, the retarded and advanced Green's function etc. at equilibrium. At finite temperature too, it thus suffices to know the retarded Green's function, from which we can obtain other propagators like the Feynman propagator. At non-equilibrium the situation is different - we cannot deduce from the retarded Green's function, for instance, the Feynman propagator which will have independent dynamics. Nevertheless, all Green's functions can be expressed in terms of two independent, real observables - the \emph{spectral function} and the \emph{statistical function}, which we briefly review now.

The spectral component (or spectral function) of bosonic Green's functions (in $d$ spatial dimensions) can be defined as the Wigner transform (i.e. the Fourier transform in the relative coordinate $\mathbf{r}$ and time difference $t_r$) of the commutator  
\begin{equation}\label{specb}
\mathcal{A}(\omega, \mathbf{k}, \mathbf{x}, t) =\int d^d r \, dt_{r} \, e^{i(\omega t_r - \mathbf{k}\cdot \mathbf{r})} \Big{\langle}\Big[\Phi\Big(\mathbf{x} + \frac{\mathbf{r}}{2}, t + \frac{t_r}{2}\Big), \Phi\Big(\mathbf{x} - \frac{\mathbf{r}}{2}, t - \frac{t_r}{2}\Big)\Big] \Big{\rangle} .
\end{equation}
Similarly in case of fermionic fields, we can define the spectral component as the Wigner transform of the anti-commutator 
\begin{equation}\label{speccf}
\mathcal{A}(\omega, \mathbf{k}, \mathbf{x}, t) = \int d^d r \, dt_{r} \, e^{i(\omega t_r - \mathbf{k}\cdot \mathbf{r})}\Big{\langle} \Big{\{ }\Psi\Big(\mathbf{x} + \frac{\mathbf{r}}{2}, t + \frac{t_r}{2}\Big), \overline{\Psi}\Big(\mathbf{x} - \frac{\mathbf{r}}{2}, t - \frac{t_r}{2}\Big)\Big{\} } \Big{\rangle}.
\end{equation}
In both the equations above $\langle ....\rangle$ denotes expectation value in a non-equilibrium state. The fermionic spectral function is :
\begin{equation}\label{specf}
A(\omega, \mathbf{k}, \mathbf{x} ,t) = \text{Tr}\Big(\gamma^t \mathcal{A}(\omega, \mathbf{k}, \mathbf{x} ,t)\Big).
\end{equation}

The statistical function (also known as the Keldysh propagator) is defined as the Wigner transform of the anti-commutator of two bosonic fields 
\begin{equation}\label{statb}
G_\mathcal{K}(\omega, \mathbf{k}, \mathbf{x}, t) =-\frac{i}{2} \int d^d r \, dt_{r} \, e^{i(\omega t_r - \mathbf{k}\cdot \mathbf{r})} \Big{\langle}\Big{\{ }\Phi\Big(\mathbf{x} + \frac{\mathbf{r}}{2}, t + \frac{t_r}{2}\Big), \Phi\Big(\mathbf{x} - \frac{\mathbf{r}}{2}, t - \frac{t_r}{2}\Big)\Big{\} } \Big{\rangle} .
\end{equation}
or as the same of the commutator of two fermionic fields 
\begin{equation}\label{statf}
G_\mathcal{K}(\omega, \mathbf{k}, \mathbf{x}, t) =-\frac{i}{2} \int d^d r \, dt_{r} \, e^{i(\omega t_r - \mathbf{k}\cdot \mathbf{r})} \Big{\langle}\Big[ \Psi\Big(\mathbf{x} + \frac{\mathbf{r}}{2}, t + \frac{t_r}{2}\Big), \overline{\Psi}\Big(\mathbf{x} - \frac{\mathbf{r}}{2}, t - \frac{t_r}{2}\Big)\Big] \Big{\rangle} .
\end{equation}

All propagators can be expressed as appropriate linear combinations of the spectral and statistical functions. In this paper, we will be interested in the retarded correlation function in particular. It is actually more convenient to define the Wigner transform of the retarded correlator. In case of bosonic fields, this is defined as
\begin{equation}
G_R (\omega, \mathbf{k}, \mathbf{x}, t) =-i \int d^d r \, dt_{r} \, e^{i(\omega t_r - \mathbf{k}\cdot \mathbf{r})} \theta(t_r)\Big{\langle}\Big[\Phi\Big(\mathbf{x} + \frac{\mathbf{r}}{2}, t + \frac{t_r}{2}\Big), \Phi\Big(\mathbf{x} - \frac{\mathbf{r}}{2}, t - \frac{t_r}{2}\Big)\Big] \Big{\rangle} .
\end{equation}
Similarly for fermionic fields, the anti-commutator is used above. 

It is clear from the definitions of the spectral functions (\ref{specb}) and (\ref{specf}) respectively that the bosonic spectral function is related to the retarded correlator via $\mathcal{A}(\omega, \mathbf{k},\mathbf{x}, t ) = -2 \text{Im} G_R(\omega, \mathbf{k},\mathbf{x}, t )$, while for the fermionic spectral function, the relation is $A(\omega, \mathbf{k},\mathbf{x}, t ) = -2 \text{Im} (\text{Tr}(\gamma^t G_R(\omega, \mathbf{k},\mathbf{x}, t )))$. The retarded correlation function does not contain any more information than the spectral function, since it is analytic in $\omega$ for a given $\mathbf{k}$ for every $\mathbf{x}$ and $t$. Therefore,
\begin{equation}
G_R (\omega, \mathbf{k}, \mathbf{x}, t) = \int \frac{d\omega'}{2\pi}\frac{ \mathcal{A}(\omega', \mathbf{k}, \mathbf{x}, t)}{\omega -\omega' + i\epsilon}
\end{equation}
in both the bosonic and fermionic cases. 

On the other hand the Feynman propagator $G_F$ is a linear combination of both the spectral and statistical components. For both bosonic and fermionic fields, prior to Wigner transform :
\begin{equation}
G_F(\mathbf{x}, t, \mathbf{y}, t') = G_{\mathcal{K}}(\mathbf{x}, t, \mathbf{y}, t')  - \frac{i}{2}\, \mathcal{A}(\mathbf{x}, t, \mathbf{y}, t')\  \text{sign}(t-t').
\end{equation}
Since the Feynman propagator involves the statistical function which is unrelated to the spectral function algebraically out of equilibrium, we cannot deduce this propagator from the retarded function in non-equilibrium states.

At equilibrium, both the spectral and statistical functions depend only on $\omega$ and $\mathbf{k}$, i.e. they are homogeneous in $\mathbf{x}$ and $t$, owing to translational invariance. Furthermore, they are related by fluctuation-dissipation relations :
\begin{equation}
G_{\mathcal{K}}(\omega, \mathbf{k}) = -i\Big(\frac{1}{2} + n_{\text{BE}}(\omega)\Big)\mathcal{A}(\omega, \mathbf{k})
\end{equation}
for the bosonic case and
\begin{equation}
G_{\mathcal{K}}(\omega, \mathbf{k}) = -i\Big(\frac{1}{2} - n_{\text{FD}}(\omega)\Big)\mathcal{A}(\omega, \mathbf{k})
\end{equation}
for the fermionic case, with $n_{\text{BE}}(\omega) = (e^{\beta\omega}-1)^{-1}$ being the Bose-Einstein distribution and $n_{\text{FD}}(\omega) = (e^{\beta\omega}+1)^{-1}$ being the Fermi-Dirac distribution. 

Away from equilibrium, the statistical and spectral functions follow a coupled set of equations which were first found by Kadanoff and Baym \cite{reviews}. These equations are not so easily tractable in field theory even at weak-coupling, however educated guesses lead us to standard kinetic equations like the Boltzmann equation with quantum corrections. We will skip issues involving renormalization etc. and simply mention here that they can be dealt with efficiently at the level of the effective action.

The spectral function, especially for fermions, is directly measurable by ARPES like experiments. Usually it is the equilibrium spectral functions that are measured experimentally, so that we need be concerned with their dependence on frequency and momentum only. Recently however, there have been time-resolved ARPES experiments in which non-equilibrium time-dependent spectral functions have been measured in approximately spatially homogeneous situations and their dependence on frequency, momentum as well as time have been obtained (see, for example, time-resolved ARPES across the metal-insulator transition in \cite{Perfetti}). Conceptually, when integrated over frequency at a given momentum and at a given point in space-time, the spectral function gives the space-time dependent density of states. The spectral function thus reveals the non-equilibrium structure of the effective phase-space of quasi-particles (provided we do have well defined quasi-particles).

The statistical function, on the other hand, carries complementary information about how quasi-particles (whenever they can be defined) are distributed in phase-space and time and can be indirectly measured. For instance, in the case of a single species of fermions, the conserved current is
\begin{equation}
j^\mu (\mathbf{x}, t) = iq \int d\omega d^d k \, \text{Tr}\Big(\mathbf{\gamma}^\mu G_{\mathcal{K}}(\omega, \mathbf{k}, \mathbf{x}, t)\Big) + \ \text{constant}, 
\end{equation}
where $q$ is the conserved charge of the fermionic field, and the constant is independent of the state and required to provide an infinite subtraction which produces a finite result. In the so called quasi-particle approximation, we can assume that the statistical function is peaked only when $\omega$ is on-shell, so that it reduces to the standard phase-space distribution which follows the semi-classical Boltzmann equation in certain limits.

This  completes our very brief review of the spectral and statistical functions respectively. In this paper, we would like to obtain the non-equilibrium retarded function holographically. Our focus will be on the retarded function because we can compute it using linear response theory even in a non-equilibrium state. The holographic dictionary enables defining the source and expectation value of an operator in \emph{any arbitrary state}. Therefore, we can avoid issues associated with the lack of a generating functional for non-equilibrium correlation functions.

To be specific, we would like to achieve the following :
\begin{enumerate}
\item to evaluate the retarded correlation function and the spectral function in non-equilibrium states, 
\item to find space-time dependent shifts in the energy and spins of quasi-particles in the non-equilibrium medium, and
\item to obtain the space-time dependent shifts in energy per particle and spin orientation at the holographic Fermi surface.
\end{enumerate} 
With respect to the last point, we will reproduce a strongly coupled version of what is expected from Landau's Fermi liquid theory, as reviewed later. The second objective is justified on the grounds that it is known that in non-equilibrium states, the effective masses of quasi-particles become space-time dependent (via an inhomogeneous temperature distribution for instance, or an inhomogeneous distribution of the velocity field as discussed later). We will succeed in all these objectives for scalar and fermionic operators.

This paper thus finishes only half of the complete formalism required to obtain quantum kinetic theory from holography. We do not address the information contained in the statistical function and how to obtain it holographically. Work in the latter direction will appear in \cite{progress1}. These issues will be complicated by the fact that we are dealing with composite operators in holography and we leave this for future study. We note here that there has been previous work where the equilibrium statistical function has been defined holographically in a consistent manner \cite{Herzog}, based on the correspondence between the generating functional of field-theoretic correlation functions and a suitable partition function of quantum gravity. However, these cannot be readily generalized to non-equilibrium states because of the lack of a generating functional for non-equilibrium correlation functions as observed before. 

The key result in this paper will be the development of perturbation theory of scalar and fermionic fields in holographic duals of non-equilibrium backgrounds. At equilibrium, the incoming boundary condition mimics causal response in field theory and suffices to define a well-defined linear response theory holographically \cite{incoming1, incoming2}. However, the incoming wave boundary condition does not suffice to give well defined linear response theory in non-equilibrium states. This can be briefly demonstrated as follows. 

Suppose we have a non-equilibrium background in which a hydrodynamic mode with momentum $\mathbf{k}_{\text{(h)}}$ has been excited. Let the source of the operator at equilibrium be $J^{(0)}(\mathbf{x}, t)$ and the expectation value be $O^{(0)}(\mathbf{x},t)$ which can be read-off from the profile of the field $\Phi^{(0)}(r, \mathbf{x}, t)$ in the bulk. The non-equilibrium bulk contribution can be denoted as $\Phi^{(1)}(\mathbf{k}_{\text{(h)}}, r, \mathbf{x}, t)$ and this gives contribution to both the source $J^{(1)}$ and expectation value $O^{(1)}$ of the operator.
The full retarded function can be obtained from :
\begin{equation}\label{rethol}
G_R(\mathbf{x}, t ; \mathbf{y} ,t') = \mathcal{C}\frac{O^{(0)}(\mathbf{x},t) + O^{(1)}(\mathbf{k}_{\text{(h)}}, r, \mathbf{x}, t)}{J^{(0)}(\mathbf{y},t') + J^{(1)}(\mathbf{k}_{\text{(h)}}, r, \mathbf{y}, t')},
\end{equation}
where $\mathcal{C}$ is a constant which depends on the action and has been set to unity here. However, the general solution for $\Phi^{(1)}$ will have : \\
i) two homogeneous solutions which are incoming and outgoing at the horizon respectively and,\\ ii) a particular solution which will be completely determined by the hydrodynamic background perturbation and the equilibrium solution $\Phi^{(0)}$. \\
This particular solution will contribute to both $O^{(1)}$ and $J^{(1)}$, as will the homogeneous solutions. The incoming boundary condition will set the coefficient of the outgoing homogeneous solution to zero. The coefficient of the homogeneous incoming wave solution is left arbitrary. At equilibrium, this arbitrary coefficient cancels between the numerator and denominator, but at non-equilibrium we have an extra coefficient from $\Phi^{(1)}$ and therefore (\ref{rethol}) is ill-defined.  

In this paper, we show that careful treatment of regularity of the solution at the horizon implies that the coefficient of the homogenous incoming solution should also be zero in presence of background quasinormal modes. This will allow us to put forth a well-defined prescription for obtaining the non-equilibrium retarded Green's function and spectral function holographically. In fact, the prescription can be precisely stated in a manner which is independent of the non-equilibrium state. Thus, holography gives a very well-defined perturbation expansion of non-equilibrium observables which can be understood in an universal manner.

The organization of the paper is as follows. In section II, we give a general review of holographic duals of non-equilibrium states. Though most of this section is a review, the explicit metrics for charged hydrodynamics and homogeneous relaxation in section II.D in $AdS_4$ are new as far as we are aware of the literature. The key point in the discussion in section II.B however, to the best of our knowledge, is novel. Here we argue that in a non-supersymmetric theory with a gravity dual, there may exist a window of temperature and chemical potential at large $N$, in which a generic non-equilibrium 
state can be characterized by just a finitely few operators with low scaling dimensions even far away from the hydrodynamic limit. We also point out that there are surprising similarities with solutions of the Boltzmann equation on the weak coupling side, which we review in section II.A. 

In section III, we develop the formalism for obtaining non-equilibrium retarded Green's function and spectral function holographically in the approximation where the background fluctuation is linearized i.e. when the non-equilibrium state is studied in the linearized approximation. An interesting result is that we can read off the relaxation modes in the background by measuring the non-equilibrium spectral function.

In section IV, we compare our holographic approach with field theory. We also make a comparison with Landau's Fermi liquid theory regarding non-equilibrium dynamics of the Fermi surface. Furthermore, we obtain a holographic prescription to calculate space-time dependent non-equilibrium shifts in the energy and spin of the quasi-particles.

In section V, we show that our prescription for the holographic retarded Green's function readily generalizes when we take non-linearities in the dynamics of the variables characterizing the non-equilibrium state into account.

Finally, in section VI, we conclude by pointing out interesting issues that could be addressed numerically.

\section{On non-equilibrium states, their holographic duals and quasi-normal modes}
An equilibrium state can always be characterized by a few macroscopic variables related by an equation of state. The distribution functions of particles, density of states, expectation values of operators, Green's functions, etc. depend on these macroscopic variables. We also know, in principle, how to calculate the equation of state relating the macroscopic variables of equilibrium states. Most importantly, we know in principle how to calculate the dependence of the observables in the underlying field theory on these variables characterizing equilibrium states.

The most pressing problem in dealing with non-equilibrium states is that, typically even at the coarse-grained level, we need an infinite number of macroscopic variables to characterize them. These variables also depend on space and time. Aside from taking recourse to a kinetic approximation, which is typically uncontrolled (but intuitively well-motivated) from the point of view of the exact field theory, we usually do not know how to obtain the equations of motion of these macroscopic variables (thereby generalizing the notion of equation of state applicable at equilibrium). It is also not clear how to relate observables in the field theory to the macroscopic coarse-grained  non-equilibrium variables.

Here, we will address these issues from the point of view of holography. Firstly, we will identify a special sector of non-equilibrium states which can be described in terms of a finite number of operators of low scaling dimensions in kinetic theories. These states exist for any value of the coupling at least in the kinetic approximation. Then we will argue holographically that these states also exist in the exact field theory and are generic at strong coupling and large $N$ after a microscopic time-scale, irrespective of the initial condition. We will further discuss how solutions in gravity describe such non-equilibrium states.

\subsection{Conservative states in the kinetic approximation}

Let us first look at the kinetic approximation in some details. In particular let us analyze the Boltzmann limit which is valid typically when, $n l_{mfp}^d$  is small, where $n$ is the typical number density, $l_{mfp}$ is the mean free path and $d$ is the number of spatial dimensions.

Boltzmann equation describes the dynamics of particle-distributions in phase space. It can be reduced to local dynamics of the infinite number of moments of the phase-space distribution of particles $f^{(s)}(\mathbf{x},\mathbf{p},t)$ of a given species $s$. These moments are 
\begin{equation}
f^{(s)}_{\mu_1 \mu_2 ... \mu_n} (\mathbf{x},t)= \int \frac{d^d p}{p^{(s)0}}
\ p_{\mu_1}p_{\mu_2}....p_{\mu_n}f^{(s)}(\mathbf{x},\mathbf{p},t)\end{equation}
where $p^\mu$ is the $d+1$-momentum with $p^{0}$ being on-shell energy for each species $s$.

A conserved current (for instance the baryon number current) is given by :
\begin{equation}
j_{\mu}(\mathbf{x}, t) = \sum_s q_s  \int \frac{d^d p}{p^{(s)0}}
\ p_{\mu} f^{(s)}(\mathbf{x},\mathbf{p},t),
\end{equation}
where $q_s$ is the charge (for instance baryon charge) of the $s-$th species.

The energy-momentum tensor is given by
\begin{equation}
t_{\mu\nu}(\mathbf{x}, t) = \sum_s  \int \frac{d^d p}{p^{(s)0}}
\ p_{\mu}p_{\nu} f^{(s)}(\mathbf{x},\mathbf{p},t).
\end{equation}
Thus we see that the energy-momentum tensor and conserved currents are parametrized by a weighted sum of first few moments of the quasi-particle distribution functions.

Three comments are in order here :
\begin{enumerate}
\item The Boltzmann equation has no dependence on temperature or non-equilibrium parameters. The latter parametrize the solutions. The thermal Bose-Einstein or Fermi-Dirac distributions are exact solutions of the Boltzmann equation. In absence of external fields, Boltzmann's H-theorem indicates all solutions finally equilibrate into thermal Bose-Einstein or Fermi-Dirac distribution.

\item The integrals involved in collision terms on the right hand side of the Boltzmann equation (see eq. (\ref{bolt}) for weakly interacting electrons) have divergences coming from phase-space volume. To regulate these divergences one can put a IR-cutoff corrsponding to the thermal mass of the quarks and gluons with temperature being the final equilibrium temperature \cite{Arnold}. The dispersion relations are also accordingly modified.

\item In the dilute limit the effect of the interactions is taken into account via an effective thermal mass. Thus the energy-momentum tensor takes a free particle form with an effective thermal mass.
\end{enumerate}

It can be shown that the higher velocity moments parametrize the flow of the flow, the flow of the flow of the flow, etc. of charge, energy and momentum. For example, if we define :
\begin{equation}
S_{\mu\nu\rho}(\mathbf{x}, t) = \sum_s   \int \frac{d^d p}{p^{(s)0}}
\ p_{\mu}p_{\nu}p_\rho f^{(s)}(\mathbf{x},\mathbf{p},t),
\end{equation}
then the heat-current is $S_\mu = S_{\mu\nu\rho}\eta^{\nu\rho}$.

The Boltzmann equation can have solutions where the partial conserved currents are $j^{(s)\mu}$ are all proportional to each other. This happens precisely when chemical equilibrium is achieved, and in fact any arbitrary solution achieves chemical equilibrium after sufficiently long time. In that case, we can define a four-velocity field $u^\mu$ and charge density $\rho$ such that :
\begin{equation}
j^{(s)}_\mu = \rho^{(s)} u_\mu, \quad \rho = \sum_s \rho^{(s)}, \quad j_\mu = \sum_s j^{(s)}_\mu = \rho u_\mu.
\end{equation}
The energy-density $\epsilon$ is :
\begin{equation}
\epsilon = t_{\mu\nu}u^\mu u^\nu.
\end{equation}
The hydrodynamic variables are $\epsilon$, $\rho$ and $u^\mu$. We can define temperature $T$ and chemical potential $\mu$ fields in terms of $\epsilon$ and $\rho$ by using the equation of state of the full system at thermal and chemical equilibrium locally.

There are special solutions of the full non-linear Boltzmann equation, known as \textit{normal solutions} in the literature, which are purely hydrodynamic \cite{Chapman}. These solutions are such that all the moments $f^{(s)}_{\mu_1 .... \mu_n}$ of the phase-space quasi-particle distributions of various species are algebraic functions of just the hydrodynamic variables $u_\mu$, $T$ and $\mu$, and their \textit{spatial} derivatives in the local inertial frame co-moving with $u^\mu$. The full phase-space distributions can thus be characterized uniquely by the hydrodynamic variables. Furthermore, any arbitrary solution of the Boltzmann equation can be approximated by an appropriate normal solution after a sufficiently long time. 

The hydrodynamic equations can be derived from the Boltzmann equation; these are the Navier-Stokes equation, charge diffusion equation and Fourier's law of energy transport with systematic higher derivative corrections. The shear viscosity, charge diffusion constant, thermal conductivity and all the higher order transport coefficients can be obtained from the relevant Boltzmann equation specified by the dominant collision processes.

These solutions can be further generalized to what were named \textit{conservative solutions} \cite{myself1}. In such solutions, the various moments $f^{(s)}_{\mu_1 .... \mu_n}$ are algebraic functionals of $\rho$, $u_\mu$ (or equivalently the conserved current $j_\mu$) and the energy-momentum tensor $t_{\mu\nu}$, and their spatial derivatives in a local inertial frame co-moving with $u^\mu$. Thus the full solution can be specified by $t_{\mu\nu}$ and $j_\mu$. In such solutions the energy-momentum tensor is not necessarily hydrodynamic. Furthermore, any solution of the Boltzmann equation reduces to an appropriate \textit{conservative solution} after sufficiently long time, and the latter reduces to an appropriate  \textit{normal solution} after the relaxational time scale. The first claim follows from the fact that the independent dynamical parts of higher moments of the quasi-particle distributions decay faster compared to the non-hydrodynamic relaxational mode of the energy-momentum tensor \cite{Grad}. 

The energy-momentum tensor $t_{\mu\nu}$ and the conserved current $j_\mu$ (or equivalently the charge density $\rho$ and the velocity $u_\mu$) follow a closed system of equations in conservative solutions of the Boltzmann equation. This gives a systematic generalization of phenomenology beyond hydrodynamics to include processes like relaxation. These phenomenological equations have been obtained in \cite{myself1, myself2}.

Obviously, the existence of normal and conservative solutions of the Boltzmann equation can be seen at the linearized level and provides a method to obtain good approximations to the transport coefficients and relaxation parameters.

Thus, \textit{in the semi-classical kinetic limit captured by the Boltzmann equation, an arbitrary non-equilibrium state can be approximated by a conservative state whose dynamics is given by the conserved current and the energy-momentum tensor even away from the hydrodynamic limit.} This approximation is reliable after a microscopic time-scale which is shorter than the leading non-hydrodynamic relaxation mode, i.e. the time scale of local thermalization. 

The quasi-particle distribution is said to have locally thermalized when it can be characterized well by space-time dependent parameters of equilibrium distribution. Afterwards, hydrodynamics takes over and the system equilibrates globally.  In a generic solution of the Boltzmann equation, we thus have three time scales. The first time-scale is the time for chemical equilibration $t_{chem}$ after which inelastic collisions effectively cease, the second time scale is  $t_{cons}$ after which an approximation by an appropriate conservative solution becomes valid, and the third time scale is after which the hydrodynamic approximation is valid and is also the time scale of thermalization $t_{therm}$. The hierarchy is
\begin{equation*}
t_{chem} < t_{cons} < t_{therm}.
\end{equation*}
The conservative solutions of Boltzmann equation describe the dynamics of both thermalization and hydrodynamics in an unified framework in the Boltzmann limit.

We note that there is no scale which parametrically separates the dynamics of the non-hydrodynamic part of the energy-momentum tensor and conserved currents from that of other relaxation modes. Thus we may argue that even if conservative states exist beyond the Boltzmann limit, they may not be typical non-equilibrium states after microscopic times as in the Boltzmann equation. The typicality is just a special feature of the Boltzmann limit.

In fact, once we go away from the dilute limit necessary for the Boltzmann equation to be reliable or consider genuine quantum dynamics (not just quantum statistics), the typicality of conservative states will no longer be preserved. The conserved currents and energy-momentum tensor do not seem to capture generic dynamics beyond the hydrodynamic limit. Conservative solutions may exist beyond the Boltzmann approximation, but only in the purely hydrodynamic limit can they approximate a generic state. 

We will argue that if a theory has a holographic dual, then in certain phases in the large $N$ limit, the dynamics can indeed be captured by just the conserved current and energy-momentum tensor generically, after a microscopic time-scale which is much shorter than the time-scale for local thermalization. In such cases, the conservative state can indeed capture generic non-equilibrium dynamics even far away from the hydrodynamic limit.

\subsection{Holographic duals of non-equilibrium states and typicality at strong-coupling}

Holography maps a field theory to a quantum theory of gravity in one extra spatial dimension. It further states that in the large $N$ and strongly coupled limit, the dual theory of gravity reduces to a classical theory. Therefore, in this limit states of the field theory are dual to solutions of the classical theory of gravity which are regular in an appropriate sense. Furthermore, every operator is dual to a field and the expectation value of an operator in a state can be obtained from the asymptotic behavior of the dual field in the corresponding gravity solution.
 
The question of which operators matter in characterizing states in the large $N$ and strong coupling limit can be seen from the masses of the dual fields. The mass of the field is related to the scaling dimension of the dual operator. 

The large $N$ limit in the ($D$ dimensional) field theory side is the limit when the scale $l$, corresponding to asymptotic curvature radius of the ($D+1$ dimensional) space-time, is large compared to the effective Planck scale $l_P$ (in $D+1$ dimensions) on the quantum gravity (string theory) side of the holographic correspondence. The strong coupling limit on the field-theory side is the limit when the length of the fundamental string $l_s$ is small compared to the asymptotic curvature radius $l$ on the quantum gravity side. The first condition $l_P << l$ allows us to consider the classical limit of gravity. The second condition $l_s << l$ allows us to ignore the massive stringy fields corresponding to higher excitations of the fundamental string.

Nevertheless, string theory is a theory in 10 dimensions. So, there has to be a compact space of $9-D$  dimensions on top of the $D+1$ dimensional non-compact coordinates. The condiitons $l_s << l$ and $l_P << l$, i.e. strong coupling  and large $N$ limit in the field theory side allows us to decouple the massive stringy modes whose masses scale like $l_s^{-1}$ when $l_s$ and $l_P$ are small compared to $l$. Thus from the ten-dimensional viewpoint we are left with just the massless fields which include the graviton and gauge fields. However, the compactification over the compact $9-D$ dimensions still creates a tower of Kaluza-Klein fields which are dual to operators with possibly small scaling dimensions if the typical size of the compact dimensions is of the same order as the asymptotic curvature radius $l$.

In a supersymmetric set-up \cite{Aharony}, the typical radius of the $9-D$ dimensional compact space is indeed of the same order as the $D+1$ dimensional asymptotic curvature radius $l$. Therefore, in the strong coupling and large $N$ field-theoretic limit, the Kaluza-Klein spectrum still plays a role in characterizing states. In fact, these Kaluza-Klein fields are dual to chiral primary operators and their descendants. Therefore, a prediction of the holographic correspondence is that at large $N$ the scaling dimensions of the chiral primary operators do not deviate much from the weak coupling limit. 

Despite the presence of the Kaluza-Klein spectrum, it is known that almost all known supergravity theories in 10 dimensions admit consistent truncation at the classical level to gauged supergravity in $D+1$ dimensions when dimensionally reduced over the appropriate $9-D$ dimensional compact space. The $D+1$ dimensional graviton is dual to the energy-momentum operator on the field-theory side and the $D+1$ dimensional gauge fields are dual to the conserved currents with the global symmetry groups being gauged in the gravity side.

One can also show that all solutions of $D+1$ dimensional gauged supergravities which thermalize to black branes with regular future horizons can be characterized uniquely by the expectation values of the energy-momentum tensor and conserved currents of the dual states \footnote{Despite these not being Cauchy data from the gravity point of view, this holds if the geometry corresponds to regular perturbations of a black brane at late time \cite{myself4}. We also note that the consistent truncation to pure gravity does not involve separation of scales. This simply reflects the fact that the conservative states are not typical states in these examples.}. 
These solutions thus correspond to special non-equilibrium states - namely the strongly coupled version of the conservative states which can be characterized by the energy-momentum tensor and conserved currents alone. The parameters of phenomenological equations for the energy-momentum tensor and conserved currents which generalize hydrodynamics should now be obtained from gravity and not from kinetic theories valid at weak coupling \cite{myself1, myself2, myself3}.
Evidence that the solutions of pure gravity in particular, which have regular future horizons, can be interpreted as conservative states has been presented in \cite{myself2} for the special case of homogeneous relaxation. It has been proved that regularity at the horizon gives an equation of motion for the non-hydrodynamic energy-momentum tensor with precise coefficients for this case.

Furthermore, such conservative states should also exist holographically away from the strong coupling and large N limit, since the dual solutions in gravity can be constructed by perturbatively correcting the gauged supergravity solutions in $l_s^2/l^2$ and $1/N^2$. Nevertheless, in the known supersymmetric cases these solutions are always special and not typical even in the strong coupling and large $N$ limit, because the intrinsic dynamics of Kaluza-Klein modes are absent in these solutions.

The situation can be expected to be very different in non-supersymmetric cases. There is no analogue of chiral primary operators and typically we do not expect that quantum corrections to scaling dimensions of operators will be small at strong coupling, unless these are suppressed because of symmetries.

In order to use our intuition obtained from well studied examples with the field theory being conformal and supersymmetric, we will need to focus only on a certain window of temperatures and chemical potentials, such that :
\begin{enumerate}
\item the effective coupling is strong,
\item the beta function is vanishing or approximately so, i.e. the system is close to a critical point, and
\item there are no new emergent symmetries at the critical point other than the (exact or approximate) full conformal symmetry.
\end{enumerate}
 Furthermore, we also require that the large $N$ approximation is valid, or useful for qualitative understanding. 
Probably, all these requirements could be satisfied for the fireball at RHIC near temperatures of 175 MeV and small baryon charge densities as supported by lattice data \cite{Gavai}. 
We can also hope that the strange metallic phase of strongly correlated electron systems will satisfy these requirements in a window of temperatures and chemical potentials.

We note that certain examples of non-supersymmetric holography have been proposed in the literature \cite{nonsusyhol}. However, in these special examples, infinite number of gauge symmetries appear in the bulk at large $N$, implying infinite number of global symmetries in the dual field theory. Our observations below will not be necessarily true in such cases \footnote{The examples \cite{nonsusyhol} are also not stringy and so far well defined only in the large $N$ limit, i.e. only when the theory of gravity is classical.}.

In case of a typical non-supersymmetric theory with a gravity dual, at temperatures and chemical potentials such that the system is close to a strongly coupled critical point, we expect there will be a few operators whose scaling dimensions will be small. We observe that the scaling dimensions depend on the  scale through the coupling and hence also on the phase of the theory being considered which is parametrized by the temperature and chemical potential. The relevant operators with small scaling dimensions in the window of temperature and chemical potentials considered here can be expected to be  
\begin{enumerate}
\item the energy-momentum tensor,
\item the conserved currents, and
\item order parameters of spontaneous symmetry breaking.
\end{enumerate}
Therefore, the operators dual to the Kaluza-Klein modes of gravity are expected to have large scaling dimensions very simlar to those dual to the stringy modes. If this expectation is true, the typical scale of the compact dimensions should be of the same order as $l_s$ and not $l$.

For instance, in the case of QCD, the relevant operators with small scaling dimensions in the conditions of RHIC can be  expected to be 
\begin{enumerate}
\item energy-momentum tensor,  
\item the baryon number current,  \item the approximately conserved $SU(3)_L \times SU(3)_R$ flavor symmetry of the light quarks, and 
\item the order parameter of chiral symmetry breaking having zero baryon number, transforming as $(3_L, 3_R)$ under the flavor symmetry group and with scaling dimension approximately $3$. 
\end{enumerate}
The dual massless fields on the gravity side should be 
\begin{enumerate}
\item the graviton, 
\item a U(1) abelian gauge field,
\item $SU(3)_L \times SU(3)_R$ non-Abelian gauge fields, and
\item a neutral scalar field transforming in the $(3_L , 3_R)$ representation of the non-Abelian gauge group and with mass approximately given by $m^2 = - 3/l^2$ \footnote{As the chiral symmetry breaking order parameter is $\langle \overline{q}^i q^j\rangle$, it has approximate mass dimension of $3$. Moreover, QCD being asymptotically free, the dual boundary condition will be approximately $AdS_5$-like as well. Then we can use the standard relation for $AdS_5$ for mass of the field $m$ and the scaling dimension of the dual operator $\Delta$ which gives $m^2 = -3/l^2$  when $\Delta = 3$.}. 
\end{enumerate}
Such a holographic model for QCD has already been proposed in \cite{hardwall}. However, our arguments above show that such models can be considered more seriously in the conditions of RHIC. In fact, for RHIC conditions we also do not need the hardwall cut-off proposed in these models to achieve confinement, as the mass gap is expected to become very mild at temperatures close to 175 MeV and for small baryon number densities.

Furthermore, if the temperature is higher than 125 MeV, chiral symmetry is expected to be restored, so that the profile of the bulk scalar field dual to the chiral symmetry breaking order parameter will be stabilized by a potential. Therefore, only the conserved currents and energy-momentum tensor can characterize non-equilibrium dynamics at large $N$ and large 't Hooft coupling $\lambda$ for temperatures above 125 MeV. The other fields in the holographic dual should have masses which grow like $1/l_s$ i.e. $1/\lambda^{\frac{1}{4}}$, and thus are expected to be effectively decoupled from the classical theory. 

The correlation functions of the non-Abelian gauge fields in the gravity backgrounds which thermalize to a black brane are all we need to construct quantum kinetic theories of production and freeze-out of axial and vector mesons (and resonances) in the expanding fireball holographically. The interpretation of poles of correlation functions of these gauge fields in terms of mesons has been given in \cite{hardwall}. Using the methods to be described later, we can obtain the non-equilibrium corrections to these mesonic poles systematically. 

Let us estimate the relevant time scale at strong coupling after which the conservative solutions become relevant. This in the dual gravity description is given by the mass of the lightest stringy field or Kaluza-Klein mode. According to the discussion above, the time scale should be $O(\lambda^{-\frac{1}{4}})$ in a non-susy conformal theory at strong coupling. After such a time-scale, we may expect that the massive fields in gravity will decay and the relevant dynamics will be described by the metric, gauge fields and the light fields dual to order parameters of symmetry breaking relevant at the critical point. Thus decay of a massive field in gravity can be interpreted as transition to a conservative state at strong coupling where the dynamics is governed by the energy-momentum tensor, conserved currents and order parameters alone.

We conclude \textit{in a typical non-supersymmetric theory which has a holographic dual, in a window of temperature and chemical potentials such that the dynamics is strongly coupled and approximately conformal, all non-equilibrium states can be characterized by just the energy-momentum tensor and conserved currents (and order parameters of spontaneous symmetry breaking if any), irrespective of the initial conditions, after a microscopic time-scale which scales with the coupling $\lambda$ like $1/\lambda^{\frac{1}{4}}$ in the large $N$ limit}. In other words, conservative states are typical states irrespective of the initial conditions after a microscopic time-scale much smaller than the time-scale of thermalization in the strongly coupled and nearly conformal phase at large $N$. 

If the above arguments are indeed relevant for QCD and strange metals in a window of temperature and chemical potentials, we have a unique 
opportunity to understand non-equilibrium dynamics with only a finitely few operators in this special phase of these theories. As conservative states will be typical non-equilibrium states, we can use general phenomenological equations for non-equilibrium dynamics as proposed in \cite{myself1, myself2}, and also hope to construct a general theory of kinetics and fluctuations to connect to experiments as we want to do here and more completely in the future. 

If the above arguments fail, the reasons should certainly be deep. In that case, we also need to know how to generalize non-equilibrium holography beyond the sector of conservative states sufficiently so that we can describe a typical non-equilbrium state.

\subsection{Quasinormal modes}

The thermal states in the field theory at large $N$ and strong coupling are captured by black brane solutions of classical gravity holographically. In the linearized limit, the non-equilibrium fluctuations are captured by the linearized equations of motion of gauge field and the metric fluctuations about the black brane background. These fluctuations are dual to perturbations of the energy-momentum tensor and conserved currents about thermal equilibrium. Furthermore, these fluctuations should satisfy the incoming boundary condition at the horizon and Dirichlet boundary condition asymptotically \cite{incoming1}. Thus they are quasinormal modes capturing intrinsic fluctuations in the dual field theory which can exist in absence of sources and provide good approximation to a typical non-equilibrium state close to equilibrium at strong coupling and large $N$.

There is, however, a significant difference between the linearized Boltzmann limit and the quasinormal mode approximation of solutions of gravity. Instead of a finitely few decay modes on top of the hydrodynamic mode, we have an infinite tower of quasi-normal modes. The reason that we do not have an infinite tower of modes for the energy-momentum tensor perturbations in the Boltzmann equation
is that it has only one time derivative (which in a Lorentz-invariant language is the derivative along the local velocity field).  Quantum corrections to the Boltzmann equation are known to result in an infinite number of time derivatives, and it is not hard to see this will produce an infinite number of decay modes as well.

We will now obtain the phenomenological form of the non-equilibrium energy-momentum tensor and conserved cuurent. Instead of stating in a Lorentz-invariant way, we will state the form of the energy-momentum tensor in the frame where the dual thermal state is at rest, i.e. the laboratory frame. It is convenient to define the velocity perturbation $\delta \mathbf{u}(\mathbf{x},t)$ such that the velocity field is co-moving with the energy-flow, instead of the charge-flow as done usually in the Boltzmann limit.  Thus the non-equilibrium energy-momentum tensor thus takes the Landau-Lifshitz form in the global co-moving frame :
\begin{eqnarray}
t_{00} &=& \epsilon(T, \mu) + \frac{\partial\epsilon(T, \mu)}{\partial T}\delta T(\mathbf{x}, t) + \frac{\partial\epsilon(T, \mu)}{\partial \mu}\delta \mu(\mathbf{x}, t),\nonumber\\
t_{0i} &=& t_{i0} = \Big(\epsilon(T, \mu) +p(T, \mu)\Big)\delta u_i(\mathbf{x},t),\nonumber\\
t_{ij} &=& p(T, \mu) \delta_{ij}+\Big( \frac{\partial p(T, \mu)}{\partial T}\delta T(\mathbf{x}, t)+\frac{\partial p(T, \mu)}{\partial \mu}\delta \mu(\mathbf{x}, t)\Big) \delta_{ij} + \pi_{ij}(\mathbf{x}, t).
\end{eqnarray}
Above $p$ is the pressure and $\pi_{ij}$ is the shear-stress tensor. The shear-stress tensor can thus be defined as the dissipative part of the energy-momentum tensor or the spatial components of the energy-momentum tensor not in local equilibrium in the co-moving frame. The conserved current takes the form :
\begin{eqnarray}
j_0 &=& \rho (T, \mu) + \frac{\partial\rho(T, \mu)}{\partial T}\delta T(\mathbf{x}, t) + \frac{\partial\rho(T, \mu)}{\partial \mu}\delta \mu(\mathbf{x}, t) + \nu_0(\mathbf{x},t),\nonumber\\
j_i &=& \rho (T, \mu) \delta u_i (\mathbf{x}, t) + \nu_i (\mathbf{x}, t).
\end{eqnarray} 
Above $\nu_i$ is the dissipative part of the consevred current or the spatial components of the current away from local equilibrium in the co-moving frame. However, as the co-moving frame is aligned with the energy flow, the charge can have a non-equilibrium part by itself. This is $\nu_0$.

In order to have conformal invariance, we should further have
\begin{equation}
\epsilon(T, \mu) = d \ p(T, \mu), \quad \delta\epsilon = d \ \delta p, \quad\pi_{ij}\delta_{ij} = 0,
\end{equation}
with $d$ being the number of spatial dimensions in the field theory. Above $\delta\epsilon$ and $\delta p$ denote change in energy density and pressure due to change in temperature and chemical potential. From now onwards, we will be interested in the specific case when the field theory is conformal, so that on the gravity side we will be using asymptotically $AdS$ boundary conditions.

The shear-stress tensor and the dissipative part of the current can be split into  hydrodynamic parts $\pi_{ij}^{\text{(h)}}$ and $\nu_i^{\text{(h)}}$ respectively which are functions of the hydrodynamic fields $\delta T$ and $\delta \mathbf{u}$, and  non-hydrodynamic parts $\pi_{ij}^{\text{(nh)}}$ and $\nu_{i}^{\text{(nh)}}$ respectively which cannot be parametrized by hydrodynamic variables alone. On the other hand, $\nu_0$ does not have any purely hydrodynamic part. 

In the case of a conformal field theory, at the linearized level,
\begin{eqnarray}\label{phentj}
\pi_{ij} &=& \pi_{ij}^{\text{(h)}} + \pi_{ij}^{\text{(nh)}}, \quad \nu_i = \nu_{i}^{\text{(h)}} + \nu_{i}^{\text{(nh)}}, \nonumber\\
\pi_{ij}^{\text{(h)}} &=& - \eta(T, \mu)\Big(\partial_{i}\delta u_j +\partial_j \delta u_i - \frac{2}{d}(\partial\cdot \delta u)\delta_{ij}\Big) + ... , \nonumber\\
\nu_i^{\text{(h)}} &=& - \mathcal{D}(T, \mu) \Big(\frac{\partial\rho(T, \mu)}{\partial T}\partial_i \delta T + \frac{\partial\rho(T, \mu)}{\partial \mu}\partial_i \delta \mu\Big) + ..., \nonumber\\
\pi_{ij}^{\text{(nh)}} &=& \sum_{n=1}^{\infty}\ a_{(n)ij} \ e^{i(\mathbf{k}\cdot\mathbf{x}-\omega_{(n)}(\mathbf{k})t)}, \ \ \text{with $a_{(n)ij} \ \delta_{ij} = 0$ for all $n$, }\nonumber\\
\nu_{i}^{\text{(nh)}} &=& \sum_{n=1}^{\infty}\ b_{(n)i} \ e^{i(\mathbf{k}\cdot\mathbf{x}-\tilde{\omega}_{(n)}(\mathbf{k})t)},\nonumber\\
\nu_{0}^{\text{(nh)}} &=& \sum_{n=1}^{\infty}\ c_{(n)} \ e^{i(\mathbf{k}\cdot\mathbf{x}-\check{\omega}_{(n)}(\mathbf{k})t)}.
\end{eqnarray}
Above, $\pi_{ij}^{\text{(h)}}$ and $\nu_{i}^{\text{(h)}}$ have been expanded in the derivative expansion, which is an expansion in the scale of variation of hydrodynamic variables over the mean free path. We also require $\delta u_i$ and $\delta T$ to be small uniformly for the linearized approximation to be valid. Furthermore, $\eta$ is the shear viscosity and $\mathcal{D}$ is the charge diffusion constant. On the other hand $a_{(n)ij}$, $b_{(n)i}$  and $c_{(n)}$ parametrize the dissipative non-hydrodynamic modes of the energy-momentum tensor and conserved current. The $n$ here represents the various non-hydrodynamic branches of quasinormal mode perturbations which dissipate because their dispersion relations $\omega_{(n)}(\mathbf{k})$, $\tilde{\omega}_{(n)}(\mathbf{k})$ and $\check{\omega}_{(n)}(\mathbf{k})$ have negative imaginary parts. We require $a_{(n)ij}/p$, $b_{(n)i}/\rho$ and $c_{(n)}/\rho$ to be small for the linearized approximation to be valid.

We note the separation of $\pi_{ij}$ and $\nu_i$ into hydrodynamic and non-hydrodynamic parts can also be done at the non-linear level. This is so because even at the non-linear level the hydrodynamic parts $\pi_{ij}^{\text{(h)}}$ and $\nu_i^{\text{(h)}}$ are solutions by themselves - from the perspective of kinetic theories this follows from existence of \textit{normal solutions} as discussed before and from the point of view of gravity they give regular metrics via fluid/gravity correspondence. For any $\pi_{ij}$ and $\nu_i$, the non-hydrodynamic parts $\pi_{ij}^{\text{(nh)}}$ and $\nu_{i}^{\text{(nh)}}$ are just whatever remains after subtracting out the purely hydrodynamic parts $\pi_{ij}^{\text{(h)}}$ and $\nu_i^{\text{(h)}}$ constructed algebraically from the profile of the hydrodynamic variables in the full solution of the energy-momentum tensor and conserved currents. 

In order to obtain the hydrodynamic modes at the linearized level, we simply put all $a_{(n)ij}$ and $b_{(n)i}$ to zero in (\ref{phentj}) and impose the conservation of energy, momentum and charge : 
\begin{equation}
\partial^\mu t_{\mu\nu} = 0, \quad \partial^\mu j_\mu = 0.
\end{equation}
We then obtain three modes, the sound mode, the shear mode and the charge diffusion mode. In the sound mode, 
\begin{eqnarray}
\delta \mathbf{u}(\mathbf{k}) \ &\text{is parallel to}& \ \mathbf{k}, \nonumber\\
\omega &=& \pm \frac{1}{\sqrt{d}} \mid \mathbf{k}\mid  - i \Big(\frac{d-1}{d}\Big)\frac{\eta(T, \mu)}{\epsilon(T, \mu) + p(T, \mu)}\mid \mathbf{k} \mid^2 + ... , \nonumber\\
\delta \epsilon(\mathbf{k}) &= d \,\delta p(\mathbf{k}) =&  \pm \sqrt{d} \mid \delta\mathbf{u}(\mathbf{k})\mid  \Big(\epsilon(T,\mu)+ p(T,\mu)\Big)+ ...,\nonumber\\ \frac{\delta \rho(\mathbf{k})}{\rho} &=& \frac{\delta\epsilon(\mathbf{k})}{\epsilon(T, \mu) + p(T,\mu)} + ... \ .
\end{eqnarray}
Above $(...)$ refers to higher derivative corrections in powers of $\mathbf{k}$. Using thermodynamic relations locally, one can obtain $\delta T(\mathbf{k})$ and $\delta \mu (\mathbf{k})$ from $\delta\epsilon (\mathbf{k})$ and $\delta\rho (\mathbf{k})$.

In the shear mode, 
\begin{eqnarray}\label{dis1}
\delta \mathbf{u}(\mathbf{k}) \ &\text{is orthogonal to}& \ \mathbf{k}, \nonumber\\
\omega &=& - i \frac{\eta(T) }{\epsilon(T) + p(T)} \mid \mathbf{k}\mid^2 + .... \ ,\nonumber\\
\delta \epsilon(\mathbf{k}) &= \delta p(\mathbf{k}) = \delta \rho(\mathbf{k}) = & 0.
\end{eqnarray}

In the charge-diffusion mode
\begin{eqnarray}
\delta\epsilon(\mathbf{k}) =0, \quad \delta p(\mathbf{k}) = 0, \quad \delta \mathbf{u}_i(\mathbf{k}) = 0, \nonumber\\
\omega = - i \mathcal{D}(T,\mu)\mid\mathbf{k}\mid^2.
\end{eqnarray}

The quasinormal modes of the metric and gauge fields contains these hydrodynamic modes as the only branches in which $\omega$ and $\mathbf{k}$ can go simultaneously to zero. We can also obtain the transport coefficients by using the incoming boundary condition at the horizon. We will be interested in the shear mode in particular. The shear-viscosity is given by \cite{fluidgravity1}:
\begin{equation}\label{dis2}
\frac{\eta(T, \mu)}{s(T, \mu)} = \frac{T\eta(T, \mu)}{\epsilon(T, \mu) + p(T, \mu)} = \frac{1}{4\pi}.
\end{equation}
Above, $s$ is the entropy density and we have used the thermodynamic identity $s = (\epsilon +p)/T$.

In order to obtain the simplest non-hydrodynamic modes we need to set the perturbations of the hydrodynamic variables $\delta u_i$, $\delta T$ and $\delta\mu$ in (\ref{phentj}) to zero. Also we look for spatially homogeneous perturbations so that the momentum $\mathbf{k}$ is zero. Nevertheless, unlike the case of hydrodynamic modes, the frequency $\omega_{(n)}$ do not vanish when $\mathbf{k}$ goes to zero. In such a configuration, for arbitrary $a_{(n)ij}$, it is easy to see that energy and momentum is conserved because $\partial^\mu t_{\mu\nu}$ vanishes identically. When the chemical potential is set to zero, the quasi-normal modes in five dimensional gravity in $AdS_5$ give \cite{Starinets} :
\begin{eqnarray}\label{freqh}
\omega_{(n)}(\mathbf{k} = 0) &=& \pi T \ \Big[ \pm 1.2139 - 0. 7775 \ i \pm 2n (1\mp i)\Big], \ \text{for large n}.
\end{eqnarray}

Clearly, the conservation equations are not enough to reproduce all the quasi-normal modes. We need extra phenomenological equations. Such phenomenological equations can be derived from kinetic theories like Boltzmann equation at weak coupling or gravity at strong coupling. However, we can also write them on general phenomenological grounds. At present, these will not be important for us, we merely mention these have been found in the most general form in \cite{myself1, myself2}.

We will be interested in the spectral function in this class of non-equilibrium states, whose dynamics is determined by the non-equilibrium fluctuations of energy-momentum tensor and conserved currents only. If we want to obtain these spectral functions holographically, we need the explicit metric and gauge field corresponding to the non-equilibrium state. It will be important for us to write the metric and gauge field fluctuation about the equilibrium black-brane background explicitly in terms of $\delta u_i$, $\delta T$, $\delta\mu$, $\pi_{ij}^{\text{(nh)}}$, $\nu_0$ and $\nu_i^{\text{(nh)}}$. As we will show in the next section, the spectral function in the dual states will depend explicitly just on these non-equilibrium variables.

Later in section V, we will discuss what happens when we take into account non-linearities in the dynamics of $\delta u_i$, $\delta T$, $\pi_{ij}^{(nh)}$, etc.

\subsection{Explicit examples of backgrounds}

We will be interested in strongly coupled conformal field theories in three space-time dimensions in the large $N$ limit. Therefore, as discussed earlier, we will be concerned with solutions of Einstein-Maxwell equations which are asymptotically $AdS_4$ and are quasi-normal mode fluctuations about a Reissner-Nordstorm black brane with both mass and charge. 

As discussed earlier, on the gravity side we will need the Einstein-Maxwell action :
\begin{equation}
S = \frac{1}{2\kappa^2} \int d^4 x \Big(R + \frac{6}{l^2} - \frac{l^2}{4}F_{MN}F^{MN}\Big).
\end{equation}
Above $l$ sets the scale of asymptotic (negative) curvature via a (negative) cosmological constant. This is required so that the asymptotic isometry of the spacetime is the same as the conformal group in 3 dimensions. We will use $\kappa$ to denote the effective Newton's constant in four-dimensional gravity in lieu of Planck length $l_P$. 

The metric of the Reissner-Nordstorm black brane in $AdS_4$ is :

\begin{eqnarray}\label{metric0}
ds^2 &=& \frac{l^2}{r^2} \frac{dr^2}{f\left(\frac{r r_0}{l^2}\right)} +   \frac{l^2}{r^2}\Big(-f\left(\frac{r r_0}{l^2}\right)dt^2 + dx^2 + dy^2 \Big),
\end{eqnarray}
where $f$ is the so-called blackening function given by :
\begin{eqnarray}
f(s) &=& 1 - \Big(3\frac{r_*^4}{r_0^4}+1\Big)s^3 + 3\frac{r_*^4}{r_0^4}s^4.
\end{eqnarray}
In case of the gauge field, it is convenient to use the gauge $A_r = 0$. The only non-zero component of the gauge field is $A_t$ and is given by :
\begin{equation}\label{gf0}
A_t = \frac{2\sqrt{3} r_*^2}{l^2 r_0} \left(1 -\frac{r r_0}{l^2} \right).
\end{equation}

The boundary of $AdS_4$ in these coordinates is at $r=0$ and the outer horizon is at $r = l^2/r_0$. The total mass $M$ and charge $Q$ of the black hole are given by :
\begin{equation}
Q =\sqrt{3}r_*^2, \quad M = r_0^3 + 3\frac{r_*^4}{r_0}.
\end{equation}
Using the standard holographic dictionary we can relate the two parameters $r_*$ and $r_0$ of the geometry and the Newton's constant $\kappa$ in to the energy density $\epsilon$, charge density $\rho$ and entropy density $s$ as below :
\begin{equation}
\epsilon = 2p = \frac{r_0^3}{\kappa^2 l^4}\Big(3 \frac{r_*^4}{r_0^4}+1\Big), \quad \rho = \frac{\sqrt{3}}{\kappa^2}\Big(\frac{r_*}{l}\Big)^2, \quad s = \frac{2\pi r_0}{\kappa^2 l^2}.
\end{equation}  
The thermodynamic relation
\begin{equation}
d\epsilon = T ds + \mu d\rho
\end{equation}
gives the temperature and chemical potential as below :
\begin{equation}
T = \frac{3r_0}{4\pi l^2}\Big(1 - \Big(\frac{r_*}{r_0}\Big)^4\Big), \quad \mu =  \frac{2\sqrt{3} r_*^2}{l^2r_0}.
\end{equation}

The first example of a non-equilibrium background we will describe is that with a hydrodynamic shear-mode turned on. The velocity perturbation will be denoted as $\delta \mathbf{u}(k_{\text{(h)}})$ with $k_{\text{(h)}}$ being the three-momentum of the fluctuation. We recall that $\mathbf{k}_{\text{(h)}}\cdot \delta\mathbf{u}(\mathbf{k}_{\text{(h)}}) = 0$, as the shear wave perturbation is transverse. 

It is a well-defined problem to find a given metric and gauge field perturbation in the bulk corresponding to a definite energy-momentum tensor and conserved current fluctuation about the equilibrium at the boundary, when the Dirichlet boudary condition is imposed for the bulk perturbations at the boundary. The latter is needed so that the dual field theory lives in flat space and is influenced by an externally fixed chemical potential. Regularity at the horizon fixes the transport coefficients appearing in the energy-momentum tensor and conserved currents.

This procedure can be readily implemented in Fefferman-Graham coordinates \cite{myself4}. A similar procedure can be implemented in Schwarzchild-like coordinates as well because the Schwarzchild radial coordinate and the Fefferman-Graham radial coordinate are only functions of each other when the temperature remains unperturbed. Then it follows \cite{myself4} that :
\begin{eqnarray*}
\delta g_{ij} \ &\text{will be proportional to}& \ (k_{\text{(h)}i} \ \delta u_j (\mathbf{k}_{\text{(h)}})+k_{\text{(h)}j} \ \delta u_i (\mathbf{k}_{\text{(h)}}))e^{i (\mathbf{k}_{\text{(h)}} \cdot \mathbf{x} - \omega_{\text{(h)}} t)}, \ \text{and}, \nonumber\\
\delta g_{i0} \ &\text{will be proportional to}& \ \delta u_i (\mathbf{k}_{\text{(h)}})e^{i (\mathbf{k}_{\text{(h)}} \cdot \mathbf{x} - \omega_{\text{(h)}} t)}.
\end{eqnarray*}
It can be also shown that in the radial gauge, $A_r = 0$, the fluctuation in the gauge field is also proportional to the fluctuation in the conserved current, i.e. proportional to :
\begin{equation*}
\delta u_i (\mathbf{k}_{\text{(h)}})e^{i (\mathbf{k}_{\text{(h)}} \cdot \mathbf{x} - \omega_{\text{(h)}} t)}.
\end{equation*}

The explicit metric is given by :
\begin{eqnarray}\label{metric1}
ds^2 &=& \frac{l^2}{r^2} \frac{dr^2}{f\left(\frac{r r_0s}{l^2}\right)} +   \frac{l^2}{r^2}\Big(-f\left(\frac{r r_0}{l^2}\right)dt^2 + dx^2 + dy^2  -2\Big(1-f\left(\frac{r r_0}{l^2}\right)\Big) \delta u_i (\mathbf{k}_{\text{(h)}}) e^{i (\mathbf{k}_{\text{(h)}} \cdot \mathbf{x} - \omega_{\text{(h)}} t)}dtdx^i \Big) \nonumber\\
&& +\frac{2l^2}{r^2} \Big(-i \frac{l^2}{3r_0} \ k_{\text{(h)}i} \ \delta u_j (\mathbf{k}_{\text{(h)}})e^{i (\mathbf{k}_{\text{(h)}} \cdot \mathbf{x} - \omega_{\text{(h)}} t)}  \ h\left(\frac{r r_0}{l^2}\right)dx^i dx^j \Big) + O(\epsilon^2),
\end{eqnarray}
where,
\begin{eqnarray}\label{h}
h(s) &=& 3\int_0^s d\tilde{s}  \frac{\tilde{s}^2}{(1+\tilde{s}+\tilde{s}^2 - 3\frac{r_*^4}{r_0^4}\tilde{s}^4)(1-\tilde{s})},
\end{eqnarray} 
and
\begin{equation}\label{dispsh}
\omega_{\text{(h)}} = - i \frac{\mathbf{k}_{\text{(h)}}^2}{4\pi T}  + O(\epsilon^3), \quad \eta = \frac{r_0^2}{2\kappa^2 l^2} = 4\pi s.
\end{equation}

In the radial gauge $A_r = 0$, the gauge field takes the form
\begin{eqnarray}\label{gf1}
A_t =\frac{2\sqrt{3} r_*^2}{l^2 r_0} \left(1 -\frac{r r_0}{l^2} \right) + O(\epsilon^2),\quad
A_i = -\frac{2\sqrt{3} r_*^2}{l^2 r_0} \left(1 -\frac{r r_0}{l^2} \right)\delta u_i (\mathbf{k}_{\text{(h)}}) e^{i (\mathbf{k}_{\text{(h)}} \cdot \mathbf{x} - \omega_{\text{(h)}} t)} + O(\epsilon^2).
\end{eqnarray}
Above $\epsilon$ denotes the parameter of derivative expansion in hydrodynamics.

It is to be noted that we have written the full metric and gauge field in a global frame co-moving with the equilibrium part of the energy-momentum tensor and conserved currents, i.e. in the laboratory frame. We can readily make the metric and gauge field Lorentz-covariant by boosting such that the unperturbed velocity field is a four-velocity vector $u^\mu$ \cite{Sayantani}. However, this will be unnecessary for the purposes of this paper as we will be interested in the results in the laboratory frame.

Also one can readily realize that the metric is singular at the outer horizon $r= l^2/r_0$. This is however only an artifact of the coordinate system.  We can systematically change coordinates order by order in the derivative expansion so that the metric and gauge fields are manifestly regular at the horizon \cite{myself4}. In our coordinates, the radius of convergence of the derivative expansion is of the order of the effective mean-free path or the inverse of the effective temperature at a given radius given by $T_{eff} (r) = T/\sqrt{f(r r_0/l^2)}$. Therefore, we have a finite radius of convergence of the derivative expansion a finite distance away from the horizon. Furthermore, we will be interested in calculating boundary correlators which are independent of  the choice of bulk coordinate system.

The metric (\ref{metric1}) and gauge field (\ref{gf1}) in manifestly regular coordinates are given in appendix A.

The second example which we will be concerned with will be a homogeneous non-hydrodynamic perturbation of the energy-momentum tensor, i.e. with one $a_{(n)ij}$ in (\ref{phentj}) turned on. The momentum of this perturbation is zero on account of homogeneity,  but its frequency is non-zero and complex like in (\ref{freqh}). The metric can be obtained following \cite{myself2} in the Fefferman-Graham coordinate and re-expressed in the Schwarzchild coordinate used here by simply changing the radial coordinate. Again, as the temperature remains unperturbed, up to linear order the change of coordinate involves transformation of one variable. It can be shown that the metric perturbation is proportional to
\begin{equation*}
a_{(n)ij}e^{-i  \omega_{(n)} t} .
\end{equation*}

Explicitly the perturbed metric is : 
\begin{eqnarray}\label{metricnh}
ds^2 = \frac{l^2}{r^2} \frac{dr^2}{f\left(\frac{r r_0}{l^2}\right)} +   \frac{l^2}{r^2}\Big(-f\left(\frac{r r_0}{l^2}\right)dt^2 + dx^2 + dy^2  \Big)  +\frac{2l^2}{r^2} \Big(a_{(n)ij}e^{-i  \omega_{(n)} t}  \ \tilde{h}\left(\frac{r r_0}{l^2}, \omega_{(n)}\right)dx^i dx^j \Big) + O(\delta^2),
\end{eqnarray}
with $\delta$ being the parameter of non-hydrodynamic amplitude expansion. Furthermore, $\tilde{h}(s, \omega_{(n)})$ follows the equation of motion :
\begin{eqnarray}
\frac{d^2 \tilde{h}(s,\omega_{(n)})}{ds^2} - \frac{\Big(2 + (1 + 3\frac{r_*^4}{r_0^4})s^3 - 6\frac{r_*^4}{r_0^4}s^4\Big)}{sf(s)}\frac{d\tilde{h}(s,\omega_{(n)})}{ds} + \frac{\omega_{(n)}^2 l^4}{r_0^2}\Bigg(\frac{1}{f^2(s)} \Bigg)\tilde{h}(s,\omega_{(n)})= 0.
\end{eqnarray}
We will also require that :
\begin{equation}
\tilde{h}(s, \omega_{(n)}) = s^3 + O(s^4) \ \text{as $s\rightarrow 0$.}
\end{equation}
This is the asymptotic boundary condition and determines $\tilde{h}$ uniquely as it puts the coefficient of the non-normalizable to zero and the coefficient of the normalizable mode to be unity so that the boundary energy-momentum tensor fluctuation is as given by (\ref{phentj}). Though the equation for $\tilde{h}$ cannot be analytically solved, the solution can be readily expanded in a power series in $\omega_{(n)}$.

Furthermore, the gauge field remains unperturbed from the black brane profile.

The metric above is also not manifestly regular at the horizon, but once again it is just an artifact of the choice of coordinates. One can again translate the metric systematically to Eddington-Finkelstein coordinates to see manifest regularity \cite{myself2}. The regularity is manifest only when we sum over all orders in $\omega_{(n)}$. This is to be expected because, although the amplitude of the non-hydrodynamic perturbation $a_{(n)ij}$ is small, it's rate of change in time is not small (unlike the hydrodynamic modes) since $\omega_{(n)}$ is of the same order as the temperature. 

Though we will not discuss the details here, we can construct the explicit metrics in the case of both hydrodynamic and non-hydrodynamic perturbations even at the non-linear level \cite{Sayantani, myself2}. The metric is regular at each order in the derivative expansion for hydrodynamic perturbations and for each order in the amplitude expansion for non-hydrodynamic perturbations, provided all time-derivatives (or covariantly speaking convective derivatives) are summed over at each order in the latter case \cite{myself2}.

\section{The holographic prescription for the non-equilibrium spectral function}

As discussed in the Introduction, the spectral function is given by the imaginary part of the retarded propagator which can be obtained from causal response of an operator to it's source. A convenient way to obtain the spectral function is to calculate the retarded propagator using linear response theory first and then isolate its imaginary part.

In this section, we will consider single trace scalar and fermionic operators in field theory whose back-reaction to the metric is suppressed by $O(1/N^2)$. As we have argued in section II.B, the possibly interesting scalar operators in the strong coupling and large $N$ limit are order parameters of symmetry breaking. If we are in a range of  temperature and chemical potentials, where such symmetry breaking does not occur, the profile of the scalar fields dual to these operators vanishes in the background classically. Therefore, the backreaction is indeed $O(1/N^2)$ suppressed. This observation may be applied to study pion correlations in the quark-gluon plasma at RHIC.

In popular holographic models of strongly correlated systems, the electron is thought to couple to a composite operator made out of strongly interacting fractionalized degrees of freedom (for a clear exposition please see \cite{semihol}). The holographic dual is thought to capture the dynamics of the fractionalized degrees of freedom. The strongly interacting fractionalized degrees of freedom are $O(N^2)$, but the coupling of the electron to the composite operator of the strongly coupled theory is $O(1)$. The spectral function obtained from photo-electron spectroscopy (ARPES) will receive corrections from the spectral function of the composite fermionic operator of the strongly coupled sector. As the coupling of the electron to this operator is $O(1)$, we can ignore the backreaction of the fermionic field dual to this operator on the geometry representing the dual state, at the leading order. If this picture is qualitatively viable, our set-up will be relevant for describing non-equilibrium features of non-Fermi liquids described by such models.

Holographically, causal response implies the incoming boundary condition at the horizon. The event horizon separates space-time into two causal parts, one that is inside and ends at a singularity, and the other that is outside and stretches all the way to the boundary.  No light ray can come out of the inside region to the outside region, though light rays can propagate from the outside to the inside. Therefore, the perturbations which respect the causal structure of the space-time are those which are purely incoming at the horizon, having no component which propagates from the inside to the outside. 

The event horizon is not only a feature of the eternal static black hole, but also of the perturbed black hole (for instance, the black hole with the quasi-normal mode fluctuations of the metric and gauge fields). The event horizons of these non-equilibrium geometries are also perturbed from their equilibrium location and their positions can be calculated in a perturbative expansion \cite{horizon}. Equilibration in this context means that the event horizon will have uniform surface gravity (the gravitational analogue of temperature) everywhere and it happens only far in the future.

Though the incoming boundary condition is insufficient for a well defined perturbation theory in non-equilibrium geometries as noted in the Introduction, we expect regularity at the future horizon to be a sufficient condition. It turns out that it is sufficient to impose the regularity condition only far in the future, that is in the asymptotic static black brane geometry. This has been observed before in \cite{Sayantani, myself2} in another context - while constructing time-dependent non-linear solutions of gravity with regular future horizons perturbatively. In such solutions it indeed suffices to impose regularity of the perturbations at the final equilibrium location of the horizon. In fact, the incoming boundary condition is itself tied up to regularity \cite{Horowitz} \footnote{See also \cite{myself2} for an explicit proof in a non-hydrodynamic context.}. In this section we will find a precise non-equilibrium generalization of the incoming boundary condition for   bosonic and fermionic field configuratons in non-equilibrium geometries.

For purposes of illustration, let us consider the non-equilibrium state which is the simplest to analyze from the gravity point of view - it is the AdS black brane with a linearized hydrodynamic shear mode perturbation of spatial momentum $\mathbf{k}_{\text{(h)}}$. The advantage of this geometry is that it can be shown that the event horizon do not fluctuate up to first order in the derivative expansion (i.e. up to first order in $\mathbf{k}_{\text{(h)}}/ T$) essentially because the temperature field does not fluctuate as discussed in section II. We will first demonstrate how we can develop a prescription for obtaining the holographic spectral function in such a non-equilibrium state. Our aim will be to obtain the correction to the equilibrium spectral function up to first order in derivative expansion, i.e. up to first order in $\mathbf{k}_{\text{(h)}}/T$.

The explicit metric and gauge field of the black brane with the hydrodynamic shear mode perturbation is given in (\ref{metric1}) and (\ref{gf1}) respctively up to first order in the derivative expansion. We will work explicitly with four space-time dimensions in gravity, as we will be interested primarily in a three space-time dimensional dual strongly coupled field theory. This is because we are interested in applications to strongly correlated electron systems at finite density living in two spatial dimensions. As argued in section II.B, our analysis may apply to the strange metallic phase in a qualitative manner. 

\begin{figure}[ht]
\begin{center}
\includegraphics[width = 80mm]{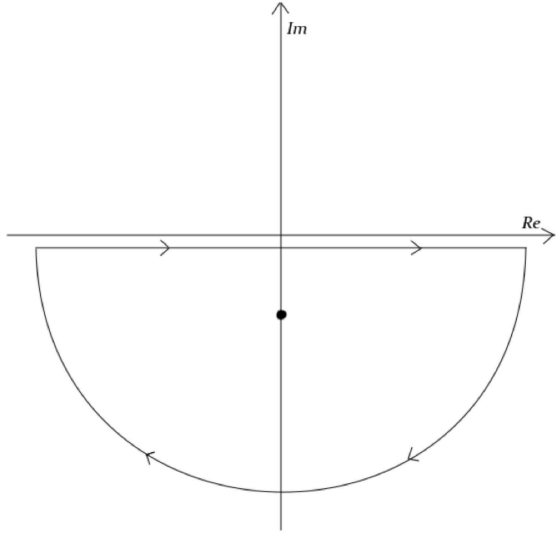}
\caption{Contour for integration over $\omega_{\text{(h)}}$, with pole at negative imaginary axis}
\label{contour}
\end{center}
\end{figure}

An elegant way to solve the equations of motion of scalar and fermionic fields is by using the Fourier transform in all the  field-theory (i.e. boundary) coordinates. Obviously, in order to express the equations of motion of the fields in Fourier space, it is necessary to do the Fourier transform of the background perturbation first, i.e. we need to do the Fourier transform of the velocity field fluctuation $\delta u_i$. The dispersion relation for this fluctuation is as given by eqs. (\ref{dis1}) and (\ref{dis2}). We see that the frequency  given by the dispersion relation is strictly (negative) imaginary, while the frequency related to Fourier transform is strictly real. Furthermore, the negative imaginary frequency given by the dispersion relation makes $\delta u_i$ decay in the future but grow in the past as a function of time. A Fourier transform of such a function needs to be defined with care. In order to distinguish from the frequency and momenta associated with the scalar/fermionic field, we will denote the frequency and momenta of $\delta u_i$ as $\omega_{\text{(h)}}$ and $\mathbf{k}_{\text{(h)}}$ respectively. The correct Fourier transform which reproduces the hydrodynamic dispersion relation is :
\begin{equation}\label{fourier}
\delta u_i (\omega_{\text{(h)}}, \mathbf{k}_{\text{(h)}}) = -\Bigg(\frac{1}{2\pi i}\Bigg)\frac{\delta u_i(\mathbf{k}_{\text{(h)}})}{\omega_{\text{(h)}} + i \frac{\mathbf{k}_{\text{(h)}}^2}{4\pi T}}.
\end{equation} 
To check the above, one can try to reproduce the time dependence by doing the inverse Fourier transform. This needs to be done with a specific contour prescription for integration over $\omega_{\text{(h)}}$ as shown in Fig.\ref{contour}.This contour is the usual contour associated with the retarded propagator in field theory - it runs from $-\infty$ to $\infty$ infinitesimally below the real axis and then closes itself through the circle at infinity. This contour picks up contribution only from the negative imaginary pole reproducing the correct time dependence of $\delta u_i$ at given $\mathbf{k}_{\text{(h)}}$.

It will be easier to solve the scalar/fermionic field equations after doing the Fourier transform of $\delta u_i$, however we need to finally integrate over $\omega_{\text{(h)}}$ with the above contour prescription in order to obtain the observed behavior in real time.

For demonstrative purposes, we will analyze the scalar field equations first and then the fermionic field equations. Finally, we will see how we can apply our prescription for the non-equilibrium retarded Green's function when the background contains other quasinormal modes of the metric and gauge field.

\subsection{Scalar field equation and the non-equilibrium spectral function} 

We will be interested in the non-equilibrium holographic spectral function for a scalar operator first. This requires us to solve the equation of motion of the dual scalar field in the non-equilibrium background; in particular we need to understand how the equilibrium part determines the non-equilibrium part completely. Without this, as we have mentioned before in the Introduction, the spectral function cannot be determined.

We will need to specify the equilibrium part of the solution first. We can assume, without loss of generality, that the equilibrium solution is in a specific $(\omega,\mathbf{k})$ mode and obtain the non-equilibrium correction for each such mode. Using the fact that our field equation is linear, we can then linearly superimpose the solutions with the non-equilibrium correction for each equilibrium mode to obtain the most general solution. 

The background in which the scalar field propagates is the $AdS_4$ Reissner-Nordstorm black hole with the hydrodynamic shear-mode perturbation. This hydrodynamic mode is given by the velocity perturbation $\delta u_i$ in a specific momentum $\mathbf{k}_{\text{(h)}}$ but its dependence on $\omega_{\text{(h)}}$ is given by (\ref{fourier}). We have to consider the background first in a definite $\omega_{\text{(h)}}$ perturbation and then integrate over $\omega_{\text{(h)}}$ finally with the contour prescription discussed before. The scalar field while propagating in the background will pick up a $(\omega + \omega_{\text{(h)}}, \mathbf{k}+\mathbf{k}_{\text{(h)}})$ mode. The profile of the scalar field, will therefore be of the following form :
\begin{widetext}
\begin{eqnarray}
\Phi(\mathbf{x},t, r) = \Phi^{(0)}(\omega, \mathbf{k}, r) e^{-i(\omega t- \mathbf{k}\cdot \mathbf{x})} + \Phi^{(1)}(\omega, \mathbf{k}, \omega_{\text{(h)}}, \mathbf{k}_{\text{(h)}}, r)e^{-i((\omega+\omega_{\text{(h)}}) t- (\mathbf{k}+\mathbf{k}_{\text{(h)}})\cdot \mathbf{x})}. 
\end{eqnarray}
\end{widetext}
The equilibium part of the solution is $\Phi^{(0)}(\omega, \mathbf{k}, r)$ and the non-equilibrium part is $ \Phi^{(1)}(\omega, \mathbf{k}, \omega_{\text{(h)}}, \mathbf{k}_{\text{(h)}}, r)$. The non-equilibrium part does not depend on the combination $\omega+\omega_{\text{(h)}}$ and $\mathbf{k} + \mathbf{k}_{\text{(h)}}$ as the space-time translational invariances of the equilibrium background are broken explicitly by the hydrodynamic quasinormal modes.

If the scalar field $\Phi$ is minimally coupled to gravity, and its mass and charge are $m$ and $q$ respectively, the equation of motion of the equilibrium part is simply
\begin{equation}
\Box^{ARN}_{\omega', \mathbf{k}'}\delta(\omega'-\omega)\delta^2(\mathbf{k}'-\mathbf{k})\Phi^{(0)}(\omega, \mathbf{k}, r) = 0,
\end{equation}
where $\Box^{ARN}_{\omega, \mathbf{k}}$
is the (gauge-invariant) Laplacian in the AdS Reissner-Nordstorm background metric (\ref{metric0}) and gauge field (\ref{gf0}) as given by :
\begin{eqnarray}
\Box^{ARN}_{\omega, \mathbf{k}} &=& r^2 f\Big(\frac{rr_0}{l^2}\Big)\partial_r^2 +r\Big[-2 f\Big(\frac{rr_0}{l^2}\Big) +\frac{rr_0}{l^2}f'\Big(\frac{rr_0}{l^2}\Big)\Big] \partial_r \nonumber \\
&+& r^2l^2\Big[\frac{(\omega+q\mu \left(1 -\frac{r r_0}{l^2} \right))^2}{f\Big(\frac{rr_0}{l^2}\Big)}-\mathbf{{k}^2}\Big] + m^2 l^2.
\end{eqnarray}
Again, $f$ is the blackening function of the AdS Reissner-Nordstorm black brane which vanishes at the horizon located at $r = l^2/r_0$. 

With the metric and gauge field in presence of hydrodynamic shear perturbation given by (\ref{metric1}) and (\ref{gf1}) respectively, the equation of motion for the non-equilibrium part up to first order in the hydrodynamic momenta $\mathbf{k}_{\text{(h)}}$ is :
\begin{widetext}
\begin{eqnarray}\label{noneqeom}
\Box^{ARN}_{\omega', \mathbf{k}'}\delta(\omega'-\omega-\omega_{\text{(h)}})\delta^2(\mathbf{k}'-\mathbf{k}-\mathbf{k}_{\text{(h)}})\Phi^{(1)} (\omega, \omega_{\text{(h)}}, \mathbf{k},\mathbf{k}_{\text{(h)}},r) = V(\omega, \omega_{\text{(h)}}, \mathbf{k},\mathbf{k}_{\text{(h)}},r)\Phi^{(0)}(\omega, \mathbf{k},r),
\end{eqnarray}
with
\begin{eqnarray}\label{sterms}
V &=& V_1+V_2 , \nonumber\\
V_1 &=& \frac{2r^2}{f\Big(\frac{rr_0}{l^2}\Big)}\Big(\omega\Big(1- f\Big(\frac{rr_0}{l^2}\Big)\Big)+q\mu \left(1 -\frac{r r_0}{l^2} \right)\Big)\delta \mathbf{u}(\omega_{\text{(h)}}, \mathbf{k}_{\text{(h)}})\cdot \mathbf{k}, \nonumber \\
V_2 &=&i \frac{2l^2 r^2}{3r_0}  h\Big(\frac{rr_0}{l^2}\Big)k_ik_j k_{\text{(h)}i} \delta u_j(\omega_{\text{(h)}}, \mathbf{k}_{\text{(h)}}).
\end{eqnarray}
\end{widetext}
Above, $h$ gives the hydrodynamic correction to the background metric which is proportional to $k_{\text{(h)}i}\delta u_j + (i\leftrightarrow j)$ as in (\ref{h}). 

The behavior of the general solution of $\Phi^{(0)}(\omega, \mathbf{k}, r)$ near the horizon is well-known. It can be split into an incoming and outgoing wave as below :
\begin{widetext}
\begin{eqnarray}
\Phi^{(0)}(\omega, \mathbf{k}, r) \approx A^{in}(\omega, \mathbf{k})\Bigg(1 - \frac{rr_0}{l^2}\Bigg)^{-i\frac{\omega}{4\pi T}} + A^{out}(\omega, \mathbf{k})\Bigg(1 - \frac{rr_0}{l^2}\Bigg)^{i\frac{\omega}{4\pi T}} \  \ \text{near $r = \frac{l^2}{r_0}$}.
\end{eqnarray}
\end{widetext}
In order to select the incoming wave, we should put 
\begin{equation}
A^{out}(\omega, \mathbf{k}) = 0.
\end{equation} 
We can also normalize the overall solution by choosing 
\begin{equation}\label{solh1}
A^{in}(\omega, \mathbf{k}) = \mathcal{C},
\end{equation}
with $\mathcal{C}$ being a numerical constant. This overall normalization will play no role in the Green's functions. 

The behavior of the general non-equilibrium part of the solution near the horizon is :
\begin{widetext}
\begin{eqnarray}\label{solh2}
\Phi^{(1)}(\omega, \omega_{\text{(h)}}, \mathbf{k},\mathbf{k}_{\text{(h)}}, r) &\approx& A^{in}(\omega, \omega_{\text{(h)}}, \mathbf{k}, \mathbf{k}_{\text{(h)}})\Bigg(1 - \frac{rr_0}{l^2}\Bigg)^{-i\frac{\omega+\omega_{\text{(h)}}}{4\pi T}} + A^{out}(\omega, \omega_{\text{(h)}},\mathbf{k}, \mathbf{k}_{\text{(h)}})\Bigg(1 - \frac{rr_0}{l^2}\Bigg)^{i\frac{\omega+\omega_{\text{(h)}}}{4\pi T}}\nonumber\\
&&+ i\mathcal{C} \Bigg(\frac{4\pi T l^2}{r_0}\Bigg)^2 \Bigg(\frac{2}{9\left(1-\frac{r_*^{4}}{r_0^4}\right)^2}\Bigg)\frac{\omega \delta \mathbf{u}(\omega, \mathbf{k}_{\text{(h)}})\cdot \mathbf{k}}{(2\omega+\omega_{\text{(h)}})\omega_{\text{(h)}}} \Bigg(1 - \frac{rr_0}{l^2}\Bigg)^{-i\frac{\omega}{4\pi T}}  \  \ \text{near $r = \frac{l^2}{r_0}$}.
\end{eqnarray}
\end{widetext}
The first two terms on the RHS above are the homogeneous incoming and outgoing solutions for frequency mode $\omega + \omega_{\text{(h)}}$. The third term is the particular solution which is determined completely by the equilibrium solution. The above behavior at the horizon is exact up to first order in $\mathbf{k}_{\text{(h)}}$. In fact the full general solution which reproduces the above can be given elegantly in an integral representation as in appendix B. 

Obviously, we need to impose the incoming boundary condition again. Therefore,
\begin{equation}
A^{out}(\omega, \omega_{\text{(h)}},\mathbf{k}, \mathbf{k}_{\text{(h)}}) = 0.
\end{equation}

We will now show that in order to impose regularity at the horizon, we also need to dispose of the ingoing non-equilibrium homogeneous solution at the horizon. We recall that finally we need to integrate over $\omega_{\text{(h)}}$. 

In order to be consistent with the derivative expansion, $A^{in}(\omega, \omega_{\text{(h)}},\mathbf{k}, \mathbf{k}_{\text{(h)}})$ must take the form as follows. It is proportional to components of $\delta\mathbf{u}$ at the linear order as it should vanish in absence of the background perturbation. It's dependence on $\omega_{\text{(h)}}$ and $\mathbf{k}_{\text{(h)}}$ can be expanded systematically in terms of rotationally invariant scalars like $\delta\mathbf{u}\cdot \mathbf{k}$, $k_i k_j k_{\text{(h)}i}\delta u_j$, $\omega_{\text{(h)}}k_i k_j k_{\text{(h)}i}\delta u_j$, etc. Up to first order in the derivative expansions only the first two scalars will apear. The coefficients of these scalars should be functions of $\omega$ and $\mathbf{k}$ only, as the depenedence on $\omega_{\text{(h)}}$ and $\mathbf{k}_{\text{(h)}}$ can be absorbed in coefficients of the scalars appearing at higher orders in the derivative expansion. Thus, up to first order in derivative expansion, we should have :
\begin{widetext}
\begin{equation}
A^{in}(\omega, \omega_{\text{(h)}},\mathbf{k}, \mathbf{k}_{\text{(h)}}) = A^{in}_1(\omega, \mathbf{k}) \   \delta\mathbf{u}(\omega_{\text{(h)}}, \mathbf{k}_{\text{(h)}})\cdot \mathbf{k} + A^{in}_2(\omega, \mathbf{k})\  k_i k_j k_{\text{(h)}i}\delta u_j(\omega_{\text{(h)}}, \mathbf{k}_{\text{(h)}}).
\end{equation}
\end{widetext}
We recall for the hydrodynamic shear mode $\delta\mathbf{u}\cdot\mathbf{k}_{\text{(h)}} =0$, so there is no more possible terms up to first order in $\mathbf{k}_{\text{(h)}}$. When we integrate over $\omega_{\text{(h)}}$, the Fourier transform of $\delta \mathbf{u}$ as given by (\ref{fourier}) will give a pole contribution. Taking this into account the behavior of the ingoing non-equilibrium mode at the horizon will be :
\begin{equation}
\Bigg(1 - \frac{rr_0}{l^2}\Bigg)^{-i\frac{\omega}{4\pi T} - \frac{\mathbf{k}_{\text{(h)}}^2}{16 \pi^2 T^2}}. 
\end{equation}
Therefore, we find the ingoing homogeneous non-equilibrium mode diverges at the horizon as $\mathbf{k}_{\text{(h)}}^2/ (16\pi^2 T^2)$ is strictly positive. This divergence is not an artifact of the coordinate system because we are studying the behavior of a scalar field. The only way this divergence can be removed is by putting
\begin{equation}
A^{in}(\omega, \omega_{\text{(h)}},\mathbf{k}, \mathbf{k}_{\text{(h)}}) = 0, \ \text{i.e.} \ A_1^{in}(\omega, \mathbf{k}) = A_2^{in}(\omega, \mathbf{k}) = 0.
\end{equation}
The particular solution at the horizon as defined as the third term in (\ref{solh2}) produces no divergence after we do the integral over $\omega_{\text{(h)}}$. It is regular at and outside the horizon. 

Summing up, the full solution with the non-equilibrium correction is the following :
\begin{widetext}
\begin{eqnarray}\label{solhe}
\Phi(\mathbf{x}, t, r) &\approx& \mathcal{C}\Bigg(\Bigg(1 - \frac{rr_0}{l^2}\Bigg)^{-i\frac{\omega}{4\pi T} }e^{-i(\omega t-\mathbf{k}\cdot\mathbf{x})} +i \Bigg(\frac{4\pi T l^2}{r_0}\Bigg)^2 \Bigg(\frac{2}{9\left(1-\frac{r_*^{4}}{r_0^4}\right)^2}\Bigg)\frac{\omega \delta \mathbf{u}(\omega, \mathbf{k}_{\text{(h)}})\cdot \mathbf{k}}{(2\omega+\omega_{\text{(h)}})\omega_{\text{(h)}}} \Bigg(1 - \frac{rr_0}{l^2}\Bigg)^{-i\frac{\omega}{4\pi T}} \nonumber\\&& e^{-i((\omega+\omega_{\text{(h)}})t -(\mathbf{k}+\mathbf{k}_{\text{(h)}})\cdot \mathbf{x})}\Bigg), \ \ \text{near $r = \frac{l^2}{r_0}$.}
\end{eqnarray}
\end{widetext}
The above behavior when specified near the horizon uniquely fixes the full non-equilibrium solution aside for an overall normalization $\mathcal{C}$. 

We can numerically extrapolate the full solution all the way to the boundary $r=0$.  As the background is asymptotically $AdS$, we should have the following behavior :
\begin{eqnarray}
\Phi(\mathbf{x}, t, r) &\approx&  J(\mathbf{x},t) r^{3-\Delta} + O(\mathbf{x}, t) r^{\Delta} 
\quad \text{near $r=0$.}
\end{eqnarray}
By the holographic dictionary, $J$ is indeed the source and $O$ is the expectation value of the dual operator in the dual non-equilibrium state \footnote{When $- 9/4 < m^2 l^2 < -5/4$, we can do an alternate quantization where $J$ can be interpreted as the expectation value and $O$ as the source \cite{AQ}. This requires the scaling dimension of the operator to be $\Delta = 3/2 - \sqrt{9/4 + m^2 l^2}$. The partition functions of the two theories are related by a Legendre transform.}. Also, $\Delta$ is the scaling dimension of the dual operator given by the mass of the scalar field as below :
\begin{equation}
\Delta = \frac{3}{2} + \sqrt{\frac{9}{4} + m^2 l^2}.
\end{equation}
The positivity of the Hamiltonian requires $m^2l^2 > -9/4$ \cite{BF}.

Furthermore, near $r= 0$, the equilibrium and non-equilibrium parts of the solution individually have the same behavior, so
\begin{eqnarray}
\Phi^{(0)}(\omega, \mathbf{k}, r) &\approx& J^{(0)}(\omega,\mathbf{k})r^{3-\Delta}+ O^{(0)}(\omega, \mathbf{k})r^{\Delta}, \nonumber\\
\Phi^{(1)}(\omega,\omega_{\text{(h)}}, \mathbf{k},\mathbf{k}_{\text{(h)}}, r) &\approx& J^{(1)}(\omega,\omega_{\text{(h)}}, \mathbf{k}, \mathbf{k}_{\text{(h)}})r^{3-\Delta} +O^{(1)}(\omega, \omega_{\text{(h)}}, \mathbf{k}, \mathbf{k}_{\text{(h)}})r^{\Delta}.
\end{eqnarray}
Therefore,
\begin{eqnarray}
J(\mathbf{x},t) &=& J^{(0)}(\omega, \mathbf{k}) e^{-i(\omega t - \mathbf{k}\cdot\mathbf{x})} +\int_{-\infty}^{\infty} d\omega_{\text{(h)}} J^{(1)}(\omega, \omega_{\text{(h)}}, \mathbf{k}, \mathbf{k}_{\text{(h)}}) e^{-i((\omega+\omega_{\text{(h)}})t -(\mathbf{k}+\mathbf{k}_{\text{(h)}})\cdot\mathbf{x})},
\nonumber\\
O(\mathbf{x},t) &=& O^{(0)}(\omega, \mathbf{k}) e^{-i(\omega t - \mathbf{k}\cdot\mathbf{x})} + \int_{-\infty}^{\infty} d\omega_{\text{(h)}} O^{(1)}(\omega, \omega_{\text{(h)}}, \mathbf{k}, \mathbf{k}_{\text{(h)}}) e^{-i((\omega+\omega_{\text{(h)}})t -(\mathbf{k}+\mathbf{k}_{\text{(h)}})\cdot\mathbf{x})}.
\end{eqnarray}
The unique solution of $\Phi^{(1)}$ with our prescribed behavior near the horizon (\ref{solhe}) gives us the precise non-equilibrium contributions to both the operator and the source in the following form :
\begin{eqnarray}\label{opso}
O^{(1)}(\omega, \omega_{\text{(h)}}, \mathbf{k}, \mathbf{k}_{\text{(h)}}) &=& O^{(1)}_A\Big(\omega,\mathbf{k}\Big) \, \delta\mathbf{u}(\omega_{\text{(h)}}, \mathbf{k}_{\text{(h)}})\cdot \mathbf{k} \, +\, O^{(1)}_B\Big(\omega, \mathbf{k}\Big)\, k_ik_j k_{\text{(h)}i}\delta u_j(\omega_{\text{(h)}}, \mathbf{k}_{\text{(h)}}),\nonumber\\
J^{(1)}(\omega, \omega_{\text{(h)}}, \mathbf{k}, \mathbf{k}_{\text{(h)}}) &=& J^{(1)}_A\Big(\omega, \mathbf{k}\Big) \, \delta\mathbf{u}(\omega_{\text{(h)}}, \mathbf{k}_{\text{(h)}})\cdot \mathbf{k} \, + \,J^{(1)}_B\Big(\omega, \mathbf{k}\Big)\, k_ik_j k_{\text{(h)}i} \delta u_j(\omega_{\text{(h)}}, \mathbf{k}_{\text{(h)}}).
\end{eqnarray}
The explicit forms of $O^{(1)}_A$, $O^{(1)}_B$, $J^{(1)}_A$ and $J^{(1)}_A$ can be obtained as in appendix B. The integration over $\omega_{\text{(h)}}$ then will be given by the contribution from the pole in $\delta u$. 

The non-equilibrium retarded correlator is \footnote{At equilibrium, this prescription has been proposed in \cite{incoming1}. As noted in the Introduction, we can apply this prescription also at non-equilibrium using the validity of linear response theory. }:
\begin{widetext}
\begin{eqnarray}
G_{R}(\mathbf{x}_1, t_1, \mathbf{x}_2, t_2) &=&\frac{ O(\mathbf{x}_1, t_1)}{J(\mathbf{x}_2,t_2)}
=  e^{-i\omega(t_1 -t_2)}e^{i\mathbf{k}\cdot(\mathbf{x}_1 -\mathbf{x}_2)}\frac{O^{(0)}(\omega, \mathbf{k}) + O^{(1)}(\omega, \mathbf{k}, \mathbf{k}_{\text{(h)}}) e^{i\mathbf{k}_{\text{(h)}}\cdot\mathbf{x}_1}e^{-\frac{\mathbf{k}_{\text{(h)}}^2}{4\pi T}t_1}}{J^{(0)}(\omega, \mathbf{k}) + J^{(1)}(\omega, \mathbf{k}, \mathbf{k}_{\text{(h)}}) e^{i\mathbf{k}_{\text{(h)}}\cdot\mathbf{x}_2}e^{-\frac{\mathbf{k}_{\text{(h)}}^2}{4\pi T}t_2}}\nonumber\\
&\approx &   e^{-i\omega(t_1 -t_2)}e^{i\mathbf{k}\cdot(\mathbf{x}_1 -\mathbf{x}_2)}\frac{O^{(0)}(\omega, \mathbf{k})}{J^{(0)}(\omega, \mathbf{k})} \nonumber\\&&\Bigg(1 +\Bigg(\frac{ O^{(1)}(\omega, \mathbf{k}, \mathbf{k}_{\text{(h)}})}{O^{(0)}(\omega, \mathbf{k})}e^{i\mathbf{k}_{\text{(h)}}\cdot\mathbf{x}_1}e^{-\frac{\mathbf{k}_{\text{(h)}}^2}{4\pi T}t_1}-\frac{J^{(1)}(\omega, \mathbf{k}, \mathbf{k}_{\text{(h)}})}{J^{(0)}(\omega, \mathbf{k})}e^{i\mathbf{k}_{\text{(h)}}\cdot\mathbf{x}_2}
e^{-\frac{\mathbf{k}_{\text{(h)}}^2}{4\pi T}t_2}\Bigg)\Bigg),
\end{eqnarray}
where
\begin{eqnarray}\label{opso1}
O^{(1)}(\omega, \mathbf{k}, \mathbf{k}_{\text{(h)}}) &=& O^{(1)}_A\Big(\omega,  \mathbf{k}\Big) \delta\mathbf{u}(\mathbf{k}_{\text{(h)}})\cdot \mathbf{k} \, +\, O^{(1)}_B\Big(\omega, \mathbf{k}\Big)\,k_ik_j k_{\text{(h)}i}\delta u_j(\mathbf{k}_{\text{(h)}}),\nonumber\\
J^{(1)}(\omega, \mathbf{k}, \mathbf{k}_{\text{(h)}}) &=& J^{(1)}_A\Big(\omega, \mathbf{k}\Big) \delta\mathbf{u}(\mathbf{k}_{\text{(h)}})\cdot \mathbf{k} \, +\, J^{(1)}_B\Big(\omega, \mathbf{k}\Big)\,k_ik_j k_{\text{(h)}i}\delta u_j(\mathbf{k}_{\text{(h)}}).
\end{eqnarray}
The difference of the above from (\ref{opso}) is that in $\delta \mathbf{u}$ which has no dependence in $\omega_{\text{(h)}}$. The latter has been integrated over. This integration produces the contribution from the diffusion pole and the residue has been obtained from (\ref{fourier}).

Clearly, the choice of overall normalization of the solution given by $\mathcal{C}$ in (\ref{solhe}) does not matter as mentioned before. It cancels between the numerator and denominator in the retarded correlator. To readily compare with experimental data, we have to do the Wigner transform of the retarded correlator, as discussed before. We find

\begin{eqnarray}\label{fund1}
G_R(\omega, \mathbf{k}, \mathbf{x}, t) &=& \int d\omega_0 \int d^2 k_0 \Bigg[\frac{O^{(0)}(\omega_0, \mathbf{k}_0)}{J^{(0)}(\omega_0, \mathbf{k}_0)}\delta(\omega- \omega_0)\delta^2 (\mathbf{k} -\mathbf{k}_0) \nonumber\\&&- \frac{O^{(0)}(\omega_0, \mathbf{k}_0)}{J^{(0)}(\omega_0, \mathbf{k}_0)}\frac{1}{2\pi i}
\Bigg(\frac{ O^{(1)}(\omega_0, \mathbf{k}_0, \mathbf{k}_{\text{(h)}})}{O^{(0)}(\omega_0, \mathbf{k}_0)}\delta^2\Big(\mathbf{k} - \mathbf{k}_0 -\frac{\mathbf{k}_{\text{(h)}}}{2}\Big)\frac{1}{\Big(\omega - \omega _0 + i \frac{\mathbf{k}_{\text{(h)}}^2}{8\pi T}\Big)}\nonumber\\
&&\qquad\quad-\frac{ J^{(1)}(\omega_0, \mathbf{k}_0, \mathbf{k}_{\text{(h)}})}{J^{(0)}(\omega_0, \mathbf{k}_0)}\delta^2\Big(\mathbf{k} - \mathbf{k}_0 +\frac{\mathbf{k}_{\text{(h)}}}{2}\Big)\frac{1}{ \Big(\omega - \omega _0 - i \frac{\mathbf{k}_{\text{(h)}}^2}{8\pi T}\Big)}\Bigg)e^{i \mathbf{k}_{\text{(h)}}\cdot\mathbf{x}}e^{- \frac{\mathbf{k}_{\text{(h)}}^2}{4\pi T}t}\Bigg].
\end{eqnarray}
\end{widetext} 
The first term above is just the equilibrium retarded propagator. The second and third terms are the non-equilibrium contributions.
The non-equilibrium contributions have an explicit space-time dependence \textit{which is co-moving with the velocity perturbation in the background.}

The spectral function can be obtained from the imaginary part of the retarded propagator by using $\mathcal{A}(\omega, \mathbf{k},\mathbf{x}, t ) = -2\text{Im} G_R(\omega, \mathbf{k},\mathbf{x}, t )$.

\subsection{Fermionic field equations and the non-equilibrium spectral function}

We will now extend the prescription to obtain the non-equilibrium fermionic spectral function. We begin by constructing the equation of motion for a Dirac spinor explicitly in the same non-equilibrium background, which is $AdS_4$ Reissner-Nordstorm black hole with a hydrodynamic 
shear-mode perturbation.

We recall that the Dirac equation for a Dirac spinor of mass m and charge q in curved space is :
\begin{equation}
\left(e^{M}_{A}\Gamma^{A}\left(\partial_{M} +\frac{1}{8} \omega_{M}^{BC}[\Gamma_{B}, \Gamma_{C}] +i q A_{M} \right) + m\right) \Psi =0,
\end{equation}
where $M$ are the space-time indices, and $A$, $B$ and $C$ are the tangent space indices collectively. We will denote tangent space indices with underlines as in $(\underline{r},\underline{t}, \underline{x}, \underline{y})$ or more compactly as $(\underline{r}, \underline{\mu})$ to distinguish from the space-time indices which will not be underlined as in $(r, t, x, y)$ or $(r, \mu)$.

In order to work with the holographic dictionary, it is convenient to choose the following representation for Gamma matrices \cite{incoming2}:
\begin{eqnarray}
\Gamma^{\underline{r}} = \begin{pmatrix} \mathbf{1} & 0\\ 0 & -\mathbf{1}\end{pmatrix}, \ \ \Gamma^{\underline{\mu}} = \begin{pmatrix} 0 & \gamma^{\underline{\mu}}\\ \gamma^{\underline{\mu}} & 0\end{pmatrix},
\end{eqnarray}
where $\gamma^{\underline{\mu}}$s are the  $2+1$ dimensional Gamma mtrices in a chosen representation. We will choose the latter in the following representation :
\begin{eqnarray}
\gamma^{\underline{t}} = i\sigma^{3}, \ \gamma^{\underline{x}} = \sigma^1, \ \gamma^{\underline{y}} = \sigma^2.
\end{eqnarray}
It is also useful to decompose the $3+1$ space-time dimensional Dirac spinor as eigenvectors of $\Gamma_{\pm}$ defined as :
\begin{equation}
\Gamma_{\pm} = \frac{1}{2}\Big(1 \pm \Gamma^{\underline{r}} \Big),
\end{equation}
so that
\begin{equation}
\Psi = \Psi_+ + \Psi_-, \quad \Psi_{\pm} = \Gamma_{\pm}\Psi .
\end{equation}
The advantage of this decomposition is that both $\Psi_+$ and $\Psi_-$ transform as 3 space-time dimensional Dirac spinors when the Gamma matrices are in the representation above.

It might be puzzling as to how a Dirac spinor in the bulk maps to two Dirac spinors in the boundary, but we note unlike the scalar field equation, the Dirac equation is first order. Therefore, as in the case of the scalar field we have two independent boundary data, corresponding to $\Psi_+$ and $\Psi_-$ each. Eventually, we will see how these two boundary data maps to source and expectation value of the dual operator, and further how they get related to each other by regularity in the bulk giving us the dual fermionic retarded propagator.

Just as in the case of the scalar field, the space-time profile of the Dirac spinor also has an equilibrium and non-equilibrium part. We can first assume that the equilibrium part is in a specific $(\omega, \mathbf{k})$ mode and determine the non-equilibrium correction to this. Later, we can obtain the most general solution by superimposing the full solutions corresponding to various equilibrium modes. The space-time profile of the Dirac spinor thus takes the following form :
\begin{widetext}
\begin{eqnarray}
\Psi(\mathbf{x},t, r) = \Psi^{(0)}(\omega, \mathbf{k}, r) e^{-i(\omega t- \mathbf{k}\cdot \mathbf{x})} + \Psi^{(1)}(\omega, \mathbf{k}, \omega_{\text{(h)}}, \mathbf{k}_{\text{(h)}}, r)e^{-i((\omega+\omega_{\text{(h)}}) t- (\mathbf{k}+\mathbf{k}_{\text{(h)}})\cdot \mathbf{x})}, 
\end{eqnarray}
\end{widetext}
where $\Psi^{(0)}$ is the equilibrium part, $\Psi^{(1)}$ is the non-equilibrium part, and $(\omega_{\text{(h)}}, \mathbf{k}_{\text{(h)}})$ correspond to the frequency and momenta of the velocity field perturbation in the background. From now on, we will denote $(\omega, \mathbf{k})$ collectively as $k$, and $(\omega_{\text{(h)}}, \mathbf{k}_{\text{(h)}})$ collectively as $k_{\text{(h)}}$.

The equations of motion for $\Psi$ can be written as two coupled first order PDEs for $\Psi_{\pm.}$ It will be convenient for us to decouple these PDEs and write a second order PDE for $\Psi_+$. It will turn out that $\Psi_-$ will be then algebraically determined by $\Psi_+$. For the equilibrium $AdS_4$ Reissner-Nordstorm black brane background, this has been done in \cite{Cubrovic}. Following this, we write the equations of motion for $\Psi_\pm^{(0)}$ as below :
\begin{eqnarray}\label{eom0}
\Bigg(\frac{\partial^2}{\partial r^2} + P(k, r) \frac{\partial}{\partial r} + Q(k, r)\Bigg)\ \Psi_+^{(0)}(k,r) &=& 0,\nonumber\\
\Psi_-^{(0)}(k,r)&=&  - \frac{\slashed{T}_k}{T_k^2} \ \Bigg(\frac{\partial}{\partial r}+\mathcal{A}^+\Bigg)\ \Psi_+^{(0)}(k,r),
\end{eqnarray}
where
\begin{eqnarray}\label{supp}
P(k, r) &=& \mathcal{A}^ + +  \mathcal{A}^-  - \frac{r_0}{l^2}\ \slashed{T}_k' \frac{\slashed{T}_k}{T_k^2}, \nonumber\\
Q(k, r) &=& \mathcal{A}^+ \mathcal{A}^- + \frac{r_0}{l^2}\mathcal{A}^{+'} - \frac{r_0}{l^2}\ \slashed{T}_k' \frac{\slashed{T}_k}{T_k^2}\ \mathcal{A}^+ + T_{k}^2 ,
\end{eqnarray}
and
\begin{eqnarray}
\mathcal{A}^{\pm} &=& -\frac{1}{2\,r}\, \Big[3-\frac{r\,f'\Big(\frac{r\,r_0}{l^2}\Big)}{2\,f\Big(\frac{r\,r_0}{l^2}\Big)}\, \frac{r_0}{l^2}\Big]\,\pm\,\frac{l}{r\, \sqrt{f\Big(\frac{r\,r_0}{l^2}\Big)}} m , \nonumber \\
\slashed{T}_k &=& \frac{i}{f\Big(\frac{r\,r_0}{l^2}\Big)}\Big[(-\omega + q\,A_t^{(0)})\,\gamma^{\underline{t}} \, + \,  \sqrt{f\Big(\frac{r\,r_0}{l^2}\Big)}\,k_i\,\gamma^{\underline{i}}\Big] ,
\end{eqnarray}
with $'$ denoting differentiation w.r.t. $rr_0/l^2$, $A_t^{(0)}$ representing the equilibrium configuration of the gauge field and $\mathcal{T}_k^2$ is $\slashed{T}_k\slashed{T}_k$.
 
In order to obtain the equations of motion for $\Psi^{(1)}_\pm$ we need to obtain the non-equilibrium first order corrections to the vielbeins and spin connections in the derivative expansion. These are given in details in appendix C with the metric being (\ref{metric1}) corresponding to the black brane perturbed by the hydrodynamic shear mode.

In order to simplify calculations, we will choose (without losing any generality) the momentum of the velocity field perturbation $\mathbf{k}_{\text{(h)}}$ in the background to be in the $x$ direction; therefore the velocity perturbation $\delta \mathbf{u}$ being transverse should then be in the $y$ direction. Later, we can make the results manifestly rotationally covariant by rotating, and also Lorentz covariant by boosting to an arbitrary frame.  The momentum of the equilibrium part of $\Psi$ of course can have arbitrary components in both $x$ and $y$ directions if we have to retain full generality.

The equations of motion of $\Psi^{(1)}$ are as follows :
\begin{widetext}
\begin{eqnarray}\label{eom1}
\Bigg(\frac{\partial^2}{\partial r^2} + P(k', r) \frac{\partial}{\partial r} + Q(k', r)\Bigg)\delta^3 (k'- k - k_{\text{(h)}})\ \Psi_+^{(1)}(k,k_{\text{(h)}}, r) &=& \Bigg(\frac{\partial}{\partial r} + \mathcal{A}^- -  \frac{r_0}{l^2}\slashed{T}_{k+k_{\text{(h)}}}' \frac{\slashed{T}_{k+k_{\text{(h)}}}}{T_{k+k_{\text{(h)}}}^2}  \Bigg)\ \mathcal{S}_+(k, k_{\text{(h)}}, r)  \nonumber\\&&-\slashed{T} _{k+k_{\text{(h)}}}\ \mathcal{S}_-(k, k_{\text{(h)}}, r)
,\nonumber\\
\delta^3(k' -k-k_{\text{(h)}})\Psi_-^{(1)}(k, k_{\text{(h)}}, r) &=&  - \frac{\slashed{T}_{k'}}{T_{k'}^2} \ \Bigg(\frac{\partial}{\partial r}+\mathcal{A}^+\Bigg)\delta^3(k'-k-k_{\text{(h)}})\ \Psi_+^{(1)}(k, k_{\text{(h)}},r) 
\nonumber\\&&+ \frac{\slashed{T}_{k+k_{\text{(h)}}}}{T_{k+k_{\text{(h)}}}^2}  \mathcal{S}_+(k, k_{\text{(h)}} ,r),
\end{eqnarray}
where
\begin{eqnarray}
\mathcal{S}_{+}(k, k_{\text{(h)}}, r) &=& - \mathcal{X}_{+}(k_{\text{(h)}}, r)\Psi_{+}^{(0)}(k,r) - \mathcal{Y}(k_{\text{(h)}}, r)\Psi_{-}^{(0)}(k,r) \nonumber \\
\mathcal{S}_{-}(k, k_{\text{(h)}},r) &=& - \mathcal{X}_{-}(k_{\text{(h)}}, r)\Psi_{-}^{(0)}(k,r) - \mathcal{Y}(k_{\text{(h)}} ,r)\Psi_{+}^{(0)}(k,r)
\end{eqnarray}
with 
\begin{eqnarray}
\mathcal{X}_{\pm}(k_{\text{(h)}}, r) &=& \mp\frac{1}{2}\Big(\mathcal{E}(k_{\text{(h)}} ,r)\,\gamma^{\underline{t}}\,\gamma^{\underline{y}}-\mathcal{F}(k_{\text{(h)}} ,r)\,\gamma^{\underline{x}}\gamma^{\underline{y}}\Big), \nonumber\\
\mathcal{Y}(k_{\text{(h)}}, r) &=& \frac{1}{2}\Big(\mathcal{B}(k_{\text{(h)}}, r)\,\gamma^{\underline{y}}-\mathcal{C}(k_{\text{(h)}} ,r)\,\gamma^{\underline{t}}\gamma^{\underline{x}}\gamma^{\underline{y}}\Big)+\mathcal{G}(k_{\text{(h)}},r)\,\gamma^{\underline{t}} +\mathcal{H}(k_{\text{(h)}}, r)\gamma^{\underline{x}}.
\end{eqnarray}
$\mathcal{B}$, $\mathcal{C}$, $\mathcal{E}$, $\mathcal{F}$, $\mathcal{G}$,$\mathcal{H}$ are given in terms of the inverse vielbeins and the spin connections as
\begin{eqnarray*}
\mathcal{B}(k_{\text{(h)}},r) &=& \frac{l}{r\sqrt{f\Big(\frac{rr_0}{l^2}\Big)}}\Big(2 i\, q\, e^y_{\underline{y}}\, A_y + 2 i\, q\, e^t_{\underline{y}}\, A_t + e^t_{\underline{t}} \,\omega_t^{\underline{t}\underline{y}} +2\,i\,\omega\,e^t_{\underline{y}}- 2\,i\,k_x\,e^x_{\underline{y}}\Big)^{(1)}\nonumber \\
\mathcal{C}(k_{\text{(h)}}, r) &=& \frac{l}{r\sqrt{f\Big(\frac{rr_0}{l^2}\Big)}}\Big(-e^t_{\underline{t}} \,\omega_t^{\underline{x}\underline{y}}-e^x_{\underline{x}} \,\omega_x^{\underline{t}\underline{y}}+e^y_{\underline{y}} \,\omega_y^{\underline{t}\underline{x}}\Big)^{(1)}\nonumber \\
\mathcal{E}(k_{\text{(h)}}, r) &=& \frac{l}{r\sqrt{f\Big(\frac{rr_0}{l^2}\Big)}}\Big(-e^t_{\underline{t}} \,\omega_t^{\underline{y}\underline{r}}-e^y_{\underline{y}} \,\omega_y^{\underline{t}\underline{r}}+e^r_{\underline{r}} \,\omega_r^{\underline{t}\underline{y}}-e^t_{\underline{y}} \,\omega_t^{\underline{t}\underline{r}}-e^y_{\underline{t}} \,\omega_y^{\underline{y}\underline{r}}\Big)^{(1)}
\end{eqnarray*}
\begin{eqnarray}
\label{coeff}
\mathcal{F}(k_{\text{(h)}}, r) &=& \frac{l}{r\sqrt{f\Big(\frac{rr_0}{l^2}\Big)}}\Big(e^x_{\underline{x}}\,\omega_x^{\underline{y}\underline{r}} -e^y_{\underline{y}} \,\omega_y^{\underline{x}\underline{r}}-e^x_{\underline{y}} \,\omega_x^{\underline{x}\underline{r}}+e^y_{\underline{x }}\,\omega_y^{\underline{y}\underline{r}}\Big)^{(1)}\nonumber \\
\mathcal{G}(k_{\text{(h)}}, r) &=&\frac{l}{r\sqrt{f\Big(\frac{rr_0}{l^2}\Big)}}\Big( -i\,k_y\, e^y_{\underline{t}}\Big)^{(1)} \nonumber \\
\mathcal{H}(k_{\text{(h)}}, r) &=& \frac{l}{r\sqrt{f\Big(\frac{rr_0}{l^2}\Big)}}\Big(-i\,k_y\, e^y_{\underline{x}}\Big)^{(1)}
\end{eqnarray}
\end{widetext}
Here $( \cdots)^{(1)}$ means that we are extracting only those parts of the full expression which is first order (i.e. linear) in $\mathbf{k}_{\text{(h)}}$. Once again we mention that the exact expressions of the inverse vielbeins (or einbeins) and spin connections appearing above are given in appendix C exactly up to first order in $\mathbf{k}_{\text{(h)}}$.

The most important observation regarding the equation of motion for $\Psi^{(1)}$ is that just as in the case of $\Psi^{(0)}$, as evident from (\ref{eom1}), $\Psi^{(1)}_+$ can be determined first by solving a second order ODE and $\Psi^{(1)}_-$ can be determined algebraically in terms of the solution for $\Psi_+$. Therefore to uniquely specify $\Psi^{(1)}$ it is sufficient to uniquely specify $\Psi^{(1)}_+$. Moreover, the differential operator on the LHS of the equation of motion (\ref{eom1}) for $\Psi^{(1)}_+$ is the same as that for $\Psi^{(0)}_+$ in (\ref{eom0}) with $k$ replaced by $k+k_{\text{(h)}}$. Therefore, the homogeneous solutions of $\Psi^{(1)}_+$ will be the same as those of $\Psi^{(0)}_+$ with $k$ replaced by $k+k_{\text{(h)}}$. 

The general behavior of the equilibrium part of the solution $\Psi^{(0)}_+$ at the horizon $r = l^2/r_0$ is 
\begin{eqnarray}
\Psi^{(0)}_+(\omega, \mathbf{k}, r) &\approx& A^{in}_+(\omega, \mathbf{k})\Big(1-\frac{rr_0}{l^2}\Big)^{-i\frac{\omega}{4\pi T}-\frac{1}{4}} + A^{out}_+(\omega, \mathbf{k})\Big(1-\frac{rr_0}{l^2}\Big)^{i\frac{\omega}{4\pi T}-\frac{1}{4}}.
\end{eqnarray}
Both $A^{in}_+$ and $A^{out}_+$ are arbitrary linear combinations of 
\begin{equation*}
\begin{pmatrix}1\\0\end{pmatrix} \ \text{and} \ \begin{pmatrix}0\\1\end{pmatrix}.
\end{equation*}
The incoming wave boundary condition requires us to impose 
\begin{equation}
A^{out}_+(\omega, \mathbf{k}) = 0.
\end{equation}
Furthermore, the choice of $A^{in}_+(\omega, \mathbf{k})$ will not matter in the final answer for the retarded propagator, so we will choose
\begin{equation}
A^{in}_+(\omega, \mathbf{k}) = \begin{pmatrix}\mathcal{K}\\0\end{pmatrix}
\end{equation}
with $\mathcal{K}$ being a constant. The behavior of $\Psi_-^{(0)}$ near the horizon can be obtained via the second algebraic equation of (\ref{eom0}) as below :
\begin{equation}
\Psi^{(0)}_-(\omega, \mathbf{k}, r) \approx -\gamma^{\underline{t}} \Big(1-\frac{rr_0}{l^2}\Big)^{-i\frac{\omega}{4\pi T}-\frac{1}{4}}\begin{pmatrix}\mathcal{K}\\0\end{pmatrix}.
\end{equation}
Thus $\Psi_-^{(0)}$ is also incoming at the horizon and is $\Psi_+^{(0)}$ times a specfic function of the frequency and momenta. 

It is to be noted that the incoming wave solution of the fermion diverges at the horizon as well. That this divergence is not an artifact of choice of coordinates can be seen by computing the scalar $\overline{\Psi}\Psi$ at the horizon. In fact, it is believed that the fermion backreaction at the horizon is strong enough to change the near horizon geometry of the black brane \cite{Tong}. As mentioned in the beginning of this section, we will assume here that the backreaction is suppressed by a factor of $O(1/N^2)$ \footnote{At order $O(1/N^2)$ we cannot ignore the backreaction even in the linearized limit. This is because the scalar and fermionic fields have non-trivial profiles even in the background due to Hawking radiation. Particularly, the Hawking radiated fermions forms a Fermi-sea in the near-horizon region of the $AdS$ Reissner-Nordstorm black hole.}. 

We now turn our attention to the non-equilibrium part of the solution. From, the first equation in (\ref{eom1}) we obtain that near the horizon, $\Psi^{(1)}_+$ behaves as :
\begin{widetext}
\begin{eqnarray}\label{bnh1}
\Psi^{(1)}_+(\omega, \mathbf{k}, \omega_{\text{(h)}}, \mathbf{k}_{\text{(h)}}, r) &\approx& A^{in}_+(\omega, \mathbf{k}, \omega_{\text{(h)}}, \mathbf{k}_{\text{(h)}})\Big(1-\frac{rr_0}{l^2}\Big)^{-i\frac{\omega+\omega_{\text{(h)}}}{4\pi T}-\frac{1}{4}} + A^{out}_+(\omega, \mathbf{k},\omega_{\text{(h)}}, \mathbf{k}_{\text{(h)}})\Big(1-\frac{rr_0}{l^2}\Big)^{i\frac{\omega + \omega_{\text{(h)}}}{4\pi T}-\frac{1}{4}}\nonumber\\&&
+\alpha(\omega, \mathbf{k}, \omega_{\text{(h)}}, \mathbf{k}_{\text{(h)}})\Big(1-\frac{rr_0}{l^2}\Big)^{-i\frac{\omega}{4\pi T}-\frac{3}{4}}\begin{pmatrix}\mathcal{K}\\0\end{pmatrix},
\nonumber\\
\alpha(\omega, \mathbf{k}, \omega_{\text{(h)}}, \mathbf{k}_{\text{(h)}})&=& \sqrt{\frac{r_0}{\pi T}}\frac{ \left(\omega_{\text{(h)}}(3 i \pi T -\omega +\omega_{\text{(h)}})-2(\pi^2 T^2 +\omega^2)\right)}{8 \left(3 \pi T + i \omega\right)\left(7 \pi T + i \omega\right)} \delta u_y (k_{\text{(h)}}) \gamma ^{\underline{t}}\gamma^{\underline{y}}.
\end{eqnarray}
\end{widetext}
when we have chosen the incoming wave boundary condition and our normalization for $\Psi_+^{(0)}$. Thus we have again two 
arbitrary coefficients for the incoming and outgoing homogeneous solutions at $2+1$ momenta $k+k_{\text{(h)}}$, and then we have a particular solution completely determined by the source term. 

We now apply a similar logic as in the case of the scalar field. We put $A^{out}$ in (\ref{bnh1}) to be zero again to satisfy the incoming boundary condition. In order to be consistent with the derivative expansion, $A^{in}$ has to linear combinations of $\delta\mathbf{u}(\omega_{\text{(h)}},\mathbf{k}_{\text{(h)}})\cdot \mathbf{k}$ and $k_i k_j k_{\text{(h)}i}\delta u_j(\omega_{\text{(h)}},\mathbf{k}_{\text{(h)}})$ with coefficients which are functions of $\omega$ and $\mathbf{k}$ only. The integration over $\omega_{\text{(h)}}$ in presence of $\delta\mathbf{u}$ will give contribution from the diffusion pole which will cause a further singularity in the behavior of the fermionic field. This singularity involves an extra factor of
\begin{equation*}
\Bigg(1 - \frac{rr_0}{l^2}\Bigg)^{- \frac{\mathbf{k}_{\text{(h)}}^2}{16 \pi^2 T^2}}. 
\end{equation*}
So we put $A^{in}$ to be zero too. There is however, a difference in the behavior of the particular solution near the horizon from the scalar case, as evident from (\ref{bnh1}). It diverges at the horizon with an extra factor of
\begin{equation*}
\Bigg(1 - \frac{rr_0}{l^2}\Bigg)^{-\frac{1}{2}}. 
\end{equation*}
The situation, therefore admittedly is confusing as both the incoming homogeneous solution and the particular solution are divergent by an extra power. Moreover, for sufficiently small hydrodynamic momenta $\mathbf{k}_{\text{(h)}}$, the divergence of the particular solution leads over that of the incoming homogeneous solution.

Nevertheless, we can argue as follows. When we take the backreaction into account, the part of the non-equilibrium solution completely determined by the source can be expected to be regular, as the source involving the regular equilibrium solution in the modified background will be regular in the next order in perturbation. This feature is observed in the case of fluid/gravity correspondence or for more general time-dependent solutions in gravity - if we make the solution regular up to $n-$th order in perturbation theory, the source terms in the equations for $n+1-$th order perturbations are also regular, and the divergences at the $n+1-$th order can be removed by adjusting the homogeneous solutions only \cite{Sayantani, myself4}. 

In the present case, we will argue that the divergence of the incoming homogeneous piece coming from the integration over $\omega_{\text{(h)}}$ is there as long as the backreacted background has a horizon at the zeroth order. If indeed there is a horizon, we can define an incoming wave also through geometrical optics approximation. We can certainly construct an appropriate function of $r$ which we denote as $r_*(r)$ such that the incoming radial null geodesic at the (modified) horizon is :
\begin{equation*}
v = t - r_*(r).
\end{equation*}
Clearly $ r_*(r)$ has to increase indefinitely as $r$ moves towards the horizon because of blue-shifting. The incoming wave at the horizon will always behave like :
\begin{equation*}
\approx e^{-i(\omega+\omega_{\text{(h)}})v}
\end{equation*}
as the geometrical optics approximation is always good at the horizon due to the blue-shifting. Therefore, as long as the backreacted geometry still has a horizon, the integration over $\omega_{(h)}$ will produce a divergent factor :
\begin{equation*}
\left(r_*(r)\right)^{\frac{\mathbf{k}_{\text{(h)}}^2}{16\pi^2 T^2}}.
\end{equation*}
Above we have used the result that the hydrodynamic dispersion relation up to the leading order remains the same in the presence of backreaction as $\eta/s$ is universally $1/4\pi$ in Einstein's gravity minimally coupled to any form of matter \cite{Iqbal}. Therefore, this divergence is not removable by backreaction as long as we do not get rid of the horizon completely. 

Getting rid of the horizon is generically impossible if we demand that the solution in gravity is well behaved, as that would expose the singularity unless the latter is also removed by the backreaction. The removal of singularity by back-reaction is impossible in Einstein's gravity minimally coupled to well-behaved matter. It is also hard to argue that solutions in gravity with naked singularities could be dual to states in thermal and chemical equilibrium in the dual theory.

We conclude that the sensible thing to do is to proceed as in the case of the scalar field and put \textit{both} $A^{in}$ and $A^{out}$ to zero in the non-equilibrium part of the solution. This determines $\Psi_+^{(1)}$ completely and its behavior near the horizon is :
\begin{eqnarray}
\Psi^{(1)}_+(\omega, \mathbf{k}, \omega_{\text{(h)}}, \mathbf{k}_{\text{(h)}}, r) 
&=& \sqrt{\frac{r_0}{\pi T}}\frac{ \left(\omega_{\text{(h)}}(3 i \pi T -\omega +\omega_{\text{(h)}})-2(\pi^2 T^2 +\omega^2)\right)}{8 \left(3 \pi T + i \omega\right)\left(7 \pi T + i \omega\right)} \delta u_y (k_{\text{(h)}}) \gamma ^{\underline{t}}\gamma^{\underline{y}}\nonumber\\&& \Big(1-\frac{rr_0}{l^2}\Big)^{-i\frac{\omega}{4\pi T}-\frac{3}{4}}\begin{pmatrix}\mathcal{K}\\0\end{pmatrix}
+ \ \text{sub-leading terms.}\end{eqnarray}
Once $\Psi_+^{(1)}$ is completely specified as above, we can determine $\Psi_-^{(1)}$ readily from the second equation in (\ref{eom1}) as it is algebraic.  The behavior near the horizon is given by :
\begin{eqnarray}
\Psi^{(1)}_-(\omega, \mathbf{k}, \omega_{\text{(h)}}, \mathbf{k}_{\text{(h)}}, r) 
&=& \sqrt{\frac{r_0}{\pi T}}\frac{\left(2 \pi i T - 2 \omega + \omega_{\text{(h)}}\right)\left(19 \pi^2 T^2 +11 i \pi T \omega-2 \omega^2 +\omega_{\text{(h)}}(2 i \pi T-\omega)\right)}{8  \left(3 \pi T + i \omega\right)\left(7 \pi T + i \omega\right)\left(\omega +\omega_{\text{(h)}}\right)}\nonumber\\&& \delta u_y (k_{\text{(h)}}) \gamma^{\underline{y}}\Big(1-\frac{rr_0}{l^2}\Big)^{-i\frac{\omega}{4\pi T}-\frac{3}{4}}\begin{pmatrix}\mathcal{K}\\0\end{pmatrix} + \ \text{sub-leading terms.}\end{eqnarray}
We can integrate numerically from the horizon and find the full profile of $\Psi_\pm$ (both equilibrium and non-equilibrium parts included) all the way up to the boundary. 

At the boundary, the behavior of $\Psi_\pm$ is specified completely by the $AdS_4$ asymptotic nature of the background. When $m\geq 0$, the behavior of $\Psi_+$ at the boundary is :
\begin{equation}\label{bnb1}
\Psi_+(k, k_{\text{(h)}} ,r) \approx \Big(J^{(0)}(k) + J^{(1)}(k,k_{\text{(h)}})\Big) r^{3-\Delta} + \Big(\mathcal{M}^{(0)}(k) + \mathcal{M}^{(1)}(k,k_{\text{(h)}})\Big) r^{\Delta + 1},
\end{equation}
with $\Delta$ being the scaling dimension of the dual operator and is related to the mass of the fermionic field by :
\begin{equation}
\Delta = \frac{3}{2} + ml.
\end{equation}\label{bnb2}
Clearly $J^{(0)}$ and $\mathcal{M}^{(0)}$ are determined by $\Psi_+^{(0)}$, and  $J^{(1)}$ and $\mathcal{M}^{(1)}$ are determined by $\Psi_+^{(1)}$. Similarly, the behavior of $\Psi_-$ at the boundary for $m\geq 0$ and $m\neq 1/2l$ is :
\begin{equation}
\Psi_-(k, k_{\text{(h)}} ,r) \approx \Big(\mathcal{N}^{(0)}(k) + \mathcal{N}^{(1)}(k,k_{\text{(h)}})\Big) r^{4-\Delta} + \Big(O^{(0)}(k) + O^{(1)}(k,k_{\text{(h)}})\Big) r^{\Delta}.
\end{equation}
When $m = 1/2l$, the leading powers of the homogeneous solutions above become the same. The behavior of $\Psi_-$ at the boundary is then given by :
\begin{equation}
\Psi_-(k, k_{\text{(h)}} ,r) \approx \Big(\mathcal{N}^{(0)}(k) + \mathcal{N}^{(1)}(k,k_{\text{(h)}})\Big)r^2 \text{ln}\ r  + \Big(O^{(0)}(k) + O^{(1)}(k,k_{\text{(h)}})\Big) r^{2}.
\end{equation}

As $\Psi_-$ is determined by $\Psi_+$ algebraically, we get
\begin{equation}
O(k, k_{\text{(h)}}) = - \frac{i \gamma \cdot k}{k^2}(2m+1)\mathcal{M}(k, k_{\text{(h)}}), \ \ \mathcal{N}(k, k_{\text{(h)}}) = \frac{i\gamma\cdot k}{(2m-1)}J(k,k_{\text{(h)}}), \ \ \gamma\cdot k = \gamma^{\underline{\mu}}k_\mu, \ \ k^2 = k^\mu k_\mu,
\end{equation}
where $O= O^{(0)} + O^{(1)}$, etc. Thus we have just 
two independent boundary data corresponding to the fermionic source and expectation value of the fermionic operator dual to the field. The holographic dictionary indeed identifies $J$ as the source and $O$ as the expectation value of the operator when $m\geq 0$ \cite{incoming2}. Both these are fixed up to an overall normalization constant by the incoming boundary condition at the horizon and our regularity argument. 

Changing the sign of $m$ is equivalent to interchanging $\Psi_+$ with $\Psi_-$ \cite{incoming2}. Consequently $J$ gets interchanged with $O$, and $\mathcal{M}$ gets interchanged with $\mathcal{N}$ \footnote{When $0\leq \vert m \vert < 1/2l$ we can also do an alternate quantization in which $O$ is interpreted as the source and $J$ as the expectation value. This requires the scaling dimension of the dual fermionic operator to be  $\Delta = 3/2 - \vert m\vert l$. The partition functions of the two theories are related by a Legendre transform.}. When $m<0$, the scaling dimension of the dual operator is given by :
\begin{equation}
\Delta = \frac{3}{2} - ml.
\end{equation}

Once the solution in the bulk is determined, the source and the expectation value of the fermionic operator get related by a matrix $\mathcal{D}$ :
\begin{equation}
J\Big(\omega, \mathbf{k}, \omega_{\text{(h)}}, \mathbf{k}_{\text{(h)}}\Big) = \mathcal{D}\Big(\omega, \mathbf{k}, \omega_{\text{(h)}}, \mathbf{k}_{\text{(h)}}\Big)\, O\Big(\omega, \mathbf{k}, \omega_{\text{(h)}}, \mathbf{k}_{\text{(h)}}\Big).
\end{equation}
Clearly $\mathcal{D}$ is independent of the choice of $A^{in}_+$ for the equilibrium solution as we have claimed earlier. The retarded propagator is given by \cite{incoming2}:
\begin{equation}
G_{R}\Big(\omega, \mathbf{k}, \omega_{\text{(h)}}, \mathbf{k}_{\text{(h)}}\Big) = i\mathcal{D}\Big(\omega, \mathbf{k}, \omega_{\text{(h)}}, \mathbf{k}_{\text{(h)}}\Big)\gamma^{\underline{t}}.
\end{equation}

Furthermore, as the non-equilibrium part of the solution is completely determined by the equilibrium part of the solution, we can compute the relations :
\begin{equation}
O^{(1)}\Big(\omega, \mathbf{k}, \omega_{\text{(h)}}, \mathbf{k}_{\text{(h)}}\Big) = \mathcal{R}_A\Big(\omega, \omega_{\text{(h)}},\mathbf{k},  \mathbf{k}_{\text{(h)}}\Big)\,O^{(0)}(\omega, k), \quad J^{(1)}\Big(\omega, \mathbf{k}, \omega_{\text{(h)}}, \mathbf{k}_{\text{(h)}}\Big) = \mathcal{R}_B\Big(\omega, \omega_{\text{(h)}},\mathbf{k},  \mathbf{k}_{\text{(h)}}\Big)\,J^{(0)}(\omega, \mathbf{k}).
\end{equation}
Above $\mathcal{R}_A$ and $\mathcal{R}_B$ are fully determined by our boundary conditions on $\Psi_+^{(1)}$ at the horizon. They take the form :
\begin{eqnarray}\label{opsof}
\mathcal{R}_A\Big(\omega, \omega_{\text{(h)}},\mathbf{k},  \mathbf{k}_{\text{(h)}}\Big) &=& \mathcal{R}_{AA}\Big(\omega,\mathbf{k}\Big)\delta\mathbf{u}(\omega_{\text{(h)}}, \mathbf{k}_{\text{h}})\cdot\mathbf{k} +\mathcal{R}_{AB}\Big(\omega, \mathbf{k}\Big)k_i k_j k_{\text{(h)}i}\delta u_j(\omega_{\text{(h)}},\mathbf{k}_{\text{(h)}}), \nonumber\\
\mathcal{R}_B\Big(\omega, \omega_{\text{(h)}},\mathbf{k},  \mathbf{k}_{\text{(h)}}\Big) &=& \mathcal{R}_{BA}\Big(\omega,\mathbf{k}\Big)\delta\mathbf{u}(\omega_{\text{(h)}}, \mathbf{k}_{\text{h}}) \cdot\mathbf{k}+\mathcal{R}_{BB}\Big(\omega, \mathbf{k}\Big)k_i k_j k_{\text{(h)}i}\delta u_j(\omega_{\text{(h)}},\mathbf{k}_{\text{(h)}}).
\end{eqnarray}

By going through the steps as in the case of the scalar field, we can easily see that the generalization of the form of the bosonic non-equilibrium retarded propagator (\ref{fund1}) to the fermionic case is :
\begin{eqnarray}\label{fund2}
G_R(\omega, \mathbf{k}, \mathbf{x}, t) &=& i\int d\omega_0 \int d^2 k_0 \Bigg[\mathcal{D}^{(0)}(\omega_0, \mathbf{k}_0)\gamma^{\underline{t}}\delta(\omega- \omega_0)\delta^2 (\mathbf{k} -\mathbf{k}_0)  \nonumber\\&&-\frac{1}{2\pi }
\Bigg(\mathcal{D}^{(0)}(\omega_0, \mathbf{k}_0)\gamma^{\underline{t}}\mathcal{R}_A\Big(\omega_0,  \mathbf{k}_0, \mathbf{k}_{\text{(h)}}\Big)\,\delta^2\Big(\mathbf{k} - \mathbf{k}_0 -\frac{\mathbf{k}_{\text{(h)}}}{2}\Big)\frac{1}{\Big(\omega - \omega _0 + i \frac{\mathbf{k}_{\text{(h)}}^2}{8\pi T}\Big)}\nonumber\\
&&-\mathcal{R}_B\Big(\omega_0, \mathbf{k}_0, \mathbf{k}_{\text{(h)}}\Big)\mathcal{D}^{(0)}(\omega_0, \mathbf{k}_0)\gamma^{\underline{t}}\delta^2\Big(\mathbf{k} - \mathbf{k}_0 +\frac{\mathbf{k}_{\text{(h)}}}{2}\Big)\frac{1}{ \Big(\omega - \omega _0 - i \frac{\mathbf{k}_{\text{(h)}}^2}{8\pi T}\Big)}\Bigg)e^{i \mathbf{k}_{\text{(h)}}\cdot\mathbf{x}}e^{- \frac{\mathbf{k}_{\text{(h)}}^2}{4\pi T}t}\Bigg],
\end{eqnarray}
where
\begin{eqnarray}\label{opsof1}
\mathcal{R}_A\Big(\omega, \mathbf{k},  \mathbf{k}_{\text{(h)}}\Big) &=& \mathcal{R}_{AA}\Big(\omega,\mathbf{k}\Big)\delta\mathbf{u}(\mathbf{k}_{\text{h}})\cdot\mathbf{k} +\mathcal{R}_{AB}\Big(\omega,\mathbf{k}\Big)k_i k_j k_{\text{(h)}i}\delta u_j(\mathbf{k}_{\text{(h)}}), \nonumber\\
\mathcal{R}_B\Big(\omega, \mathbf{k},  \mathbf{k}_{\text{(h)}}\Big) &=& \mathcal{R}_{BA}\Big(\omega,\mathbf{k}\Big)\delta\mathbf{u}(\mathbf{k}_{\text{h}}) \cdot\mathbf{k}+\mathcal{R}_{BB}\Big(\omega,  \mathbf{k}\Big)k_i k_j k_{\text{(h)}i}\delta u_j(\mathbf{k}_{\text{(h)}}).
\end{eqnarray}
The first line in (\ref{fund2}) denotes the equilibrium correlator and the lines below are the non-equilibrium contributions co-moving with the background velocity perturbation. The difference between (\ref{opsof1}) and (\ref{opsof}) is that the integration over $\omega_{\text{(h)}}$ has kept only the residue of the diffusion pole in the Fourier transform of $\delta \mathbf{u}$ given by (\ref{fourier}).

Once again the spectral function can be obtained by computing the imaginary part of the retarded propagator above and using $A(\omega, \mathbf{k},\mathbf{x}, t ) = -2 \text{Im}\Big(\text{Tr}(\gamma^{\underline{t}}G_R(\omega, \mathbf{k},\mathbf{x}, t ))\Big)$.

\subsection{Generalization to backgrounds with other quasinormal modes}

The prescription we have presented so far is for the non-equilibrium retarded propagator in the hydrodynamic shear-wave background. We will now show that this prescription with it's underlying logic can be readily generalized to any background which is a quasinormal mode fluctuation of the black brane geometry.

The key observations are as follows :
\begin{enumerate}
\item Even if the horizon fluctuates in presence of the non-equilibrium energy-momentum and charge current fluctuations in the dual state, i.e. the metric and gauge field quasinormal modes in the background, in the perturbation expansion, we need to apply the incoming boundary condition and regularity only at the radial location of the horizon at late time, which in our coordinates is always at $r = l^2/r_0$.

\item The quasinormal modes always have a negative imaginary part in their dispersion relation, so the pole in the complex frequency plane of the background perturbation will always be in the lower half plane. 
\end{enumerate}

The first point above makes sure that we can always write the non-equilibrium part of the solution as the incoming and outgoing homogeneous solutions  plus a particular solution completely specified by the source at the horizon exactly as in the case of the hydrodynamic shear mode. The second point will imply that integration over the background frequency will produce a divergence at the horizon unless we put the coefficients of both the incoming and outgoing parts of the non-equilibrium 
part of the solution to zero. Therefore, the non-equilibrium part of the solution is completely determined by the equilibrium part of the solution for any background quasinormal mode.  We can thus simply repeat the exercise as we have done for the hydrodynamic shear-mode perturbation to obtain the retarded propagator for any background quasinormal perturbation.

One may wonder if our prescribed solution at the horizon involving the specific particular solution is itself regular at the horizon. We have checked this is always so for the scalar field. In case of the fermionic field, we can repeat the arguments we have made in case of the hydrodynamic shear-mode.

For instance, let us consider a quasinormal mode for metric perturbation in the tensor channel with momentum $\mathbf{k}_{(b)} = 0$. The frequency will be complex with a negative imaginary part as in  (\ref{freqh}). The explicit metric and gauge field for such a spatially homogeneous perturbation is as in (\ref{metricnh}). We can check that our prescribed non-equilibrium solution for the scalar field dies down at the horizon due to the factors :
\begin{equation*}
\Big(1- \frac{rr_0}{l^2}\Big)^n \Bigg(\ln \Big(1- \frac{rr_0}{l^2}\Big)\Bigg)^m 
\end{equation*} 
multiplying the equilibrium incoming wave solution with $n$ and $m$ being positive integers \footnote{This can checked by expanding $\tilde{h}(s, \omega_{(n)})$ in (\ref{metricnh}) in $\omega_{(n)}$. Though this expansion as noted before is dangerous for seeing manifest regularity of the metric, it does good job for analyzing the behavior of the scalar field in the perturbed background.}.

The general dispersion relation for a quasi-normal mode may be written as :
\begin{equation}
\omega_{\text{(b)}}(\mathbf{k}_{\text{(b)}}) =  \omega_{\text{R(b)}}(\mathbf{k}_{(b)}) - i \omega_{\text{I(b)}}(\mathbf{k}_{\text{(b)}}), \quad \text{with}\quad \omega_{\text{I(b)}}(\mathbf{k}_{\text{(b)}})>0.
\end{equation}
Also both $\omega_{\text{R(b)}}$ and $\omega_{\text{I(b)}}$ admit Taylor expansion in $\mathbf{k}_{\text{(b)}}$ (and do not vanish when $\mathbf{k}_{\text{(b)}} =0$). The bosonic retarded propagator will take the following form in such a background :
\begin{eqnarray}
G_R(\omega, \mathbf{k}, \mathbf{x}, t) &=& \int d\omega_0 \int d^d k_0\Bigg[\frac{O^{(0)}(\omega_0, \mathbf{k}_0)}{J^{(0)}(\omega_0, \mathbf{k}_0)}\delta(\omega- \omega_0)\delta^2 (\mathbf{k} -\mathbf{k}_0) \nonumber\\&&- \frac{O^{(0)}(\omega_0, \mathbf{k}_0)}{J^{(0)}(\omega_0, \mathbf{k}_0)}\frac{1}{2\pi i}
\Bigg(\frac{ O^{(1)}(\omega_0, \mathbf{k}_0, \mathbf{k}_{\text{(b)}})}{O^{(0)}(\omega_0, \mathbf{k}_0)}\delta^2\Big(\mathbf{k} - \mathbf{k}_0 -\frac{\mathbf{k}_{\text{(b)}}}{2}\Big)\frac{1}{\Big(\omega - \omega _0 -\frac{1}{2}\Big(\omega_{\text{R(b)}}(\mathbf{k}_{(b)}) - i \omega_{\text{I(b)}}(\mathbf{k}_{\text{(b)}})\Big)\Big)}\nonumber\\
&&\qquad\qquad\qquad\qquad-\frac{ J^{(1)}\Big(\omega_0, \mathbf{k}_0, \mathbf{k}_{\text{(b)}}\Big)}{J^{(0)}(\omega_0, \mathbf{k}_0)}\delta^2\Big(\mathbf{k} - \mathbf{k}_0 +\frac{\mathbf{k}_{\text{(b)}}}{2}\Big)\frac{1}{ \Big(\omega - \omega _0 + \frac{1}{2}\Big(\omega_{\text{R(b)}}(\mathbf{k}_{(b)}) - i \omega_{\text{I(b)}}(\mathbf{k}_{\text{(b)}})\Big)\Big)}\Bigg)\nonumber\\&&
\qquad\qquad\qquad\qquad
e^{i \mathbf{k}_{\text{(b)}}\cdot\mathbf{x}}e^{- i\Big(\omega_{\text{R(b)}}(\mathbf{k}_{(b)}) - i \omega_{\text{I(b)}}(\mathbf{k}_{\text{(b)}})\Big)t}\Bigg].
\end{eqnarray}
The non-equilibrium part of the source and expectation values of the dual operators $J^{(1)}(\omega, \omega_{\text{(b)}},\mathbf{k}, \mathbf{k}_{\text{(b)}})$ and $O^{(1)}(\omega, \omega_{\text{(b)}},\mathbf{k}, \mathbf{k}_{\text{(b)}})$ can be determined from the non-equilibrium part of the solution. $J^{(1)}(\omega,\mathbf{k}, \mathbf{k}_{\text{(b)}})$ and $O^{(1)}(\omega,\mathbf{k}, \mathbf{k}_{\text{(b)}})$ appearing in the retarded propagator above are the residues of $J^{(1)}(\omega, \omega_{\text{(b)}},\mathbf{k}, \mathbf{k}_{\text{(b)}})$ and $O^{(1)}(\omega, \omega_{\text{(b)}},\mathbf{k}, \mathbf{k}_{\text{(b)}})$ respectively in $\omega_{\text{(b)}}$ at $\omega_{\text{R(b)}}(\mathbf{k}_{(b)}) - i \omega_{\text{I(b)}}(\mathbf{k}_{\text{(b)}})$. These will be linear in the hydrodynamic fluctuations $\delta \mathbf{u}_i$, $\delta T$, $\delta \rho$ and the non-hydrodynamic fluctuations $\delta \pi_{ij}^{(nh)}$, $\nu_0$ and $\nu_i^{(nh)}$, and will have a systematic expansion in $\mathbf{k}_{\text{(b)}}$ \footnote{The Taylor expansion is $\mathbf{k}_{\text{(b)}}$ always make sense near equilibrium as the perturbations are slowly varying in space. However, all time derivatives need to be summed up for non-hydrodynamic perturbations at each order in the amplitude and  $\mathbf{k}_{\text{(b)}}$ as the variation of these modes in time is not small even near equilibrium.}.  

Similarly, the fermionic non-equilbrium retarded propagator will take the general form :
\begin{eqnarray}\label{retgenf}
G_R(\omega, \mathbf{k}, \mathbf{x}, t) &=& i\int d\omega_0 \int d^d k_0\Bigg[\mathcal{D}^{(0)}(\omega_0, \mathbf{k}_0)\gamma^{\underline{t}}\delta(\omega- \omega_0)\delta^2 (\mathbf{k} -\mathbf{k}_0) \nonumber\\&&- \frac{1}{2\pi }
\Bigg(\mathcal{D}^{(0)}(\omega_0, \mathbf{k}_0)\,\gamma^{\underline{t}}\mathcal{R}_A\Big(\omega_0,  \mathbf{k}_0, \mathbf{k}_{\text{(b)}}\Big)\,\delta^2\Big(\mathbf{k} - \mathbf{k}_0 -\frac{\mathbf{k}_{\text{(b)}}}{2}\Big)\frac{1}{\Big(\omega - \omega _0 -\frac{1}{2}\Big(\omega_{\text{R(b)}}(\mathbf{k}_{(b)}) - i \omega_{\text{I(b)}}(\mathbf{k}_{\text{(b)}})\Big)\Big)}\nonumber\\
&&\qquad-\mathcal{R}_B\Big(\omega_0, \mathbf{k}_0, \mathbf{k}_{\text{(b)}}\Big)\,\mathcal{D}^{(0)}(\omega_0, \mathbf{k}_0)\gamma^{\underline{t}}\,\delta^2\Big(\mathbf{k} - \mathbf{k}_0 +\frac{\mathbf{k}_{\text{(b)}}}{2}\Big)\frac{1}{ \Big(\omega - \omega _0 + \frac{1}{2}\Big(\omega_{\text{R(b)}}(\mathbf{k}_{(b)}) - i \omega_{\text{I(b)}}(\mathbf{k}_{\text{(b)}})\Big)\Big)}\Bigg)
\nonumber\\&&\qquad e^{i \mathbf{k}_{\text{(b)}}\cdot\mathbf{x}}e^{- i\Big(\omega_{\text{R(b)}}(\mathbf{k}_{(b)}) - i \omega_{\text{I(b)}}(\mathbf{k}_{\text{(b)}})\Big)t}\Bigg].
\end{eqnarray}
$\mathcal{R}_A$ and $\mathcal{R}_B$ can be determined from the non-equilibrium part of the solution via the defining relations :
\begin{equation}
O^{(1)}\Big(\omega, \mathbf{k}, \omega_{\text{(b)}}, \mathbf{k}_{\text{(b)}}\Big) = \mathcal{R}_A\Big(\omega, \omega_{\text{(b)}},\mathbf{k},  \mathbf{k}_{\text{(b)}}\Big)\, O^{(0)}(\omega, k), \quad J^{(1)}\Big(\omega, \mathbf{k}, \omega_{\text{(b)}}, \mathbf{k}_{\text{(b)}}\Big) = \mathcal{R}_B\Big(\omega, \omega_{\text{(b)}},\mathbf{k},  \mathbf{k}_{\text{(b)}}\Big)\, J^{(0)}(\omega, \mathbf{k}).
\end{equation}
 $\mathcal{R}_A(\omega,\mathbf{k}, \mathbf{k}_{\text{(b)}})$ and $\mathcal{R}_B(\omega,\mathbf{k}, \mathbf{k}_{\text{(b)}})$ appearing in the retarded propagator above are the residues of $\mathcal{R}_A(\omega, \omega_{\text{(b)}},\mathbf{k}, \mathbf{k}_{\text{(b)}})$ and $\mathcal{R}_B(\omega, \omega_{\text{(b)}},\mathbf{k}, \mathbf{k}_{\text{(b)}})$ respectively in $\omega_{\text{(b)}}$ at $\omega_{\text{R(b)}}(\mathbf{k}_{(b)}) - i \omega_{\text{I(b)}}(\mathbf{k}_{\text{(b)}})$.
Both $\mathcal{R}_A(\omega,\mathbf{k}, \mathbf{k}_{\text{(b)}})$ and $\mathcal{R}_B(\omega,\mathbf{k}, \mathbf{k}_{\text{(b)}})$ will be linear in the hydrodyanamic fluctuations $\delta \mathbf{u}_i$, $\delta T$, $\delta \rho$ and the non-hydrodynamic fluctuations $\delta \pi_{ij}^{(nh)},$ $\nu_0$ and $\nu_i^{(nh)}$, and will have a systematic expansion in $\mathbf{k}_{\text{(b)}}$. 

Thus we indeed obtain an universal form of the holographic non-equilibrium retarded propagator (and hence the spectral function) in linearized non-equilibrium backgrounds at sufficiently late time.

\section{Non-equilibrium Fermi surface and dispersion relations}

We will show here that our prescription for obtaining the non-equilibrium retarded correlator gets a lot of support from field theoretic comparisons.  We will begin with a brief review of how we obtain non-equilibrium correlation functions in field theory. Then we will show how our prescription reproduces the strongly coupled version of non-equilibrium dynamics at the Fermi surface in Landau's Fermi-liquid theory, and the non-equilibrium modifications of quasi-particle dispersion relations expected in field theory.

\subsection{Comparison with field-theoretic approach}

In field theory, there is no partition function which can play the role of a generating functional of non-equilibrium correlation functions. The way we obtain these is to construct a generalized effective action $\Gamma\left(O_l(x), G_{ll'}(x,y)\right)$ whose arguments are not only the expectation value of the operator but also the two-point correlation functions of the operators. 
Extremizing this leads us to obtain non-equilibrium correlation functions as functionals of the expectation values of the operators in equilibrium and non-equilibrium states. The crucial point is that the generalized effective action has no dependence on temperature or other equilibrium/non-equilibrium parameters \footnote{This is also true for kinetic equations, like the Boltzmann equation. These equations do not depend on temperature or non-equilibrium parameters, the latter parametrize equilibrium and non-equilibrium solutions of these equations.}. It is defined as a double Legendre transform of a vacuum observable constructed over the Schwinger/Keldysh closed real time contour as briefly reviewed in appendix D. Both equilibrium (temperature and chemical potential dependent) and non-equilibrium dynamics of expectation values of operators and their correlation functions can be derived by extremizing this generalized effective action. At equilibrium, we can take an alternative route by constructing a generating functional of thermal correlation functions as in vacuum, but in order to obtain non-equilibrium correlation functions the use of the generalized effective action is indispensable.

We would like to mention here that the generalized effective action not only allows us to obtain the non-equilibrium two-point correlation functions, but it is also sufficient to obtain the three, four and higher point correlation functions \cite{reviews}. This is possible because through the effective action, we know the two point correlation function as a functional of expectation values of operators, i.e. we know them not in one but in a manifold of states. Furthermore, the effective action technique readily ensures that we satisfy Ward identities. 
In practice, we need to make an uncontrolled but educated approximation of the effective action which allows us to obtain non-equilibrium dynamics of expectation values of operators and their correlation functions. This has been successful for instance in the case of dilute cold non-relativistic Bose gases in optical traps \cite{Rammer}, and in constructing a quantum kinetic theory of hadrons for modeling their evolution after their chemical and thermal freeze-out in the RHIC fireball \cite{Bleicher}.

The important point to note is that we can obtain the non-equilibrium correlation function by extremizing the effective action with respect to the correlation function first as below :
\begin{equation}\label{extrem1}
\frac{\delta \Gamma\left(O_l, G^{0}_{ll'}(O_l)\right)}{\delta G^{0}_{ll'}(x,y)} = 0.
\end{equation}
Thus we obtain the two point correlation functions  as functionals of expectation values of the operators. Here the time contour is the Schwinger-Keldysh closed real time contour, so this determines both the statistical function and the retarded propagator (or the spectral function). Further when we substitute the extremal forms of the two-point correlation functions in the generalized effective action, we obtain the ordinary effective action, i.e.
\begin{equation}
 \Gamma\left(O_l, G^{0}_{ll'}(O_l)\right) = \Gamma(O_l).
\end{equation}
Extremizing this further we obtain non-equilibrium dynamics of expectation values of operators.

It is certainly interesting to see if we can construct a generalized effective action to obtain non-equilibrium correlation functions in holography too. This will allow us to determine the statistical function also and not the retarded propagator alone as we have done here.
Work in this direction will appear in \cite{progress1}. However, we note two crucial points of our holographic prescription for obtaining the retarded correlator (equivalently the spectral function).
\begin{enumerate}
\item Our prescription obtains the non-equilibrium retarded propagator as a functional of the expectation value of the energy-momentum tensor and the charged current parametrized by $T$, $\rho$, $\delta T$, $\delta \mathbf{u}$, $\delta \rho$, $\nu_i^{(nh)}$, $\nu_0$ and $\pi_{ij}^{(nh)}$. 

\item The non-equilibrium part of the correlation function is determined completely by the equilibrium part through universal rules at the horizon which do not depend on the non-equilibrium state concerned. The rule simply involves putting the homogeneous pieces of the non-equilibrium part of the solution of the bulk bosonic/fermionic field to zero at the horizon.
\end{enumerate}
Putting these together, we can see a parallel with field theory. In both approaches, we do not need a specific rule for each non-equilibrium state, there is a universal rule which allows us to extract the non-equilibrium correlation functions from observables defined at equilibrium. In field theory the equilibrium temperature arises as the boundary condition appearing in the far future. The generalized effective action as mentioned before is just the double Legendre transform of an equilibrium observable, therefore non-equilibrium dynamics can be obtained from equilibrium observables in field theory as well. Furthermore, our holographic prescription has the same measure of universality as the generalized effective action to bring all non-equilibrium spectral functions under one fold at least in perturbation theory.

The advantage of the holographic approach is that the late time behavior of the non-equilibrium spectral function is reproduced automatically without any need for resummation. Thus we can do conventional perturbation theory.

\subsection{Non-equilibrium dynamics at the Fermi-surface}

It might have been a bit surprising that the logic of regularity required that we put the extra boundary condition needed to determine the non-equilibrium part of the solution completely, at the horizon instead of at the boundary. It might seem that it would have been more natural to suppose that the source does not fluctuate from it's equilibrium value, so a Dirichlet condition at the boundary would have been more justified. As we have already argued, this is not the case - the source gets screened or dressed by the collective excitations present in the non-equilibrium state also. From the holographic perspective, the horizon determines the screening/dressing of the source.

We will here give another holographic interpretation of the non-equilibrium modification of the source. This will further vindicate that we need to put the extra universal boundary condition at the horizon and not at the boundary. That we have allowed the source to fluctuate from it's equilibrium value, is what will bring out the non-equilibrium oscillation of the energy per particle at the Fermi surface and non-equilibrium shifts in the quasi-particle dispersion relations.

A hallmark of Landau Fermi-liquid theory is that the collective modes as captured by the Boltzmann equation leads to non-equilibrium dynamics at the Fermi surface. This dynamics is characterized by \textit{shifts in energy per quasi-particle at the Fermi surface} $\delta\epsilon$ at a given direction $\hat{\mathbf{n}}$ and at a given point in space and time in response to a local fluctuation in occupation numbers of quasi-particles at the Fermi surface $\delta n$. Landau postulated the following phenomenological relation \cite{no}:
\begin{equation}\label{phen}
\delta\epsilon(k_F\hat{\mathbf{n}}, \mathbf{x}, t) = \epsilon(k_F\hat{\mathbf{n}}, \mathbf{x}, t) - \epsilon_0(k_F\hat{\mathbf{n}}) =
\sum_{\hat{\mathbf{n}}'}
\,f(\hat{\mathbf{n}}, \hat{\mathbf{n}}')\,  \delta n
(k_F\hat{\mathbf{n}}', \mathbf{x}, t),
\end{equation}
where $\epsilon_0(k_F\hat{\mathbf{n}})$ is the equilibrium energy of a quasi-particle at the Fermi surface which is just $k_F^2/2m^*(T,\mu)$ with $m*(T,\mu)$ being the effective mass at the Fermi surface dependent on temperature and chemical potential. The parameters $f(\hat{\mathbf{n}}, \hat{\mathbf{n}}')$ are phenomenological inputs of the Landau model which can be obtained from field-theoretic two-point density correlation functions. These phenomenological parameters determine all thermodynamic and many non-equilibrium properties of Fermi liquids.

To obtain non-equilibrium properties one has to assume validity of Boltzmann equation for $\delta n$. The equilibrium distribution $n^{(0)}$ is the Fermi-Dirac distribution at a fixed temperature and chemical potential and is a trivial solution of the Boltzmann equation. Using (\ref{phen}) and the Boltzmann equation, it can be shown that the fluctuations $\delta n$ follows :
\begin{equation}\label{bolt}
\frac{\partial \delta n(k_F\hat{\mathbf{n}}, \mathbf{x}, t)}{\partial t} + \frac{k_F\hat{\mathbf{n}}}{m^*(T,\mu)}\cdot \frac{\partial\delta n(k_F\hat{\mathbf{n}}, \mathbf{x}, t)}{\partial\mathbf{x}} + \frac{\partial n^{(0)}(k_F \hat{\mathbf{n}}, T, \mu)}{\partial \mathbf{k}}\cdot \sum_{\hat{\mathbf{n}}'}
\,f(\hat{\mathbf{n}}, \hat{\mathbf{n}}')\,  \frac{\partial\delta n
(k_F\hat{\mathbf{n}}', \mathbf{x}, t)}{\partial\mathbf{x}} = I\Big(n^{(0)}(T,\mu), \delta n(k_F\hat{\mathbf{n}}, \mathbf{x}, t)\Big)
\end{equation}
in the linearized limit. Above $I$ captures the so-called quasi-particle collision kernel. Studying this equation we can extract all collective excitations including the zero sound, hydrodynamic shear-mode and non-hydrodynamic relaxation modes. In order to obtain the zero sound velocity,  the collision kernel is not necessary but it is so in order to obtain the viscosity and relaxation modes. Substituting a solution for $\delta n$ in (\ref{phen}) we can obtain the oscillation of the energy per particle at the Fermi surface.

The crucial point is that the oscillation is related \text{locally} to the fluctuation in the occupation number of the quasi-particles in (\ref{phen}). So, the oscillation in energy per particle at the Fermi surface is in sync with the propagating collective excitation.

We note that in non-equilibrium, we cannot obtain the change in energy at the Fermi-surface by looking at the spectral function alone. This is because the non-equilibrium change in the spectral function comes from both (i) the shift of the residue, and (ii) the shift in the pole itself. We need to identify which part of the non-equilibrium contribution comes from the shift in the residue and which part comes from the shift in the pole.  Moreover, the situation could be worse, as there can be non-equilibrium contributions which are simply analytic near the location of the equilibrium Fermi surface and be neither the shift in the residue nor shift of the pole.

In the holographic set-up, the Fermi surface(s) is related to the existence of normalizable mode(s) of the bulk fermion field at zero frequency on a fixed momentum shell \cite{Liu}. As the black brane retains rotational symmetry, the Fermi surface is spherical (circular for a $2+1$ dimensional system). We will be working in $2+1$ dimensional system (i.e. in a $3+1$ dimensional bulk) for the sake of concreteness. 

It will be worthwhile for us to first define the Fermi surface holographically in a more general background which may not preserve rotational symmetry. This will help us to readily understand non-equilibrium dynamics at the Fermi surface.

A Fermi surface picks up an internal direction in spin space. Therefore, let us represent first an arbitrary normalized complex 2-vector which picks up a direction in spin space by two real angles $\theta$ and $\phi$ as below :
\begin{equation}\label{choice}
\begin{pmatrix} \cos\theta \, e^{i\phi}\\ \sin\theta \, e^{-i\phi}\end{pmatrix}
\end{equation}
The vector above may still be multiplied by an overall phase, but this will be unimportant for us. We then note that the hermitian matrix $P$ defined as
\begin{equation}\label{P}
P(\theta, \phi) = \begin{pmatrix}\cos^2 \theta & \cos\theta\sin\theta \, e^{i2\phi}\\ \cos\theta\sin\theta \, e^{-i2\phi} & \sin^2\theta \end{pmatrix}
\end{equation}
is a matrix such that
\begin{equation}
P^2  = P, \quad P\begin{pmatrix} \cos\theta \, e^{i\phi}\\ \sin\theta \, e^{-i\phi}\end{pmatrix} = \begin{pmatrix} \cos\theta \, e^{i\phi}\\ \sin\theta \, e^{-i\phi}\end{pmatrix} , \quad P\begin{pmatrix} \sin\theta \, e^{i\phi}\\ -\cos\theta \, e^{-i\phi}\end{pmatrix}= 0.
\end{equation}
Therefore $P$ is a projection operator, and it projects in the direction (\ref{choice}) and in the orthogonal direction it has eigenvalue zero.

The holographic definition of Fermi surface at equilibrium is as follows. Let us choose a direction $\hat{\mathbf{n}}$ in momentum space. Then there exists $\theta$, $\phi$ specifying a vector in \textit{spin space} and $k_F$ for every $\hat{\mathbf{n}}$ such that :
\begin{eqnarray}\label{fdef}
\Big[P(\theta, \phi), G_R(\omega = 0, \mathbf{k} = k_F \hat{\mathbf{n}})\Big] &=& 0, \nonumber\\
P(\theta, \phi)J\Big(\omega = 0, \mathbf{k} = k_F \hat{\mathbf{n}}\Big) &=& 0.
\end{eqnarray}
where $P$ is as defined in (\ref{P})and $J$ is the source obtained from the bulk solution. The first equation above says that 
$G_R$ is diagonal in spin space in the following basis :
\begin{equation}\label{basis}
\begin{pmatrix} \cos\theta \, e^{i\phi}\\ \sin\theta \, e^{-i\phi}\end{pmatrix} , \quad \begin{pmatrix} \sin\theta \, e^{i\phi}\\ -\cos\theta \, e^{-i\phi}\end{pmatrix}
\end{equation}
which is the same basis in which $P$ is diagonal. Thus this defines $\theta$ and $\phi$. We note if we replace $\theta$ by $\theta + \pi/2$, we merely exchange the eigenbasis. Therefore, if $\theta$ is a solution, so is $\theta+\pi/2$. The second equation is equivalent to :
\begin{equation}
J\Big(\omega = 0, \mathbf{k} \equiv k_F \hat{\mathbf{n}}\Big) \equiv \begin{pmatrix}
\chi_1 \\ \chi_2 \end{pmatrix}, \quad \cos\theta e^{-i\phi}\chi_1 = \sin\theta e^{i\phi}\chi_2.
\end{equation}
Thus we have one linear complex equation to define $k_F$. Therefore $k_F$ is complex (at finite temperature) and associated with a specific direction in spin space. To get the Fermi surface associated with the orthogonal direction in spin space which is also an eigenvector of $P$ and $G_R$ we need to solve above with $\theta$ replaced by $\theta + \pi/2$, i.e. 
\begin{equation}
J\Big(\omega = 0, \mathbf{k} \equiv k_F \hat{\mathbf{n}}\Big) \equiv \begin{pmatrix}
\chi_1 \\ \chi_2 \end{pmatrix}, \quad \sin\theta e^{-i\phi}\chi_1 = -\cos\theta e^{i\phi}\chi_2.
\end{equation}
As the $AdS_4$ Reissner-Nordstorm black brane background preserves rotational invariance, $\theta$, $\phi$ and $k_F$ will be independent of $\mathbf{n}$.

More generally, the holographic Fermi surface is $k_F(\mathbf{n})$ which solves (\ref{fdef}) and is associated with a specific direction in spin space in which the retarded propagator can be diagonalized. The general definition stated here should be useful in analyzing cases where we have spontaneous symmetry breaking in the boundary, particularly when these order parameters break rotational invariance \cite{pwave}. We note that at zero temperature $k_F$ is strictly real and corresponds to the pole at $\omega = 0$, but for non-zero temperatures the pole at $\omega = 0$ is complex. The imaginary part of the pole is negative and represents smearing of the Fermi surface at finite temperature, and vanishes as the temperature is reduced to zero. Thus we can think of $k_F$ as a complex parameter whose imaginary part vanishes at zero temperature and has a small $T$ expansion. The real part of $k_F$ also has a small $T$ expansion and is the Fermi surface. There is no dependence on $\omega$ as to find the Fermi surface $\omega$ is set to zero. In the Reissner-Nordstorm black brane, the dependence of the negative imaginary part of this complex parameter $k_F$ on the temperature is given by a power law for small temperatures \cite{finiteT}. This power is controlled by the near horizon $AdS_2 \times R^2$ geometry.

It can also be shown that the retarded propagator and the spectral function also have a pole precisely when the source vanishes. Therefore, the holographic definition of the Fermi surface matches with the conventional definition which is that it is the location of pole of the spectral function in momentum space at vanishing frequency. In holographic systems we typically get a family of nested Fermi surfaces.

As an aside let us mention that the pole structure of the holographic spectral function at equilibrium is different at small frequencies from that of a conventional Fermi liquid and the scaling exponents are controlled by the near-horizon $AdS_2\times R^2$ geometry \cite{Faulkner}. This means that holographic  systems have generically non-Fermi liquid behavior.

The full non-equilibrium source is :
\begin{equation}
J\Big(\mathbf{x}, t\Big) = \int d^3x \  \Bigg(J^{(0)}\Big(\omega, \mathbf{k}\Big) + J^{(1)}\Big(\omega, \mathbf{k}, \mathbf{k}_{\text{(b)}}\Big)e^{i \mathbf{k}_{\text{(b)}}\cdot\mathbf{x}}e^{- i\Big(\omega_{\text{R(b)}}(\mathbf{k}_{(b)}) - i \omega_{\text{I(b)}}(\mathbf{k}_{\text{(b)}})\Big)t}\Bigg)e^{-i(\omega t - \mathbf{k}\cdot \mathbf{x})},
\end{equation}
We recall that the full source $J$ can be determined from the boundary behavior of our prescribed solution for $\Psi_+$ through (\ref{bnb1}). In fact we can explicitly write in case of the hydrodynamic shear-mode up to first order in the hydrodynamic momentum $\mathbf{k}_{\text{(h)}}$ :
\begin{eqnarray}\label{U}
J^{(1)}(\omega, \mathbf{k}, \mathbf{k}_{\text{(h)}}) &=& J^{(1)}_A\Big(\omega, \mathbf{k}\Big) \delta\mathbf{u}(\mathbf{k}_{\text{(h)}})\cdot \mathbf{k} \, +\, J^{(1)}_B\Big(\omega, \mathbf{k}\Big)\,k_ik_j k_{\text{(h)}i}\delta u_j(\mathbf{k}_{\text{(h)}}),
\end{eqnarray}
where $J^{(1)}_A$ and $J^{(1)}_B$ can be determined from the solution.

We will be interested in obtaining the energy oscillation $\delta\omega
(\hat{\mathbf{n}}, \mathbf{x}, t)$ at the Fermi surface by calculating shift of the frequency pole for a fixed Fermi momentum. We have to solve this perturbatively in the momentum of the collective background mode $\mathbf{k}_{\text{(b)}}$. 

Perturbatively, the energy shift on the Fermi surface $\delta \omega$ in the direction $\hat{\mathbf{n}}$ at a given point in space-time is thus obtained by solving :
\begin{eqnarray}
\delta \omega (\hat{\mathbf{n}}, \mathbf{x}, t) \Bigg(P(\theta^{(0)}, \phi^{(0)})\partial_{\omega} J^{(0)}\Big(\omega = 0, \mathbf{k}= k_F\hat{\mathbf{n}}\Big)\Bigg) &+& \delta \theta (k_F \hat{\mathbf{n}}, \mathbf{x}, t) \Bigg(\partial_\theta P(\theta^{(0)}, \phi^{(0)}) J^{(0)}\Big(\omega = 0, \mathbf{k} = k_F\hat{\mathbf{n}}\Big)\Bigg) \nonumber\\ +\delta \phi (\hat{\mathbf{n}}, \mathbf{x}, t) \Bigg(\partial_\phi P(\theta^{(0)}, \phi^{(0)}) J^{(0)}\Big(\omega = 0, \mathbf{k} = k_F\hat{\mathbf{n}}\Big)\Bigg)  &=& -\Bigg(P(\theta^{(0)}, \phi^{(0)}) J^{(1)}\Big(\omega = 0, \mathbf{k}= k_F\hat{\mathbf{n}}, \mathbf{k}_{\text{(b)}}\Big)\Bigg)\nonumber\\&& e^{i \mathbf{k}_{\text{(b)}}\cdot\mathbf{x}} \,\,\, e^{- i\Big(\omega_{\text{R(b)}}(\mathbf{k}_{(b)}) - i \omega_{\text{I(b)}}(\mathbf{k}_{\text{(b)}})\Big)t},
\end{eqnarray}
where $\theta^{(0)}$ and $\phi^{(0)}$ label the spin orientation of the equilibrium Fermi surface as discussed before and $P$ is as defined in (\ref{P}). The above amounts to two complex equations and we have four unknowns, namely real $\delta\theta$ and $\delta\phi$ giving change in the orientation in spin space and complex 
$\delta\omega$. As we have mentioned earlier, the change in orientation in spin space cannot be directly read off from the change in retarded correlator due to the ambiguity in identifying which change is due to shift in the pole and which change is due to shift in the residue. We can obtain the non-equilibrium shift in spin space at the Fermi surface from the non-equilibrium source directly.
 
The shift in the energy of the equilibrium Fermi surface with orthogonal spin orientation can be obtained by solving the above equation with $\theta^{(0)}$ replaced by $\theta^{(0)} + \pi/2$.

Clearly in the hydrodynamic shear wave background, $\delta \omega$ takes the form :
\begin{equation}
\delta \omega (\hat{\mathbf{n}}, \mathbf{x}, t) = \Bigg(\delta\omega_{A}\Big(\hat{\mathbf{n}}, k_F\Big) \delta\mathbf{u}(\mathbf{k}_{\text{(h)}})\cdot \hat{\mathbf{n}} \, +\, \delta\omega_{B}\Big(\hat{\mathbf{n}}, k_F \Big)\,\hat{n}_i \hat{n}_j k_{\text{(h)}i}\delta u_j(\mathbf{k}_{\text{(h)}})\Bigg)e^{i \mathbf{k}_{\text{(h)}}\cdot\mathbf{x}}e^{- i\frac{\mathbf{k}_{\text{(h)}}^2}{4\pi T}t}.
\end{equation}

Therefore, we find that the holographic Fermi surface indeed oscillates in space and time in sync with the background collective excitation.  Nevertheless in order to obtain the analogue of (\ref{phen}) in holography linking the spectral shift at the Fermi surface to the statistical shift (i.e. shift in occupation number) we need to obtain the statistical function holographically also. We leave this for the future.

\subsection{Non-equilibrium shifts in energy and spin of quasi-particles}

Not only the energy per particle at the Fermi surface but other normalizable modes with non-zero frequencies also receive space-time dependent shifts in energy at a given momentum in sync with the background collective excitation. This can be interpreted as the space-time dependent shifts of the dispersion relations of the quasi-particles in the non-equilibrium medium. This is certainly expected as quasi-particles receive a thermal mass and if the temperature oscillates for instance, the dispersion relations indeed become space-time dependent. This is usually a hard calculation in non-equilibrium quantum field theory, but we can readily generalize the holographic strategy discussed above to obtain non-equilibrium shifts in quasi-particle dispersion relations.

A particular quasi-particle branch can be identified via the following steps at equilibrium.
\begin{enumerate}
\item Consider the equilibrium Green's function $G_R^{(0)}(\omega, \mathbf{k})$. This can be diagonalized at  a given $\omega$ and $\mathbf{k}$ and the eigenvectors can be labelled as in (\ref{basis}) by $\theta^{(0)}(\omega, \mathbf{k})$ and $\phi^{(0)}(\omega, \mathbf{k})$. Furthermore, if $\theta^{(0)}$ is a solution, so is the $\theta^{(0)}+\pi/2$ as this merely exchanges the eigenbasis.

\item The quasiparticle pole $\omega^{(0)}(\mathbf{k})$ can be identified with a definite orientation in spin space by solving :
\begin{equation}
P(\theta^{(0)}, \phi^{(0)}) \, J^{(0)}\Big(\omega^{(0)}(\mathbf{k}), \mathbf{k}\Big) = 0.
\end{equation}
The above amounts to one complex equation which determines $\omega^{(0)}(\mathbf{k})$ with $\theta^{(0)}$ and $\phi^{(0)}$ determined in the previous step. The imaginary part of $\omega^{(0)}(\mathbf{k})$ is negative. To obtain the quasi-particle branch with opposite spin orientation, we need to solve the above with $\theta^{(0)}$ replaced by $\theta^{(0)}+ \pi/2$. 
\end{enumerate}

Once again, if there is rotational symmetry in the background, i.e. if there are no order paramaters of spontaneous symmetry breaking which breaks rotational invariance, $\theta^{(0)}$, $\phi^{(0)}$ and $\omega^{(0)}(\mathbf{k})$ can depend only on the modulus of $\mathbf{k}$.
 
The space-time dependent shift in dispersion relation is characterized by :
\begin{equation}
\omega = \omega^{(0)}(\mathbf{k}) + \delta\omega(\mathbf{k}, \mathbf{k}_{\text{(b)}},\mathbf{x}, t).
\end{equation}
The shift $\delta\omega$ can be obtained by solving :
\begin{eqnarray}
\delta \omega (\mathbf{k}, \mathbf{x}, t) \Bigg(P(\theta^{(0)}, \phi^{(0)})\partial_{\omega} J^{(0)}\Big(\omega = \omega^{(0)}(\mathbf{k}), \mathbf{k}\Big)\Bigg) &+& \delta \theta ( \mathbf{k}, \mathbf{x}, t) \Bigg(\partial_\theta P(\theta^{(0)}, \phi^{(0)})J^{(0)}\Big(\omega = \omega^{(0)}(\mathbf{k}), \mathbf{k}\Big)\Bigg) \nonumber\\ +\delta \phi (\mathbf{k}, \mathbf{x}, t) \Bigg(\partial_\phi P(\theta^{(0)}, \phi^{(0)})J^{(0)}\Big(\omega = \omega^{(0)}(\mathbf{k}), \mathbf{k}\Big)\Bigg)  &=& -\Bigg(P(\theta^{(0)}, \phi^{(0)}) J^{(1)}\Big(\omega = \omega^{(0)}(\mathbf{k}),\mathbf{k}, \mathbf{k}_{\text{(b)}}\Big)\Bigg)\nonumber\\&& e^{i \mathbf{k}_{\text{(b)}}\cdot\mathbf{x}} \,\,\, e^{- i\Big(\omega_{\text{R(b)}}(\mathbf{k}_{(b)}) - i \omega_{\text{I(b)}}(\mathbf{k}_{\text{(b)}})\Big)t}.
\end{eqnarray}
The above equation amounts to two complex equations which allows us to solve the real unknowns $\delta\theta$ and $\delta\phi$ giving shifts in spin space and the complex unknown $\delta\omega$. To obtain the non-equilibrium shift in the dispersion relation for the other equilibrium branch with orthogonal spin orientation, we need to solve the above with $\theta^{(0)}$ replaced by $\theta^{(0)}+ \pi/2$.

 The solution of $\delta\omega$ will take the form in a hydrodynamic shear-wave background, for instance, clearly takes the form :
\begin{equation}
\delta \omega (\mathbf{k}, \mathbf{x}, t) = \Bigg(\delta\omega_{A}(\mathbf{k}) \, \delta\mathbf{u}(\mathbf{k}_{\text{(h)}})\cdot \mathbf{k} \, +\, \delta\omega_{B}(\mathbf{k})\, k_i k_j k_{\text{(h)}i}\delta u_j(\mathbf{k}_{\text{(h)}})\Bigg)e^{i \mathbf{k}_{\text{(h)}}\cdot\mathbf{x}}e^{- i\frac{\mathbf{k}_{\text{(h)}}^2}{4\pi T}t}.
\end{equation}

Therefore, we see that the shift in the dispersion relation of the quasi-particle pole is also in sync with the propagating collective mode. Furthermore, though we have discussed the fermionic case explicitly here, clearly the same strategy can be applied to the bosonic field also. In fact, the source being a complex number instead of a complex 2-vector in the bosonic case, the equations will be much simpler.

The shift  $\delta\omega$ in the quasi-particle pole is generically complex. Interestingly the sign of the imaginary part of $\delta\omega$ can be both positive and negative. Thus we can get both non-equilibrium suppression or enhancement of quasi-particle decays as indeed observed in the RHIC fireball for various resonances (short-lived quasi-particles) \cite{Florkowski}.

\section{Taking into account non-linearities in the dynamics of the non-equilibrium variables}

It is known that solutions of gravity which have regular future horizons reproduce non-linear phenomenological equations for irreversible processes in the dual field theory. The best studied examples are related to fluid/gravity correspondence. The full non-linear Navier-Stokes' equation with higher derivative corrections can be reproduced from gravity and this success has also been extended to the case of charged hydrodynamics \cite{Sayantani}. As we have discussed before, gravity is expected to reproduce the general phenomenological equations which describe the full evolution of energy-momentum tensor and conserved currents which generalize hydrodynamics \cite{myself1}. This has been shown explicitly for the case of spatially homogeneous relaxation \cite{myself2}. In all cases, the regularity of the future horizon determines the phenomenological coefficients.

We would like to show that the prescriptions described so far can be readily generalized to include non-linearities in the dynamics of the energy-momentum tensor and conserved currents characterizing the non-equilibrium states.
We can systematically include these non-linearities into the retarded correlator, the shifts in the dispersion relations of quasi-particles, etc. 

The key is to see how the solutions for the bosonic and fermionic fields get determined in the perturbed background. Let us focus on the case of the hydrodynamic background. If we take into account non-linearities in $\delta \mathbf{u}(\mathbf{k}_{\text{(h)}})$ in the background, clearly these non-linearities will also appear in the Laplacian of the bosonic field. Let us consider quadratic dependence on two distinct velocity perturbations $\delta \mathbf{u}(k_{\text{(h)}})$ and $\delta \mathbf{u}(k_{\text{(h)}}')$ for instance, at a given order in the derivative expansion $m$ (i.e. at the $m$th order in the hydrodynamic momentum). The solution for $\Phi$ will receive a correction quadratic in the amplitude of velocity perturbation and at $m$th order in the derivative expansion which can be represented as:
\begin{equation}
\Phi^{(2,m)}\Big(r, k, k_{\text{(h)}}, k_{\text{(h)}}'\Big)\, e^{i(k + k_{\text{(h)}} + k_{\text{(h)}}')\cdot x}.
\end{equation}
The radial dependence above can be determined the equation of motion :
\begin{equation}
\Box^{ARN}_k \delta^3 (k -k')
\Phi^{(2,m)}(r, k', k_{\text{(h)}}, k_{\text{(h)}}')= S^{(2,m)}(r, k,k_{\text{(h)}}, k_{\text{(h)}}'),
\end{equation}
where  $\Box^{ARN}_k$ is the Laplacian for a scalar with $3$-momentum $k$ in the unperturbed $AdS_4$ Reissner-Nordstorm background and $S^{(2,m)}$ is a generic source term. For $m=1$ the source $S^{(2,1)}$ can contain terms like $(\mathbf{k}\cdot\delta\mathbf{u}(k_{\text{(h)}}))(\mathbf{k_{\text{(h)}}}\cdot\delta\mathbf{u}(k_{\text{(h)}}'))$, etc. It also contains the solutions at the lower order in the perturbation expansion for instance $\Phi^{(1, 1)}\Phi^{(1, 0)}$.

Clearly the general solution of $\Phi$ near the horizon can again be separated into two homogeneous pieces, the incoming and the outgoing modes, and a particular piece which has no arbitrary integration constant and is completely determined by the source term $S^{(2,m)}$. In order to satisfy the incoming boundary condition, we should put the coefficient of the outgoing mode to zero. Also as discussed before, the integration over the hydrodynamic frequencies $\omega_{\text{(h)}}$ and  $\omega_{\text{(h)}}'$ will produce a divergence at the horizon for the incoming mode, as for instance in the case above with dependence on two hydrodynamic shear wave background modes like :
\begin{equation}
\Big(1 - \frac{rr_0}{l^2}\Big)^{-i\frac{\omega}{4\pi T}- \frac{\mathbf{k}_{\text{(h)}}^2}{16 \pi^2 T^2}- \frac{\mathbf{k'}_{\text{(h)}}^2}{16 \pi^2 T^2} - ...}.
\end{equation}
Obviously the coefficient of the incoming mode has to depend on $\delta\mathbf{u}$ and the hydrodynamic momenta required by the order in the perturbation expansion. The contribution from the frequency pole in $\delta\mathbf{u}(\omega_{\text{(h)}}, \mathbf{k}_{\text{(h)}})$ given by the hydrodynamic shear dispersion relation produces the above divergent behavior.
In general the divergence will always be there for any quasinormal wave  background as it's dispersion relation $\omega_{\text{(h)}}(\mathbf{k}_{\text{(h)}})$ will have a negative imaginary part. Therefore, we should put the coefficients of the incoming mode at the horizon to zero too. We are just left with the particular piece which is completely determined by the source term containing the solutions at the lower orders in the perturbation expansion. Therefore, applying induction, \textit{at each order in the perturbation expansion, the non-equilibrium solution is uniquely determined by the equilibrium solution, i.e. the solution at the zeroth order in the unperturbed black brane background}. The consistency of holographic duality requires the solution at each order in the perturbation to be regular at the horizon.

As the solution is uniquely fixed at each order in the perturbation expansion, we can obtain the non-equilibrium contributions to the source and the expectation value of the dual operator by studying the asymptotic behavior of the solution at each order. This procedure can also be applied for fermionic fields.

Once the source is obtained at a given order in the perturbation expansion, it is straightforward to obtain the shift in the dispersion relation of quasi-particles. For example,  $\delta\omega^{(2,m)}(\mathbf{k}, \mathbf{x}, t)$ along with the non-equilibrium shift in the spin orientation given by $\delta\theta^{(2,m)}(\mathbf{k}, \mathbf{x}, t)$ and $\delta\phi^{(2,m)}(\mathbf{k}, \mathbf{x}, t)$ can be obtained from $J^{(2,m)}$ by solving :
\begin{eqnarray}
\delta \omega^{(2,m)} (\mathbf{k}, \mathbf{x}, t) \Bigg(P(\theta^{(0)}, \phi^{(0)})\partial_{\omega} J^{(0)}\Big(\omega = \omega^{(0)}(\mathbf{k}), \mathbf{k}\Big)\Bigg) &+& \delta \theta^{(2,m)} ( \mathbf{k}, \mathbf{x}, t) \Bigg(\partial_\theta P(\theta^{(0)}, \phi^{(0)})J^{(0)}\Big(\omega = \omega^{(0)}(\mathbf{k}), \mathbf{k}\Big)\Bigg) \nonumber\\ +\delta \phi^{(2,m)} (\mathbf{k}, \mathbf{x}, t) \Bigg(\partial_\phi P(\theta^{(0)}, \phi^{(0)})J^{(0)}\Big(\omega = \omega^{(0)}(\mathbf{k}), \mathbf{k}\Big)\Bigg)  &=& -\Bigg(P(\theta^{(0)}, \phi^{(0)}) J^{(2,m)}\Big(\omega = \omega^{(0)}(\mathbf{k}),\mathbf{k}, \mathbf{k}_{\text{(b)}}\Big)\Bigg)e^{i (\mathbf{k}_{\text{(b)}}+\mathbf{k'}_{\text{(b)}})\cdot\mathbf{x}} \nonumber\\&&  \, \, \,e^{- i\Big((\omega_{\text{R(b)}}(\mathbf{k}_{\text{(b)}}) + \omega_{\text{R(b)}}(\mathbf{k'}_{\text{(b)}}))- i (\omega_{\text{I(b)}}(\mathbf{k}_{\text{(b)}})+\omega_{\text{I(b)}}(\mathbf{k'}_{\text{(b)}})\Big)t}.
\end{eqnarray}

A consistent perturbation theory for the solution in the non-equilibrium background thus suffices to take into account non-linearities in $\delta \mathbf{u}_i$, $\delta T$, $\delta \pi_{ij}^{(nh)}$ etc. in the retarded correlation function, spectral function, non-equilibrium shift in the dispersion relations, etc.

\section{Discussion}

This paper has been devoted to developing a general holographic formalism for determining non-equilibrium retarded correlator, spectral function, shifts in dispersion relations, etc. Needless to say, we would like to use this formalism to numerically calculate these space-time dependent quantities in the specific set up of charged bosonic and fermionic fields minimally coupled to Einstein-Maxwell gravity in $AdS_4$ discussed here. In particular, the following questions require attention.
\begin{enumerate}
\item It is known that at equilibrium the temperature modifies the spectral function only in the infrared, while in the ultraviolet the spectral function remains as in the vacuum. It can be expected that we have a similar feature even in non-equilibrium - the ultraviolet behavior of the spectral function, quasi-particle dispersion relations should be independent of the state. It will be interesting to see if this is reproduced in our case. Some of the background quasi-normal modes indeed can have very high frequencies, while high frequency dependent corrections can also be generated by non-linearities. Therefore, numerical studies can help us understand how the effect of high frequency dependent background modes gets suppressed in the ultraviolet, if this is indeed the case.

\item The non-equilibrium shifts in the dispersion relations can have both positive and negative imaginary parts. If positive it leads to suppression and if negative it leads to enhancement of the decay. It will be interesting to see if one can use non-linearities to design a background in which a specific quasi-particle can be stabilized against decay to a large extent in a certain range of energies. This can allow us to observe otherwise short-lived quasi-particles. In particular, it will be interesting to see if some bound states of heavy quarks can indeed exist in the quark-gluon plasma at temperature 175 MeV.

\item The quasi-particle dispersion relations can change non-analytically with the temperature particularly if there is level crossing. It will be interesting to design a non-equilibrium background where the temperature varies in space and time over the range in which this non-analyticity can occur and study exactly how the quasi-particles behave in such backgrounds. It will be interesting to learn from such holographic examples how to describe such non-equilibrium states in field theory. 
\end{enumerate}
Work is in progress by the authors of this paper to tackle such issues numerically \cite{progress2}. Our prescription here gives an algorithm to tackle such questions in specific holographic models.

Another direction we want to pursue in the future is to study non-equilibrium spectral functions in states corresponding to a plasma undergoing boost-invariant hydrodynamic expansion as in the RHIC fireball. This will give us insights into how hadrons are produced and transported in the medium, and finally get frozen chemically and thermally.

\begin{acknowledgments}
The authors would like to thank C. Gowdigere for collaboration in the initial stages of this work. The authors would also like to thank R. Gopakumar and G. Policastro for valuable comments on the manuscript. AM would like to thank C. Bachas and G. Policastro for inspiration and useful discussions. AM would also like to thank all the members of the AdS/CMT discussion group in Paris for inspiration. The discussion in section II.B has been motivated by questions posed by D. Mateos during AM's seminar at Department de Fisica Fonamental, Universitat de Barcelona. The results of section IV.B and IV.C followed from valuable discussions with G. Policastro. AM would also like to thank the Institute for Theoretical Physics, Utrecht University, the organizers of SCGSC 2011, London and the organizers of Rencontres Th\'{e}oriciennes
held at Institut Henri Poincar\'e, Paris  for opportunities to present this work prior to publication. SB would like to thank the organizers of Asian String School 2012 for their hospitality, and Theory Center, KEK, Japan and Yukawa Intitute for Theoretical Physics, Japan for opportunities to present this work prior to publication. SB would like to thank S. Mukherji and M. Natsuume for inspiration and useful discussions which have helped us improve our manuscript. The research of AM is supported by the grant number ANR-07-CEXC-006 of L'Agence Nationale de La Recherche. RI would like to thank the Division of Biology at Calfornia Institute of Technology, where part of this work was carried out, for their hospitality.  

\end{acknowledgments}

\newpage

\section*{Appendices} 
\appendix
\section{Eddington-Finkelstein vs Schwarzchild coordinates}
In order to see regularity at the horizon manifestly in the metric (\ref{metric1}) corresponding to hydrodynamic shear-mode perturbation of the $AdS_4$ Reissner-Nordstorm black brane, we can consider the following change of coordinates following \cite{myself4} :
\begin{eqnarray}\label{trnsl}
t &=& v + \frac{l^2}{r_0}k\Big(\frac{rr_0}{l^2}\Big) + O(\epsilon^2), \nonumber\\
x^i &=& \tilde{x}^i + \frac{l^2}{r_0}k\Big(\frac{rr_0}{l^2}\Big) \ \delta u^i (k_{\text{(h)}}) e^{i(\mathbf{k}_{\text{(h)}} \cdot\tilde{x}-\omega_{\text{(h)}} v)} \nonumber\\&&- i \omega_{\text{(h)}} \frac{l^4}{r_0^2}k_1\Big(\frac{rr_0}{l^2}\Big) \ \delta u^i (k_{\text{(h)}}) e^{i(\mathbf{k}_{\text{(h)}} \cdot\tilde{x}-\omega_{\text{(h)}} v)} + O(\epsilon^2),
\end{eqnarray}
where
\begin{equation}\label{k}
k(a) = \int_0^a db \frac{1}{f(b)},
\end{equation}
and
\begin{equation}\label{k_1}
k_1(a) = \int_0^a db \ \Big(\frac{1-f(b)}{f(b)}\Big)k(b).
\end{equation}
These new coordinates $r$, $v$ and $\tilde{x}^i$ are ingoing Eddington-Finkelstein coordinates.

In these coordinates, the metric assumes the form :
\begin{eqnarray}\label{metric2}
ds^2 &=& -\frac{2l^2}{r^2}\Big(dv - \delta u_i (k_{\text{(h)}}) e^{i(\mathbf{k}_{\text{(h)}} \cdot\tilde{x}-\omega_{\text{(h)}} v) }d\tilde{x}^i\Big) dr +\frac{l^2}{r^2}\Bigg(-f\Big(\frac{rr_0}{l^2}\Big)dv^2+d\tilde{x}^2 + d\tilde{y}^2\Bigg)\nonumber\\
&& -2\delta u_i (k_{\text{(h)}}) e^{i(\mathbf{k}_{\text{(h)}} \cdot\tilde{x}-\omega_{\text{(h)}} v)}\Bigg(1-f\Big(\frac{rr_0}{l^2}\Big)+i\frac{\omega_{\text{(h)}} l^2}{r_0} f\Big(\frac{rr_0}{l^2}\Big)k\Big(\frac{rr_0}{l^2}\Big)\Bigg)dv \ d\tilde{x}^i\nonumber\\
&&-i2\frac{l^2}{r_0}k_{\text{(h)i}} \ \delta u_j (k_{\text{(h)}}) e^{i(\mathbf{k}_{\text{(h)}} \cdot\tilde{x}-\omega_{\text{(h)}} v)}\Bigg(\frac{1}{3}h\Big(\frac{rr_0}{l^2}\Big)-k\Big(\frac{rr_0}{l^2}\Big)\Bigg)d\tilde{x}^i\ d\tilde{x}^j \nonumber\\&&+ O(\epsilon^2).
\end{eqnarray}

The bulk gauge field however no longer remains in the radial gauge and takes the form :
\begin{eqnarray}\label{gf2}
 A_r &=& -i\frac{1}{f\Big(\frac{rr_0}{l^2}\Big)} \frac{\sqrt{3}g_F r_0}{l^2} \left(1 -\frac{r r_0}{l^2} \right) + O(\epsilon^2), 
\nonumber\\ A_v &=&\frac{\sqrt{3}g_F r_0}{l^2} \left(1 -\frac{r r_0}{l^2} \right) + O(\epsilon^2),\nonumber\\
A_{i} &=& -\frac{\sqrt{3}g_F r_0}{l^2} \left(1 -\frac{r r_0}{l^2} \right)\delta u_i (k_{\text{(h)}}) e^{i (\mathbf{k}_{\text{(h)}} \cdot \tilde{x}- \omega_{\text{(h)}} v)}\Big(1 -i\frac{\omega_{\text{(h)}} l^2}{r_0} k\Big(\frac{rr_0}{l^2}\Big)\Big)+O(\epsilon^2).
\end{eqnarray}
It can be checked that the gauge field is also regular at the horizon. $A_v$, $A_i$ vanish while $A_r$ is a constant at the horizon. We can bring the gauge field back to radial gauge by a regular gauge transformation.

Most importantly, the $ij$ components of the metric is regular as
\begin{eqnarray}
\frac{1}{3}h(a) - k(a) &=& \ \text{terms which are regular at the horizon (i.e. at $a=1$).}
\end{eqnarray}
So, the metric is manifestly regular up to the first order in the derivative expansion in these coordinates.
 
We can implement this change of coordinates order by order in the derivative expansion. Even beyond the fluid/gravity correspondence, such coordinate transformations can be implemented perturbatively to see manifest regularity \cite{myself2}.

\section{The general solution for the non-equilibrium profile of the scalar field}

At the zeroth order, the equilibrium solution for a given mode can be written as an arbitrary linear superposition of two linearly independent homogeneous solutions $\Phi^A (k, r)$ and $\Phi^B (k,r)$. Here $k$ denotes $(\omega, \mathbf{k})$ collectively. Thus
\begin{equation}
\Phi^{(0)}(k,r) = A^{(0)}(k)\Phi^{A}(k, r) + B^{(0)}(k)\Phi^{B}(k, r),
\end{equation}
where $A^{(0)}(k)$ and $B^{(0)}(k)$ are arbitrary.

Using the method of variation of parameters, we can write the general solution for the equation of  motion (\ref{noneqeom}) for the non-equilibrium part can be found and is as below :
\begin{eqnarray}
\Phi(k,k_{\text{(h)}},r)&=& -\Phi^{A}(k+k_{\text{(h)}},r)\int_{l_1}^{r} dr' \frac{\Phi^{B}(k+k_{\text{(h)}}, r') \Big(V_1+V_2\Big)(k, k_{\text{(h)}}, r') \Phi^{(0)}(k,r)}{W[\Phi^{A}(k+k_{\text{(h)}} ,r'),\Phi^{B}(k+k_{\text{(h)}}, r')]r^{'2} f\Big(\frac{r'r_0}{l^2}\Big)}
\nonumber\\
&&+\Phi^{B}(k+k_{\text{(h)}},r)\int_{l_2}^{r} dr' \frac{\Phi^{A}(k+k_{\text{(h)}} ,r')\Big(V_1+V_2\Big)(k, k_{\text{(h)}}, r') \Phi^{(0)}(k,r)}{W[\Phi^{A}(k+k_{\text{(h)}}, r'),\Phi^{B}(k+k_{\text{(h)}}, r')]r^{'2} f\Big(\frac{r'r_0}{l^2}\Big)}.
\end{eqnarray}
Above $k_{\text{(h)}}$ denotes $(\omega_{\text{(h)}}, \mathbf{k}_{\text{(h)}})$ collectively, $W$ denotes the Wronskian of the two homogeneous solutions, and $l_1$ and $l_2$ are arbitrary setting the range of the two integrals.

One can readily verify that the above is independent of the choice of $\Phi^{A}$ and $\Phi^{B}$ for fixed $l_1$ and $l_2$. To see the general behavior at the horizon given by (\ref{solh2}) one can set $\Phi^A$ to be $\Phi^{in}$ and $\Phi^B$ to be $\Phi^{out}$.

Furthermore, one notes that the above is consistent with the derivative expansion for any $l_1$ and $l_2$ as the dependence on $\delta u_i$ and $k_{\text{(h)}}$ comes from $V_1$ and $V_2$ directly. Comparing (\ref{opso}) with (\ref{sterms}) one gets that the explicit contribution to $O^{(1)}_A$ and $J^{(1)}_A$ comes from $V_1$, and the contribution to $O^{(1)}_B$ and $J^{(1)}_B$ comes from $V_2$. 

\section{Vielbeins and spin connections in the hydrodynamically perturbed black-brane metric}

We calculate here vielbeins, their inverses (or einbeins) and spin connections for the metric (\ref{metric1}) which corresponds to a black brane perturbed by a hydrodynamic shear mode. The notation we use here is the same as defined in section III.B of the paper. As noted there, to ease computations we will choose, without loss of generality that $\delta \mathbf{u}$ is in the $y$ direction. As  $\delta \mathbf{u}$ is transvere in the shear-mode, $\mathbf{k}_{\text{(h)}}$ will be then in the $x$ direction. On the other hand $\mathbf{k}$ can have arbitrary $x$ and $y$ components in order to retain full generality.

The non-zero vielbeins upto first order of derivative expansion are :

\begin{eqnarray}
e^{\underline{t}}_{t} &=& \frac{l}{r} \, \sqrt{f\Big(\frac{r r_0}{l^2}\Big)}, \quad
e^{\underline{t}}_{y} = \frac{l}{2 r} \, \frac{1-f \Big(\frac{r r_0}{l^2}\Big)}{\sqrt{f \Big(\frac{r r_0}{l^2}\Big)}} \, \delta u_y (k_{\text{(h)}}) \, e^{i(k_{\text{(h)x}} x -\omega_{\text{(h)}} t)}, \nonumber\\
e^{\underline{x}}_{x} &=& \frac{l}{r}, \quad 
e^{\underline{x}}_{y} = -i \frac{l}{r} \, \Big(\frac{l^2}{6r_0^2}\Big) k_{\text{(h)}x} \, \delta u_y (k_{\text{(h)}})\, e^{i(k_{\text{(h)x}} x -\omega_{\text{(h)}} t)} \,h\Big(\frac{r r_0}{l^2}\Big), \nonumber\\
e^{\underline{y}}_{t} &=& -\frac{l}{2 r} \, \Big(1-f\Big(\frac{r r_0}{l^2}\Big)\Big) \, \delta u_y (k_{\text{(h)}})\, e^{i(k_{\text{(h)x}} x -\omega_{\text{(h)}} t)}, \nonumber\\
e^{\underline{y}}_{x} &=& -i \frac{l}{r} \,\Big(\frac{l^2}{6r_0^2}\Big) \, k_{\text{(h)}x} \, \delta u_y (k_{\text{(h)}}) \, e^{i(k_{\text{(h)x}} x -\omega_{\text{(h)}} t)} \, h\Big(\frac{r r_0}{l^2}\Big),  \nonumber\\
e^{\underline{y}}_{y} &=& \frac{l}{r}, \quad
e^{\underline{r}}_{r} = \frac{l}{r} \, \frac{1}{\sqrt{f\Big(\frac{r r_0}{l^2}\Big)}}.
\end{eqnarray}

From this, one can also construct inverse vielbeins (or einbeins) which are as follows :

\begin{eqnarray}
\label{invv}
e^{t}_{\underline{t}} &=& \frac{r}{l \, \sqrt{f\Big(\frac{r r_0}{l^2}\Big)}}, \quad
e^{t}_{\underline{y}} = -\frac{r}{2 l} \, \frac{1-f\Big(\frac{r r_0}{l^2}\Big)}{f\Big(\frac{r r_0}{l^2}\Big)} \, \delta u_y (k_{\text{(h)}}) \, e^{i(k_{\text{(h)x}} x -\omega_{\text{(h)}} t)}, \nonumber\\
e^{x}_{\underline{x}} &=& \frac{r}{l}, \quad
e^{x}_{\underline{y}} = i \frac{r}{l} \, \Big(\frac{l^2}{6r_0^2}\Big) \, k_{\text{(h)}x} \, \delta u_y (k_{\text{(h)}}) \, e^{i(k_{\text{(h)x}} x -\omega_{\text{(h)}} t)} \, h\Big(\frac{r r_0}{l^2}\Big), \nonumber\\
e^{y}_{\underline{t}} &=&  \frac{r}{2 l} \, \frac{1-f\Big(\frac{r r_0}{l^2}\Big)}{\sqrt{f\Big(\frac{r r_0}{l^2}\Big)}} \, \delta u_y (k_{\text{(h)}})\, e^{i(k_{\text{(h)x}} x -\omega_{\text{(h)}} t)}, \nonumber\\
e^{y}_{\underline{x}} &=&  i \frac{r}{l} \, \Big(\frac{l^2}{6r_0^2}\Big) \, k_{\text{(h)}x} \, \delta u_y (k_{\text{(h)}}) \, e^{i(k_{\text{(h)x}} x -\omega_{\text{(h)}} t)} \,h \Big(\frac{r r_0}{l^2}\Big), \nonumber\\
e^{y}_{\underline{y}} &=& \frac{r}{l} \quad
e^{r}_{\underline{r}} = \frac{r \sqrt{f\Big(\frac{r r_0}{l^2}\Big)}}{l}.
\end{eqnarray}

In order to derive the equation of motion of the Fermions in the given background, we require the spin connections
associated with the first order metric. The non-zero components of the spin connection,\,$\omega^{\underline{A}\,\underline{B}}_M$ are as below :
\begin{eqnarray*}
\omega^{\underline{t}\underline{x}}_y &=& -\omega^{\underline{x}\underline{t}}_y = -\frac{i \,\delta u_y (k_{\text{(h)}})\, e^{i(k_{\text{(h)x}} x -\omega_{\text{(h)}} t)}\, k_{\text{(h)}x}\, \Big[3r_0^3 \Big(-1+f\Big(\frac{r r_0}{l^2}\Big)\Big) - 4\, i \, l^4 \, \eta\,  \kappa^2 \, h\Big(\frac{r r_0}{l^2}\Big)\, \omega_{\text{(h)}}\Big]}{6 r_0^3\sqrt{f\Big(\frac{r r_0}{l^2}\Big)}}, \nonumber\\
\omega^{\underline{t}\underline{y}}_t &=& -\omega^{\underline{y}\underline{t}}_t = -\frac{i \, \delta u_y (k_{\text{(h)}})\, e^{i(k_{\text{(h)x}} x -\omega_{\text{(h)}} t)}\,\Big(-1+f\Big(\frac{r r_0}{l^2}\Big)\Big)\,\omega_{\text{(h)}}}{2 \,\sqrt{f\Big(\frac{r r_0}{l^2}\Big)}},\nonumber\\
\omega^{\underline{t}\underline{y}}_x &=& -\omega^{\underline{y}\underline{t}}_x = -\frac{l^2\delta u_y (k_{\text{(h)}})\, e^{i(k_{\text{(h)x}} x -\omega_{\text{(h)}} t)} \, h\Big(\frac{r r_0}{l^2}\Big)\, k_{\text{(h)}x}\, \omega_{\text{(h)}}}{3\,r_0^2\,\sqrt{f\Big(\frac{r r_0}{l^2}\Big)}}, \nonumber\\
\omega^{\underline{t}\underline{y}}_r &=& -\omega^{\underline{y}\underline{t}}_r = -\frac{\delta u_y (k_{\text{(h)}})\,e^{i(k_{\text{(h)}} x -\omega_{\text{(h)}} t)}\, r_0\,\Big(-1+f\Big(\frac{r r_0}{l^2}\Big)\Big)f'\Big(\frac{r r_0}{l^2}\Big)}{4\, l^2\, f^{\frac{3}{2}}\Big(\frac{r r_0}{l^2}\Big)}, \nonumber\\
\omega^{\underline{t}\underline{r}}_t &=& -\omega^{\underline{r}\underline{t}}_t = \frac{-f\Big(\frac{r r_0}{l^2}\Big)}{r}\,+\,\frac{r_0\,f'\Big(\frac{r r_0}{l^2}\Big)}{2\, l^2}\nonumber\\
\omega^{\underline{t}\underline{r}}_y &=& -\omega^{\underline{r}\underline{t}}_y = \frac{\delta u_y (k_{\text{(h)}})\, e^{i(k_{\text{(h)x}} x -\omega_{\text{(h)}} t)}\Big[l^2 \,\Big(-1+f\Big(\frac{r r_0}{l^2}\Big)\Big) -r\, r_0\, f'\Big(\frac{r r_0}{l^2}\Big)\Big]}{2\, l^2\, r}, \nonumber\\
\omega^{\underline{x}\underline{y}}_t &=& -\omega^{\underline{y}\underline{x}}_t = -\frac{1}{2}\, i\, \delta u_y (k_{\text{(h)}})\, e^{i(k_{\text{(h)x}} x -\omega_{\text{(h)}} t)}\, \Big(-1+f\Big(\frac{r r_0}{l^2}\Big)\Big)\, k_{\text{(h)}x},
\nonumber\\
\omega^{\underline{x}\underline{r}}_x &=& -\omega^{\underline{r}\underline{x}}_x = -\frac{\sqrt{f\Big(\frac{r r_0}{l^2}\Big)}}{r},\nonumber\\
\omega^{\underline{x}\underline{r}}_y &=& -\omega^{\underline{r}\underline{x}}_y = \frac{i\, \delta u_y (k_{\text{(h)}})\, e^{i(k_{\text{(h)x}} x -\omega_{\text{(h)}} t)}\, \sqrt{f\Big(\frac{r r_0}{l^2}\Big)}\, k_{\text{(h)}x}\, \Big(3\, l^2\, h\Big(\frac{r r_0}{l^2}\Big) - 2\, r\, r_0\, h'\Big(\frac{r r_0}{l^2}\Big)\Big)}{6\, r\, r_0^2} ,
 \end{eqnarray*}
\begin{eqnarray}
\omega^{\underline{y}\underline{r}}_t &=& -\omega^{\underline{r}\underline{y}}_t = -\frac{\delta u_y (k_{\text{(h)}})\, e^{i(k_{\text{(h)x}} x -\omega_{\text{(h)}} t)}\,\Big[2 \,l^2 \,\Big(-1 + f\Big(\frac{r r_0}{l^2}\Big)\Big)\,f\Big(\frac{r r_0}{l^2}\Big) - r\, r_0\, \Big(1+f\Big(\frac{r r_0}{l^2}\Big)\Big) f'\Big(\frac{r r_0}{l^2}\Big)\Big]}{4\, l^2\, r\, \sqrt{f\Big(\frac{r r_0}{l^2}\Big)}}, \nonumber\\
\omega^{\underline{y}\underline{r}}_x &=& -\omega^{\underline{r}\underline{y}}_x = \frac{i\, \delta u_y (k_{\text{(h)}})\, e^{i(k_{\text{(h)x}} x -\omega_{\text{(h)}} t)}\, \sqrt{f\Big(\frac{r r_0}{l^2}\Big)}\, k_{\text{(h)x}}\, \Big(3\, l^2\, h\Big(\frac{r r_0}{l^2}\Big) -2\, r\, r_0\, h'\Big(\frac{r r_0}{l^2}\Big)\Big)}{6\, r\, r_0^2},\nonumber\\
\omega^{\underline{y}\underline{r}}_y &=& -\omega^{\underline{r}\underline{y}}_y = -\frac{\sqrt{f\Big(\frac{r r_0}{l^2}\Big)}}{r}.
\end{eqnarray}
Here prime denotes derivative with respect to $\frac{r r_0}{l^2}$.

It can be checked that the above spin connections satisfy Cartan structure equations up to first order in the derivative expansion.

\section{The generalized effective action}

We will review the formalism for bosonic operators here. The generalization to fermionic operators is straightforward.

The starting point of the construction of the generalized effective action is to generalize the partition function which is a generating functional of the vaccum correlation functions. Here on top of a source $J_l(x)$ for a single operator $\mathcal{O}_l(x)$, we add a non-local source $K_{ll'}(x,y)$ for a pair of operators $\mathcal{O}_l(x)$ and $\mathcal{O}_{l'}(y)$, and define:
\begin{widetext}
\begin{eqnarray}
Z(J_l, K_{ll'})&=& e^{i W(J_l , K_{ll'})}\nonumber\\
\phantom{G} &=& \int \mathcal{D}\Phi_s \exp \Bigg[i\Bigg(S[\Phi_s] + \int d^{D}x J_l(x)\mathcal{O}_l(x) +\frac{1}{2}\int d^D xd^Dy \ \mathcal{O}_l(x)K_{ll'}(x,y)\mathcal{O}_{l'}(y)\Bigg)\Bigg]. 
\end{eqnarray}
\end{widetext}
Above $D$ is the number of space-time dimensions in field theory.

We then define the expectation value of the operator $O_l(x)$ and the Green's function $G_{ll'}(x,y)$ through :
\begin{eqnarray}
\frac{\delta W(J_l, K_{ll'})}{\delta J_l(x)} &=& O_l (x), \nonumber\\
\frac{\delta W(J_l, K_{ll'})}{\delta K_{ll'}(x,y)} &=& \frac{1}{2}\Bigg(O_l(x)O_{l'} (y)+ G_{ll'}(x,y)\Bigg).
\end{eqnarray}
Eliminating $J_l$ and $K_{ll'}$ in favor of $O_l$ and $G_{ll'}$, we can now do a Legendre transform to define the generalized effective action :
\begin{widetext}
\begin{eqnarray}
\Gamma(O_l, G_{ll'}) &=& W(J_l, K_{ll'}) -  \int d^{D}x J_l(x)O_l(x) -\frac{1}{2}\int d^D xd^Dy \ K_{ll'}(x,y)\Bigg(O_l(x)O_{l'}(y)+G_{ll'}(x,y)\Bigg). 
\end{eqnarray}
\end{widetext}
Clearly,
\begin{eqnarray}\label{extrem}
\frac{\delta\Gamma(O_l, G_{ll'})}{\delta O_l(x)} &=& -J_l (x) - \int d^Dy \ K_{ll'}(x,y) O_{l'}(y), \nonumber\\
\frac{\delta\Gamma(O_l, G_{ll'})}{\delta G_{ll'}(x,y)} &=& - \frac{1}{2} K_{ll'}(x,y).
\end{eqnarray}
Therefore, in absence of sources, extremizing the generalized effective action $\Gamma(O_l, G_{ll'})$ gives the dynamics of both the operators and their Green's functions.

Such an effective action is usually considered for the elementary fields and their Green's functions in the literature. However, as discussed above we can construct the same for the set of gauge-invariant single trace operators in a non-Abelian gauge theory. 

\begin{figure}[ht]
\begin{center}
\includegraphics[width = 80mm]{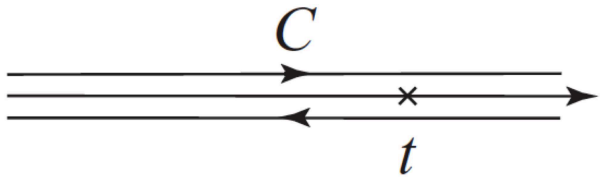}
\caption{The closed time Schwinger-Keldysh contour is as above. The forward and backward directed parts of the contour have been displaced slightly above and below the real axis just to distinguish them clearly. }
\label{Keldyshcontour}
\end{center}
\end{figure}

There is one important point in the above construction. The effective action is constructed over the so-called Schwinger-Keldysh real time contour (Fig. \ref{Keldyshcontour}), which travels from $-\infty$ to 
$\infty$ infinitesimially above the real line and then back from $\infty$ to $-\infty$ infinitesimially below the real line. It is necessary to consider this "closed-time" contour because the usual time-ordered Green's function or the Feynmann propagator do not contain the full information about the operator in presence of sources in a non-equilibrium state as mentioned in the Introduction. The closed-time contour ensures we propagate the full information of the operator in presence of the sources. In fact, the full closed-time contour ordered Green's function can be written as a combination of the commutator and the anti-commutator. For instance, if both operators are bosonic the 
\begin{eqnarray}
G_{\mathcal{C}ll'}(x,y) = \frac{1}{2}\langle\{\mathcal{O}_l(x), \mathcal{O}_{l'}(y)\}\rangle - \frac{i}{2}\langle[\mathcal{O}_l(x), \mathcal{O}_{l'}(y)]\rangle \, \text{sign}_{C}(x^{0}-y^{0}).
\end{eqnarray} 
Above $\mathcal{C}$ denotes the closed-time contour, and $x^0$ and $y^0$ are the time coordinates of the D-dimensional position vector $x$ and $y$ respectively.

In fact, as discussed in the Introduction, the spectral function $\mathcal{A}_{ll'}(x,y)$ is related to the commutator and the statistical function (or Keldysh propagator) $G_{\mathcal{K}ll'}(x,y)$ is related to the anti-commutator in the following way (for bosonic fields):
\begin{eqnarray}
\mathcal{A}_{ll'}(x,y) &=& i\langle[\mathcal{O}_l(x), \mathcal{O}_{l'}(y)]\rangle, \nonumber\\
G_{\mathcal{K}ll'}(x,y) &=& \frac{1}{2}\langle\{\mathcal{O}_l(x), \mathcal{O}_{l'}(y)\}\rangle .
\end{eqnarray}
The coupled equation of motion of the spectral and statistical functions are obtained from the generalized effective action. 

\textit{The generalized effective action has no dependence on temperature or non-equilibrium variables, it is defined as a Legendre transform of the vacuum persistence amplitude in the presence of single and double operator sources. However, thermal and non-equilibrium propagators also can be obtained as solutions which extremize this generalized effective action.} In order to obtain thermal propagators, we need to impose translational invariance, so the Wigner transformed spectral and statistical functions $\mathcal{A}_{ll'}(\omega,\mathbf{p}, \mathbf{x}, t)$ and $G_{\mathcal{K}ll'}(\omega,\mathbf{p}, \mathbf{x}, t)$ do not depend on the centre-of-mass coordinates $\mathbf{x}$ and $t$. Furthermore, they should be related by a temperature dependent fluctuation-dissipation relation.


\begin{thebibliography}{99}

\bibitem{adscft}
J.~M.~Maldacena, Int.\ J.\ Theor.\ Phys.\ {\bf 38}, 1113 (1999)]
  [arXiv:hep-th/9711200];
S.~S.~Gubser, I.~R.~Klebanov and A.~M.~Polyakov,
 Phys. Lett.  B {\bf 428}, 105 (1998)
  [arXiv:hep-th/9802109];
E.~Witten, Adv. Theor. Math. Phys.  {\bf 2}, 253 (1998)
  [arXiv:hep-th/9802150].

\bibitem{fluidgravity1}
G.~Policastro, D.~T.~Son, A.~O.~Starinets,
  JHEP {\bf 0209}, 043 (2002)
  [hep-th/0205052];
G.~Policastro, D.~T.~Son, A.~O.~Starinets,
  JHEP {\bf 0212}, 054 (2002)
  [hep-th/0210220].

\bibitem{Janik}
R.~A.~Janik and R.~B.~Peschanski, Phys. Rev.  D {\bf 73}, 045013 (2006)
  [arXiv:hep-th/0512162];
R.~A.~Janik,
 Phys. Rev. Lett. {\bf 98}, 022302 (2007)
  [hep-th/0610144];
M.~P.~Heller and R.~A.~Janik,
  Phys. Rev.  D {\bf 76}, 025027 (2007)
  [arXiv:hep-th/0703243].

\bibitem{fluidgravity2}
 R.~Baier, P.~Romatschke, D.~T.~Son, A.~O.~Starinets and
M.~A.~Stephanov,
  JHEP {\bf 0804}, 100 (2008)
  [arXiv:0712.2451 [hep-th]];
  M.~Natsuume and T.~Okamura, Phys.\ Rev.\  D {\bf 77}, 066014 (2008)
  [Erratum-ibid.\  D {\bf 78}, 089902 (2008)]
  [arXiv:0712.2916 [hep-th]].  
\bibitem{Sayantani}
 S.~Bhattacharyya, V.~E.~Hubeny, S.~Minwalla and M.~Rangamani,
 JHEP {\bf 0802}, 045 (2008)
  [arXiv:0712.2456 [hep-th]]; N.~Banerjee, J.~Bhattacharya, S.~Bhattacharyya, S.~Dutta, R.~Loganayagam and P.~Surowka,
  JHEP {\bf 1101}, 094 (2011)
  [arXiv:0809.2596 [hep-th]].

\bibitem{myself1}
  R.~Iyer, and A.~Mukhopadhyay,
 Phys. Rev.  D {\bf 81}, 086005 (2010)
  [arXiv:0907.1156 [hep-th]].

\bibitem{myself2}
R.~Iyer and A.~Mukhopadhyay,
  Phys.\ Rev.\ D {\bf 84}, 126013 (2011)
  [arXiv:1103.1814 [hep-th]].

\bibitem{myself3}
R.~Iyer and A.~Mukhopadhyay,
  PoS {\bf EPS-HEP2011}, 123 (2011)
  [arXiv:1111.4185 [hep-th]].


\bibitem{thermalization} 
  U.~H.~Danielsson, E.~Keski-Vakkuri and M.~Kruczenski,
  JHEP {\bf 0002}, 039 (2000)
  [hep-th/9912209];
  S.~B.~Giddings and A.~Nudelman,
  JHEP {\bf 0202}, 003 (2002)
  [hep-th/0112099];
   V.~Balasubramanian, A.~Bernamonti, J.~de Boer, N.~Copland, B.~Craps, E.~Keski-Vakkuri, B.~Muller and A.~Schafer {\it et al.},
  Phys.\ Rev.\ D {\bf 84}, 026010 (2011)
  [arXiv:1103.2683 [hep-th]];
  J.~Erdmenger, C.~Hoyos and S.~Lin,
  [arXiv:1112.1963 [hep-th]].

\bibitem{Florkowski}
  W.~Florkowski,
  ``\textit{Phenomenology of Ultra-Relativistic Heavy-Ion Collisions},'' World Scientific, 2010

\bibitem{Bleicher} 
  M.~Bleicher, E.~Zabrodin, C.~Spieles, S.~A.~Bass, C.~Ernst, S.~Soff, L.~Bravina and M.~Belkacem {\it et al.},
  J.\ Phys.\ G G {\bf 25}, 1859 (1999)
  [hep-ph/9909407].


\bibitem{Lee}
  S.~S.~Lee,
  Phys.\ Rev.\  D {\bf 79}, 086006 (2009)
  [arXiv:0809.3402 [hep-th]].


\bibitem{Liu}
  H.~Liu, J.~McGreevy, D.~Vegh, Phys.\ Rev.\  {\bf D83}, 065029 (2011)
  [arXiv:0903.2477 [hep-th]].

\bibitem{Cubrovic}
  M.~Cubrovic, J.~Zaanen, K.~Schalm, Science {\bf 325}, 439-444 (2009)
  [arXiv:0904.1993 [hep-th]].

\bibitem{Faulkner}
  T.~Faulkner, H.~Liu, J.~McGreevy, D.~Vegh, Phys.\ Rev.\  {\bf D83}, 125002 (2011)
  [arXiv:0907.2694 [hep-th]].


\bibitem{Hartnoll} 
  S.~A.~Hartnoll and A.~Tavanfar,
  Phys.\ Rev.\ D {\bf 83}, 046003 (2011)
  [arXiv:1008.2828 [hep-th]].



\bibitem{Sachdev} 
M.~Cubrovic, J.~Zaanen and K.~Schalm,
  JHEP {\bf 1110}, 017 (2011)
  [arXiv:1012.5681 [hep-th]];
  S.~Sachdev,
  Phys.\ Rev.\ D {\bf 84}, 066009 (2011)
  [arXiv:1107.5321 [hep-th]].

\bibitem{reviews}
 T. Kita, Prog. Theor. Phys. Vol. 123 No. 4 (2010) pp.581-658 [arXiv:1005.0393[cond-mat]]; J.~Berges, AIP Conf.\ Proc.\  {\bf 739}, 3 (2005) [arXiv:hep-ph/0409233].

\bibitem{Perfetti}
L. Perfetti et. al. , Phys. Rev. Lett. \textbf{97}, 067402 (2006)


\bibitem{progress1}
A. Mukhopadhyay, in progress


\bibitem{Herzog}
 C.~P.~Herzog and D.~T.~Son,
  JHEP {\bf 0303}, 046 (2003)
  [hep-th/0212072];
K.~Skenderis and B.~C.~van Rees, Phys.\ Rev.\ Lett.\  {\bf 101}, 081601 (2008) [arXiv:0805.0150 [hep-th]]; G.~C.~Giecold,
  JHEP {\bf 0910}, 057 (2009)
  [arXiv:0904.4869 [hep-th]].

 \bibitem{incoming1} 
  D.~T.~Son and A.~O.~Starinets,
  JHEP {\bf 0209}, 042 (2002)
  [hep-th/0205051]. 
\bibitem{incoming2}
N.~Iqbal and H.~Liu,
  Fortsch.\ Phys.\  {\bf 57}, 367 (2009)
  [arXiv:0903.2596 [hep-th]].


\bibitem{Arnold}
P.~Arnold, G.~D.~Moore and L.~G.~Yaffe, JHEP {\bf 0011}, 001 (2000)
  [arXiv:hep-ph/0010177].


\bibitem{Chapman}
S. Chapman and T. Cowling, ``\textit{The Mathematical Theory of
Non-Uniform Gases}'', Cambridge University Press, Cambridge,
England, 1960, Chapters 7, 8, 10, 15 and 17;  J. M. Stewart, \textit{Ph. D. dissertation}, University of Cambridge,
1969.
\bibitem{Grad}
H. Grad,  Comm. Pure Appl. Math., $\mathbf{2}$ (1949), 331-407.

\bibitem{Aharony} 
  O.~Aharony, S.~S.~Gubser, J.~M.~Maldacena, H.~Ooguri and Y.~Oz,
  Phys.\ Rept.\  {\bf 323}, 183 (2000)
  [hep-th/9905111].


\bibitem{myself4}
R.~K.~Gupta and A.~Mukhopadhyay,
  JHEP {\bf 0903}, 067 (2009)
  [arXiv:0810.4851 [hep-th]].

\bibitem{Gavai} 
  R.~V.~Gavai and S.~Gupta,
  Phys.\ Rev.\ D {\bf 71}, 114014 (2005)
  [hep-lat/0412035].

\bibitem{nonsusyhol}
I.~R.~Klebanov and A.~M.~Polyakov,
  Phys.\ Lett.\ B {\bf 550}, 213 (2002)
  [hep-th/0210114]; M.~R.~Gaberdiel and R.~Gopakumar,
  Phys.\ Rev.\ D {\bf 83}, 066007 (2011)
  [arXiv:1011.2986 [hep-th]].



\bibitem{hardwall} 
  J.~Erlich, E.~Katz, D.~T.~Son and M.~A.~Stephanov,
  Phys.\ Rev.\ Lett.\  {\bf 95}, 261602 (2005)
  [hep-ph/0501128];  L.~Da Rold and A.~Pomarol,
  Nucl.\ Phys.\ B {\bf 721}, 79 (2005)
  [hep-ph/0501218].


\bibitem{Starinets}
  A.~O.~Starinets,
 Phys. Rev. D  {\bf 66}, 124013 (2002).
  [hep-th/0207133].

\bibitem{semihol}
T.~Faulkner and J.~Polchinski,
  JHEP {\bf 1106}, 012 (2011)
  [arXiv:1001.5049 [hep-th]].
\bibitem{horizon} 
  S.~Bhattacharyya, V.~E.~Hubeny, R.~Loganayagam, G.~Mandal, S.~Minwalla, T.~Morita, M.~Rangamani and H.~S.~Reall,
  JHEP {\bf 0806}, 055 (2008)
  [arXiv:0803.2526 [hep-th]].

\bibitem{Horowitz}
  G.~T.~Horowitz, V.~E.~Hubeny, Phys. Rev. D {\bf 62}, 024027 (2000).
  [hep-th/9909056].


\bibitem{AQ}
V.~Balasubramanian, P.~Kraus and A.~E.~Lawrence,
  Phys.\ Rev.\ D {\bf 59}, 046003 (1999)
  [hep-th/9805171];
I.~R.~Klebanov and E.~Witten,
  Nucl.\ Phys.\ B {\bf 556}, 89 (1999)
  [hep-th/9905104].

\bibitem{BF}
P. Breitenlohner and D.Z. Freedman, Ann. Phys. $\mathbf{144}$ (1982) 249.






\bibitem{Tong} 
  S.~A.~Hartnoll, J.~Polchinski, E.~Silverstein and D.~Tong,
  JHEP {\bf 1004}, 120 (2010)
  [arXiv:0912.1061 [hep-th]]; L.~Y.~Hung, D.~P.~Jatkar and A.~Sinha,
  Class.\ Quant.\ Grav.\  {\bf 28}, 015013 (2011)
  [arXiv:1006.3762 [hep-th]].
\bibitem{Iqbal}
  N.~Iqbal and H.~Liu,
  Phys.\ Rev.\ D {\bf 79}, 025023 (2009)
  [arXiv:0809.3808 [hep-th]].

\bibitem{Rammer}
J. Rammer, ``\textit{Quantum field theory of non-equilibrium states}," Cambridge University Press, Cambridge, UK, 2007, Chapter 10.6.


\bibitem{no}
J. W. Negele and H. Orland, ``\textit{Quantum many-particle systems (Advanced Book classics)}," Westview Press, USA 1998, Chapter 6.

\bibitem{pwave}
P.~Basu, J.~He, A.~Mukherjee and H.~-H.~Shieh,
  Phys.\ Lett.\ B {\bf 689}, 45 (2010)
  [arXiv:0911.4999 [hep-th]]; M.~Ammon, J.~Erdmenger, V.~Grass, P.~Kerner and A.~O'Bannon,
  Phys.\ Lett.\ B {\bf 686}, 192 (2010)
  [arXiv:0912.3515 [hep-th]]; F.~Benini, C.~P.~Herzog and A.~Yarom,
  Phys.\ Lett.\ B {\bf 701}, 626 (2011)
  [arXiv:1006.0731 [hep-th]].


\bibitem{finiteT}
  T.~Faulkner, N.~Iqbal, H.~Liu, J.~McGreevy and D.~Vegh,
  arXiv:1101.0597 [hep-th].

\bibitem{progress2}
S. Banerjee, R. Iyer and A. Mukhopadhyay, in progress.
\end{thebibliography}
\end{document}